\documentclass[a4paper, amsfonts, amssymb, amsmath, reprint, showkeys, nofootinbib, twoside]{revtex4-2}
\usepackage[english]{babel}
\usepackage[utf8]{inputenc}
\usepackage{amsthm}
\usepackage{mathtools}
\usepackage{physics}
\usepackage{graphicx}
\usepackage[left=23mm,right=13mm,top=35mm,columnsep=15pt]{geometry} 
\usepackage{adjustbox}
\usepackage{placeins}
\usepackage[T1]{fontenc}
\usepackage{csquotes}
\usepackage{comment}
\usepackage[acronym,toc,shortcuts]{glossaries} \makeglossaries \newacronym{MRI}{MRI}{magnetic resonance imaging}
\newacronym{fMRI}{fMRI}{functional MRI}
\newacronym{rsfMRI}{rsfMRI}{resting-state fMRI}
\newacronym{EEG}{EEG}{electroencephalogram}
\newacronym{MEG}{MEG}{magnetoencephalography}
\newacronym{AD}{AD}{Alzheimer's disease}
\newacronym{PD}{PD}{Parkinson's disease}
\newacronym{CFC}{CFC}{cross-frequency coupling}
\newacronym{CF}{CF}{cross-frequency}
\newacronym{PAC}{PAC}{phase-amplitude coupling}
\newacronym{LFP}{LFP}{local field potentials}
\newacronym{NARX}{NARX}{nonlinear autoregressive with exogenous input}
\newacronym{ARX}{ARX}{autoregressive with exogenous input}
\newacronym{OLS}{OLS}{Orthogonal Least Squares}
\newacronym{iFRO}{iFRO}{iterative Forward Regression Orthogonal Least Squares}
\newacronym{PRESS}{PRESS-statistic}{prediction sum of squares statistic}
\newacronym{GFRF}{GFRF}{Generalised Frequency Response Function}
\newacronym{NOFRF}{NOFRF}{Nonlinear Output Frequency Response Function} 
\newacronym{MI}{MI}{Modulation Index}
\newacronym{PLV}{PLV}{Phase-Locking Value}
\newacronym{MVL}{MVL}{Mean Vector Length}
\newacronym{GLM-CFC}{GLM-CFC}{Generalized Linear Modelling Cross-Frequency Coupling}
\newacronym{SNR}{SNR}{signal-to-noise ratio}
\newacronym{NMM}{NMM}{neural mass model}
\newacronym{NMMs}{NMMs}{neural mass models}
\newacronym{QPC}{QPC}{quadratic phase coupling}
\newacronym{SISO}{SISO}{single-input single-output}
\newacronym{FFT}{FFT}{fast Fourier transform}
\newacronym{ECoG}{ECoG}{electrocorticography}
\newacronym{GLM}{GLM}{Generalized Linear Models}
\newacronym{ADHD}{ADHD}{Attention Deficit Hyperactivity Disorder}
\newacronym[shortplural={SSVEPs} , longplural={steady-state visual evoked potentials}]{SSVEP}{SSVEP}{steady-state visual evoked potential}
\newacronym[shortplural={ASSRs} , longplural={auditory steady-state responses}]{ASSR}{ASSR}{auditory steady-state response}

\usepackage{float}
\makeatletter
\let\newfloat\newfloat@ltx
\makeatother
\usepackage[svgnames]{xcolor}
\usepackage{algorithm}
\usepackage[rightComments=true, noEnd=false, italicComments=false, commentColor=Green, endLComment= ]{algpseudocodex}
\usepackage{mathrsfs}

\usepackage[pdftitle={Article}, pdfauthor={Author}]{hyperref} 

\begin{document}
\title{A Dynamical Systems and System Identification Framework for Phase–Amplitude Coupling Analysis}

\author{Rajintha Gunawardena}
    \affiliation{Centre for Computational Science and Mathematical Modelling, Coventry University, Coventry CV1 5FB, UK}
\author{Fei He}
    \email[Correspondence email address: ]{fei.he@coventry.ac.uk}
    \affiliation{Centre for Computational Science and Mathematical Modelling, Coventry University, Coventry CV1 5FB, UK}

\begin{abstract}
\section*{Abstract}
\Ac{PAC}, a form of cross-frequency coupling, is involved in diverse cognitive functions and neural communication, making its accurate detection and characterisation essential yet challenging. Most existing methods infer PAC from variations in instantaneous phase and amplitude profiles, but are limited by sensitivity to filter bandwidths, inconsistent performance across noise levels and data lengths, and vulnerability to spurious couplings. Here, we formulate \ac{PAC} as a nonlinear dynamical-systems phenomenon characterised by \ac{QPC}, and propose a nonlinear system identification framework that directly models the dynamics generating PAC. Rather than relying on filtered phase and amplitude fluctuations, the proposed method identifies a generative nonlinear model, enabling noise-free simulation of the estimated dynamics and model-based characterisation of coupling strength and preferred phase. It also provides dynamically grounded criteria for identifying harmonic- and intermodulation-related false detections, remains robust under high noise, and yields consistent characterisation despite changes in slow-frequency power. In simulations and rat hippocampal \ac{LFP} recordings, the proposed method produced more sharply localised and frequency-specific PAC estimates than benchmark filtering-based methods, while preserving detailed preferred-phase information.  In simulations, it rejected harmonic-related spurious coupling, remained robust at a \ac{SNR} of 2 and reasonably robust at an \ac{SNR} of 1, and reliably characterised PAC using 5-second analysis windows. These results establish nonlinear system identification as a complementary dynamical framework for detecting and characterising PAC, with particular advantages for noisy and short-duration neural recordings susceptible to spurious coupling.
\end{abstract}

\keywords{Phase-amplitude coupling, quadratic phase coupling, nonlinear dynamics, system identification, NARX}

\raggedbottom 
\maketitle
\section{Introduction} \label{sec:Intro}
%
Interactions between neural oscillations, both within and across brain regions, exhibit substantial complexity, even during resting states \cite{Deco2008, Ghosh2008, Florin2015, Florin2018}. Understanding how neural activity is coordinated across spatial and temporal scales is therefore a central question in neuroscience \cite{Buzsaki2006, Hyafil2015}. An important aspect of this coordination is that oscillatory neural activity can interact across distinct frequency bands through higher-order dynamical interactions \cite{CANOLTY2010, Tort2008, Tort2009, Deco2008, Cohen2008, Friston2000}. These interactions are commonly referred to as \ac{CFC}.

Evidence suggests that \ac{CFC} arises from the dynamical interactions between large-scale neural networks and local circuits, where activity in distributed networks and local neuronal processing mutually influence one another \cite{Buzsaki2006, Chehelcheraghi2017, Jirsa2013,JIANG2015,Nandi2019}. \ac{CFC} has therefore been proposed as a fundamental mechanism supporting communication between neural processes operating at different spatial and temporal scales and facilitating information integration across the brain \cite{Jensen2007}. Consistent with this view, studies using \ac{MEG} and \ac{EEG} have identified \ac{CFC} across a range of cognitive and sensory processes \cite{Jirsa2013, Cohen2008, Florin2015, Sadaghiani2022, Hyafil2015, Friese2013}. Furthermore, alterations in \ac{CFC} have been associated with a range of neurological and psychiatric conditions. These include neurodegenerative diseases such as \ac{AD} and \ac{PD} \cite{Wang2021, Fraga2013, Jackson2019, Muthuraman2020, KLEPL2023, ChenXi2023}, neurodevelopmental disorders such as \ac{ADHD} and autism spectrum disorder \cite{Mariscal2021, Efstratia2023, TANG2025}, and various psychiatric disorders \cite{Hirano2018, Murphy2020, Sacks2021, YAKUBOV2022, DePieri2025, Wang2025}. These findings highlight the potential of \ac{CFC} for characterising altered neural dynamics in health and disease.

Amplitude–amplitude coupling, phase–phase coupling, and phase–amplitude coupling (PAC) represent distinct forms of \ac{CFC} \cite{Sadaghiani2022}. While amplitude–amplitude and phase–phase coupling have been observed, \ac{PAC} has attracted particular interest within brain electrophysiology. As the most commonly observed form of \ac{CFC}, \ac{PAC} is directly involved in motor functions, cognitive processes, attention, sensory processing, and consciousness \cite{Lakatos2008, Tort2009, Axmacher2010, Yanagisawa2012, Roux2014, Seymour2017, CHACKO2018, FIEBELKORN2019, Dong2022, ESGHAEI2022, Aurimas2025}. Consequently, \ac{PAC} analysis offers a powerful means of probing the complex cross-frequency interactions underpinning brain activity \cite{Sadaghiani2022}, thereby deepening our understanding of the neural mechanisms driving cognitive and sensory functions, as well as the pathophysiology of neurological and psychiatric disorders \cite{Axmacher2010, FIEBELKORN2019, Sacks2021, Wang2021, Jackson2019, YAKUBOV2022, TANG2025, Wang2025}. Conceptually, \ac{PAC} refers to a cross-frequency interaction in which the phase of a low-frequency neural oscillation is associated with, or modulates, the amplitude envelope of a higher-frequency oscillation \cite{Buzsaki2006, Chehelcheraghi2017, Jirsa2013,JIANG2015,Nandi2019}.
%
\subsection{Limitations of current methods and the need for a dynamical-systems approach}
Metrics such as the \ac{MI} \cite{Tort2010}, \ac{MVL} \cite{Canolty2006}, \ac{GLM} \cite{PENNY2008} and their variants \cite{OZKURT2011,KRAMER2013,VANWIJK2015,Jurkiewicz2021} have been widely applied to detect and characterise \ac{PAC}. However, reliable detection and interpretation of \ac{PAC} signals remain challenging, and these approaches are susceptible to spurious \ac{PAC} detections \cite{KRAMER2008,DVORAK2014}. This difficulty arises from factors such as non-sinusoidal waveforms, noise, and sharp edges or spikes in the data, all of which compromise the filtering-based analysis procedures underlying many commonly used \ac{PAC} metrics \cite{KRAMER2008,DVORAK2014,ARU2015,Gerber2016,Jensen2016,Dellavale2020,GIEHL2021}. In addition, methodological considerations such as the inappropriate selection of filter bandwidths (e.g. overly narrow or excessively broad) and analysis window lengths can lead to inaccurate measurements and spurious detections \cite{CANOLTY2010,DVORAK2014,ARU2015}. These limitations motivate the development of methods that characterise the dynamics generating \ac{PAC} more directly, rather than relying solely on instantaneous phase and amplitude-envelope fluctuations. 

\ac{PAC} is thought to arise from interactions between two distinct neural populations, where the phase of slower neural oscillations in one population is linked to the amplitude or power variations of faster neural oscillations in another \cite{ONSLOW2011, Hyafil2015, DVORAK2014}. This interaction reflects a dynamical process that is characterised by a specific form of nonlinearity, known as quadratic phase coupling (\ac{QPC}) \cite{hasselmann1963, Sigl1994, WITTE2000, Hyafil2015b, Yang2016, KOVACH2018}. However, conventional \ac{PAC} metrics typically overlook these underlying nonlinear dynamics and instead focus on temporal variations in signal profiles (i.e. amplitude and phase). Consequently, accurately detecting genuine \ac{PAC} may require explicitly accounting for the nonlinear dynamics, in particular \ac{QPC}, that give rise to it \cite{KRAMER2008, VELARDE2019, Dellavale2020}. This highlights the need for data-driven methodologies, grounded in nonlinear dynamical systems theory, that can directly identify and characterise the dynamics generating the coupling. 

Higher-order spectral analyses, such as bicoherence and the bispectrum, which are capable of capturing \ac{QPC}, have been recommended as complementary tools to filtering-based metrics including \ac{MI}, \ac{MVL}, and \ac{GLM} for validating genuine \ac{PAC} \cite{KRAMER2008,DVORAK2014,Hyafil2015b,KOVACH2018}. However, while these methods provide useful higher-order spectral descriptions of \ac{PAC}-related interactions, they are not by themselves sufficient for reliable \ac{PAC} detection and characterisation, particularly when signals are short, noisy, or strongly non-stationary \cite{Elgar1988,Siu2008,KOVACH2018}. Additionally, caution must be taken to avoid spurious \ac{PAC} arising from harmonics of the fundamental frequency of non-sinusoidal slow oscillations, a phenomenon known as harmonic \ac{PAC} \cite{GIEHL2021}. These limitations motivate a model-based framework that captures the underlying nonlinear dynamics directly while retaining specificity to \ac{PAC}.
%
\subsection{Proposed framework and contributions}
The principal novelty of this study is a \ac{PAC}-specific dynamical-systems framework that uses nonlinear system identification to recover the quadratic phase-coupled structure underlying PAC directly from neural time-series data, rather than inferring coupling solely from filtered phase and amplitude profiles. Specifically, we introduce a nonlinear system-identification methodology based on the \ac{NARX} model structure \cite{Chen1989a} and an \ac{OLS}-based identification framework \cite{CHEN1989b}, particularly the \ac{iFRO} algorithm \cite{guo2015a}. The proposed approach represents \ac{PAC} as a quadratic phase-coupled interaction between slow and fast oscillations and identifies a compact generative model that approximates its basic (canonical) dynamical form. This enables noise-free simulation of the estimated coupling dynamics, model-based estimation of modulation strength and preferred phase, and improved discrimination between genuine and spurious detections, including false detections arising from harmonics and intermodulation-related effects. \ac{NARX} and related NARMAX methods are widely used for modelling nonlinear systems across diverse disciplines \cite{billings2013a,Chiras2002,WANG2024,ZAINOL2022,RITZBERGER2017,Gao2023,HE2016,HE2021, LIU2024}. As demonstrated, the proposed \ac{NARX}-based methodology provides a system-identification-based dynamical-systems framework for \ac{PAC} detection and characterisation. It performs well under high-noise conditions and with short time windows, while remaining robust to factors that commonly produce spurious detections in conventional metrics. Dynamically motivated rejection procedures \cite{Fabus2022} can be naturally incorporated and extended within this framework as the mechanisms underlying false PAC detections become better understood.

The remainder of this paper is organised as follows. Section \ref{sec:PAC_gen} reviews \ac{PAC}, focusing on the mechanisms underlying its generation, analytical and neural mass models of \ac{PAC}, current \ac{PAC} methods, and the problem of spurious coupling. Section \ref{sec:Method} develop the proposed framework by introducing \ac{QPC} as the dynamical basis of \ac{PAC}, deriving a canonical representation of \ac{PAC}, and presenting a \ac{NARX}-based system identification framework for \ac{PAC} detection and characterisation. The experimental section evaluates the proposed method against commonly used \ac{PAC} metrics using both synthetic and real neural recordings. Finally, we discuss the findings, limitations, and future directions.
\section{Phase-Amplitude Coupling: Analytical models, current methods and spurious coupling} \label{sec:PAC_gen}
\begin{figure}[!htb]
	\centerline{\includegraphics[scale=0.41]{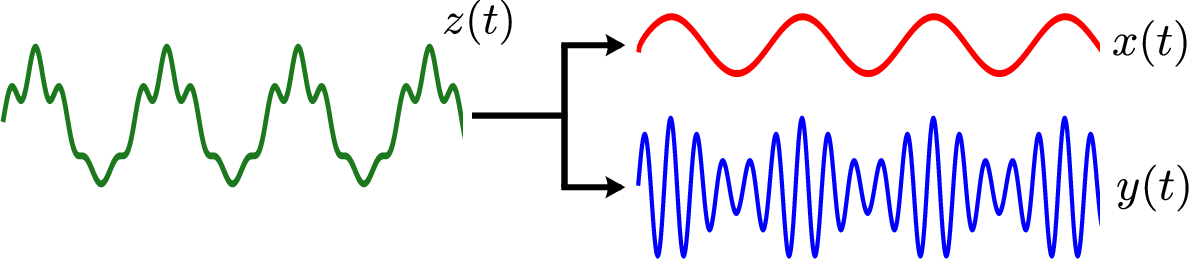}}
	\caption{\textbf{Decomposition of a general \ac{PAC} signal.} A \ac{PAC} signal, $z(t)$, can be decomposed into two components: a low-frequency oscillation, $x(t)$, and a high-frequency oscillation, $y(t)$, with a modulated amplitude. The amplitude envelope of $y(t)$, $ A_{y}(t)$, is coupled to the phase of $x(t)$, $ \varphi_{x}(t)$.}
	\label{fig:pac_decom}
\end{figure}

Global and local processes in the brain are thought to interact through \acs{CFC}, facilitating the integration of information across distributed brain regions \cite{Jensen2007}. This interaction is inherently dynamic, with oscillatory components at different frequencies influencing one another \cite{Jensen2007, Jirsa2013, Chehelcheraghi2017}. Among the various forms of \ac{CFC}, \ac{PAC} is one of the most widely studied. In a \ac{PAC} signal $z(t)$, the amplitude envelope (power fluctuations, $A_y(t)$) of the higher-frequency oscillation $y(t)$ is coupled to the instantaneous phase ($\varphi_x(t)$) of the lower-frequency oscillation $x(t)$ \cite{CANOLTY2010, ARU2015, GIEHL2021}, such that $A_y(t)$ varies systematically with $\varphi_x(t)$. Figure \ref{fig:pac_decom} illustrates this decomposition into its basic components: a slow (low-frequency) oscillation, $x(t)$, and an amplitude-modulated fast (high-frequency) oscillation, $y(t)$. 
\begin{figure}[!htb]
	\centerline{\includegraphics[scale=0.41]{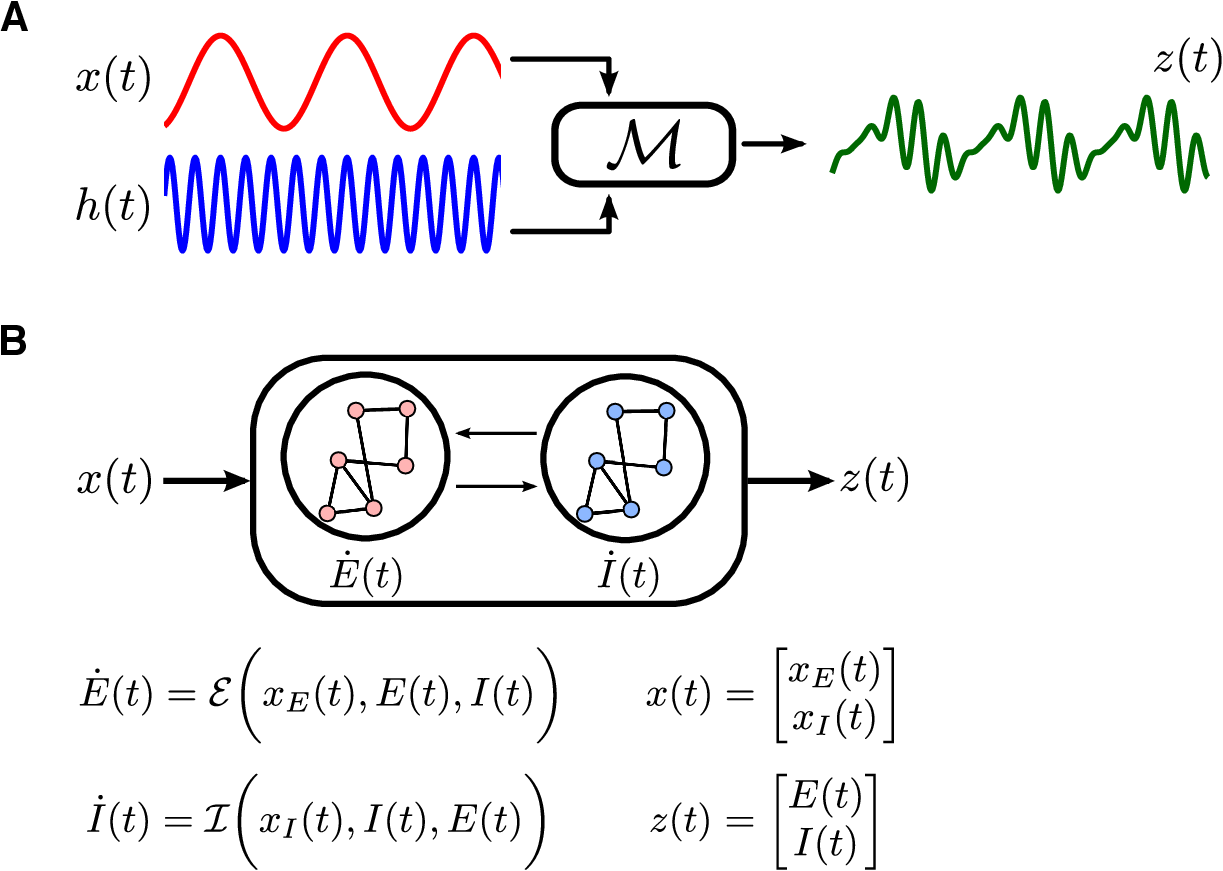}}
	\caption{\textbf{Mechanisms and models for generating \ac{PAC}.} \textbf{A} and \textbf{B} illustrate two model classes capable of generating \ac{PAC} dynamics. \textbf{A} shows a simple analytical model of a \ac{PAC} signal. In these models, a slow oscillation, $x(t)$, and a fast oscillation, $h(t)$, are combined through a nonlinear function in model $\mathcal{M}$ to produce the \ac{PAC}, $z(t)$. \textbf{B} illustrates coupled differential equations that generate \ac{PAC} signals resembling those observed in neural activity. Here, the mean activity of two interacting neural populations, $E(t)$ and $I(t)$, is modelled using differential equations, $\mathcal{E}\left( \ \right)$ and $\mathcal{I}\left( \ \right)$, respectively. $x_E(t)$ and $x_I(t)$ denote the slow oscillation inputs to the neural populations $E$ and $I$. The fast oscillation dynamics are generated internally due to the activity between $E(t)$ and $I(t)$. An explicit link can be established between these two types of models as described in Section \ref{sec:NMM}.}
	\label{fig:gen_pac}
\end{figure}

The \ac{PAC} phenomenon has been observed in brain signals recorded using \ac{LFP}, \ac{EEG}, and \ac{MEG}, across regions such as the auditory and prefrontal cortices and the hippocampus \cite{Voloh2015, Bragin1995, Axmacher2010}. It has been linked to cognitive and sensory processing as well as altered neural dynamics in neurological and psychiatric disorders \cite{Axmacher2010, Roux2014, Seymour2017, ESGHAEI2022, Aurimas2025, YAKUBOV2022, TANG2025, Wang2025}.

\ac{PAC} has been investigated using models of varying complexity, ranging from simple nonlinear functions \cite{Hyafil2015b, JIANG2015} (Fig. \ref{fig:gen_pac}A) to coupled differential equations \cite{Deco2008, Onslow2014, Chehelcheraghi2017} (Fig. \ref{fig:gen_pac}B). Although their underlying mechanisms differ, the resulting \ac{PAC} signal can generally be decomposed into a slow oscillation and a fast oscillation whose amplitude is modulated by the slow component, as illustrated in Figure \ref{fig:pac_decom}. Here, the constituent oscillations may themselves be non-sinusoidal, non-stationary, or broadband. The following subsections therefore focus on analytical and neural mass models of PAC to identify the dynamical features most relevant to the proposed dynamical-systems framework. This is followed by a brief review of commonly used PAC metrics and spurious \ac{PAC}.
\subsection{Simple analytical models of PAC}\label{sec:PAC_models}
Simple analytical models are useful because they explicitly describe the dynamical, temporal and spectral properties of the amplitude modulation involved in \ac{PAC}, in particular \ac{QPC}, modulation depth, and the preferred phase of the coupling.
\begin{figure*}
  \centering
  \includegraphics[width=\textwidth]{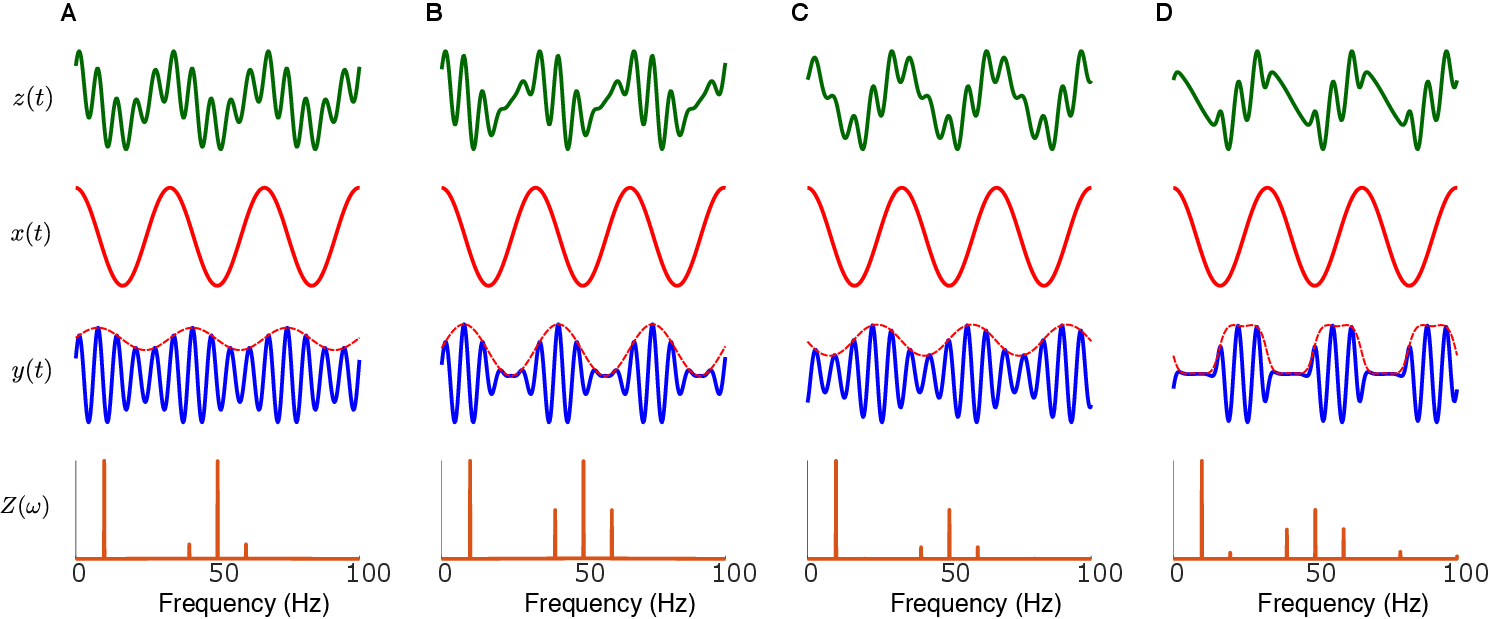} 
  \caption{\textbf{Effects of varying modulation parameters in simple \ac{PAC} models.} The generated \ac{PAC} signals, $z(t)$ are shown in green, the $10$ Hz slow oscillation $x(t)$ in red, the amplitude-modulated $50$ Hz fast oscillation $y(t)$ in blue, its envelope as dotted red lines, and the magnitude spectrum $Z(\omega)$ in orange. \textbf{A} and \textbf{B} illustrate the effect of coupling strength in the basic \ac{PAC} model (equations \eqref{eq:PAC_math} and \eqref{eq:simple_PAC}) for $m=0.5$ and $m=1$, respectively. Similarly, \textbf{C} and \textbf{D} show the effect of varying $\alpha$ in the more complex \ac{PAC} model (equations \eqref{eq:PAC_math} and \eqref{eq:nonsine_PAC}), with $\alpha=1$ in \textbf{C}, $\alpha=6$ in \textbf{D}, and $c=10^{-6}$ in both. In all cases, immediate sidebands appear at $40$Hz and $60$Hz. Comparison of \textbf{A} to \textbf{B} and \textbf{C} to \textbf{D} shows that greater modulation depth corresponds to larger sideband (intermodulation) magnitudes. The more complex modulation in \textbf{D} produces a non-sinusoidal envelope and additional sidebands.}
  \label{fig:simple_gen_model}
\end{figure*}
\begin{figure*}
  \centering
  \includegraphics[width=\textwidth]{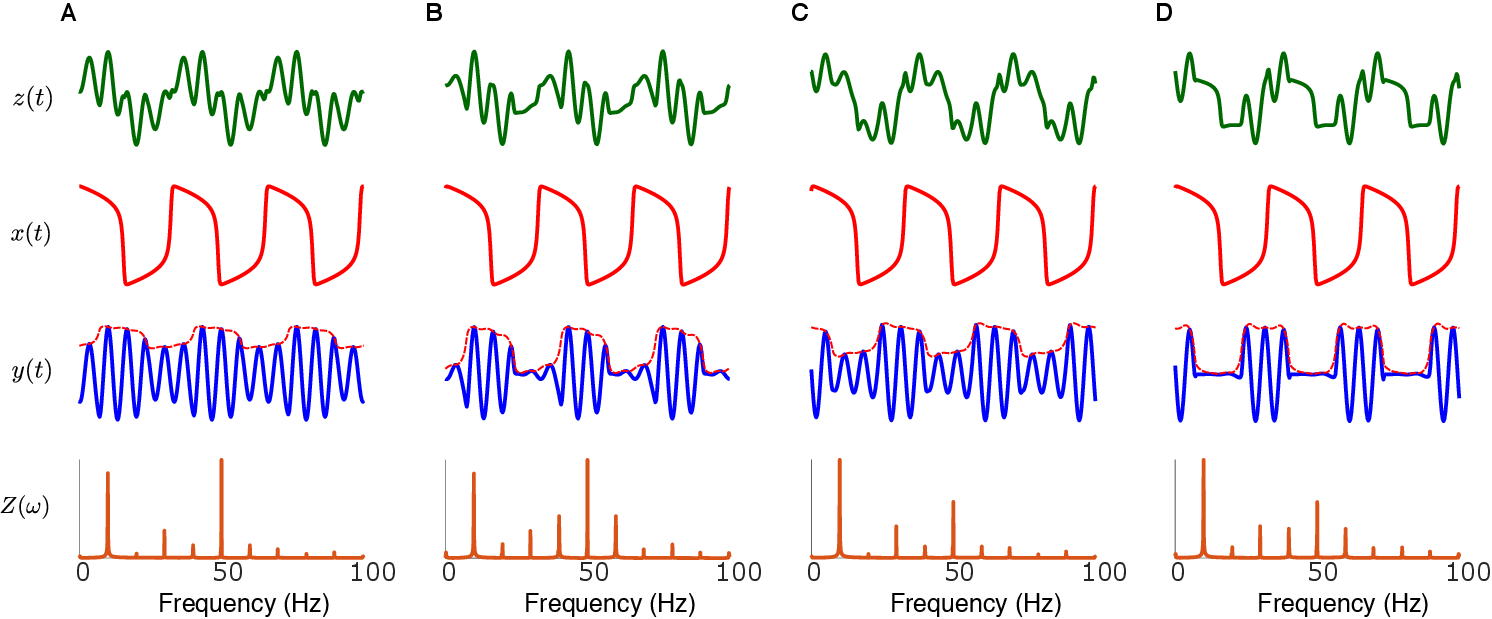} 
  \caption{\textbf{Effects of non-sinusoidal input and modulation parameters in simple \ac{PAC} models.} The generated \ac{PAC} signals, $z(t)$ are shown in green, the $10$ Hz slow oscillation $x(t)$ in red, the amplitude-modulated $50$ Hz fast oscillation $y(t)$ in blue, its envelope as dotted red lines, and the magnitude spectrum $Z(\omega)$ in orange. The figure uses the same modulation parameters as Figure \ref{fig:simple_gen_model}, but with a non-sinusoidal slow oscillation. \textbf{A} and \textbf{B} show the basic \ac{PAC} model (equations \eqref{eq:PAC_math} and \eqref{eq:simple_PAC}) for $m=0.5$ and $m=1$, respectively. Compared with Figure \ref{fig:simple_gen_model}A--B, harmonics in the non-sinusoidal slow oscillation generate additional frequencies, and hence extra intermodulations; the effect of the latter is more pronounced in \textbf{B} because modulation depth is proportional to intermodulation magnitude. \textbf{C} and \textbf{D} show the more complex \ac{PAC} model (equations \eqref{eq:PAC_math} and \eqref{eq:nonsine_PAC}) for $\alpha=1$ and $\alpha=6$, respectively, with $c=10^{-6}$ in both.}
  \label{fig:simple_gen_model_vdp}
\end{figure*}
\begin{figure*}
  \centering
  \includegraphics[width=\textwidth]{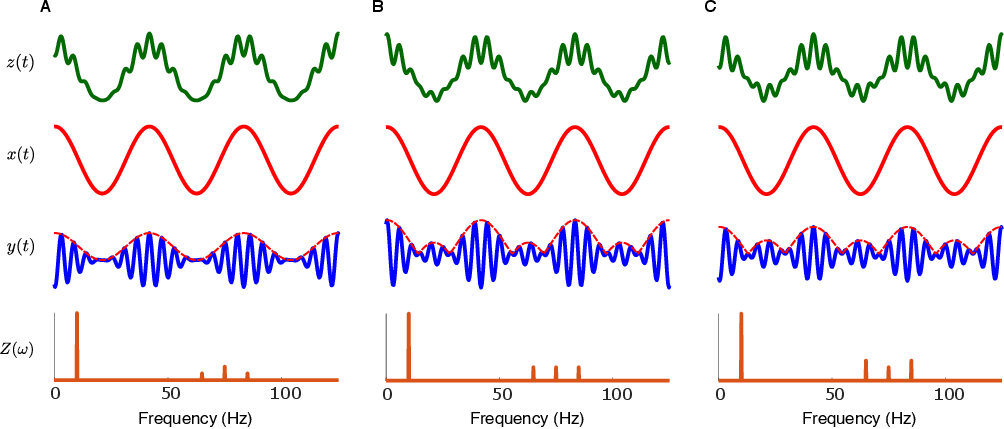} 
  \caption{\textbf{Monophasic and biphasic \ac{PAC}.} This figure illustrates the transition from monophasic (\textbf{A}) to biphasic (\textbf{B} and \textbf{C}) \ac{PAC} using the basic \ac{PAC} model, equations \eqref{eq:PAC_math} and \eqref{eq:simple_PAC}. The generated \ac{PAC} signals, $z(t)$, are depicted in green, while the $10$Hz sinusoidal slow oscillation, $x(t)$, is shown in red. The corresponding amplitude-modulated $75$Hz fast oscillations, $y(t)$, with the amplitude envelope shown in dotted red lines, are depicted in blue. The magnitude spectrum of $z(t)$, $Z(\omega)$ is shown in orange. The modulation parameter, $m$, in equation \eqref{eq:simple_PAC} is increased from its level in \textbf{A}, $m=1$, to transition from monophasic to biphasic \ac{PAC} in \textbf{B}, $m=2$. After this point, a further increase in $m$  increases the depth of modulation at the second phase. This is shown in \textbf{C}, where $m=3$.}
  \label{fig:simple_gen_model_biphasic}
\end{figure*}

Simple \ac{PAC} models (Fig. \ref{fig:gen_pac}A) can be represented as
\begin{equation}\label{eq:PAC_math}
    \begin{cases}
        z(t) = x(t) + y(t + \delta t) \\[5pt]
        y(t) = f\Big( x(t), h(t) \Big),
    \end{cases}
\end{equation}
where $z(t)$ is the \ac{PAC} signal, $x(t)$ is the slow oscillation and $y(t)$ is the amplitude modulation of the fast oscillation $h(t)$. $f( \ )$ denotes a nonlinear function that modulates the amplitude of $h(t)$ according to the phase of $x(t)$. One of the simplest forms is
\begin{equation}\label{eq:simple_PAC}
    y(t) = \Big( 1 + m \times x(t) \Big) h(t),
\end{equation}
where the parameter $m$ controls the modulation depth, and hence the effective coupling strength, between the slow and fast oscillations. A more complex form,
\begin{equation}\label{eq:nonsine_PAC}
    y(t) = \left( 1  - \frac{1}{ 1 + \exp\Big( -\alpha\Big( x(t) - c \Big) \Big)} \right) h(t),
\end{equation}
adapted from \cite{JIANG2015}, can generate non-sinusoidal amplitude modulations \cite{Prendergast2010}. The parameters $\alpha$ and $c$ jointly determine the form and strength of the modulation. The preferred phase can be shifted by introducing a delay of $\delta t$ seconds in $y(t)$ relative to $x(t)$ \cite{JIANG2015}, as shown in equation \eqref{eq:PAC_math}. Here, the preferred phase denotes the phase of the slow oscillation at which fast oscillatory activity tends to peak \cite{Tort2010, CHACKO2018}.

Figures \ref{fig:simple_gen_model} and \ref{fig:simple_gen_model_vdp} illustrate the corresponding temporal and spectral properties of these models. The 10Hz non-sinusoidal slow oscillation in Figure \ref{fig:simple_gen_model_vdp} is generated using the Van der Pol oscillator, 
\begin{equation}\label{eq:van_der_pol}
    \frac{\mathrm{d}^2 x}{\mathrm{d} t^{2}}(t) - \mu(1-x^2)\frac{\mathrm{d} x}{\mathrm{d} t}(t) + x(t) = 0,
\end{equation}
which should be simulated with $\mu = 116.5$, and initial conditions ($t=0$), $\mathrm{d}^2 x / \mathrm{d} t^{2} = 2$ and $\mathrm{d} x / \mathrm{d} t = 1$.

In general, amplitude modulation in \ac{PAC} produces sidebands (i.e. intermodulation components) around the fast oscillation frequency, $h(t)$ (see $Z(\omega)$ in Fig. \ref{fig:simple_gen_model}A and B), reflecting the nonlinear interaction between $h(t)$ and $x(t)$ that generates $y(t)$. Modulation depth, or coupling strength, quantifies the degree of variation in the amplitude envelope of $y(t)$ \cite{Tort2010}. As the coupling parameters increase ($m$ and $\alpha$ in equations \eqref{eq:simple_PAC} and \eqref{eq:nonsine_PAC}, respectively), the modulation depth increases, accompanied by an increase in the magnitude of the corresponding sidebands relative to the fast-frequency component (compare Fig. \ref{fig:simple_gen_model}A--B and \ref{fig:simple_gen_model}C--D). The sidebands immediately adjacent to the high-frequency component are referred to as immediate sidebands, and their magnitudes relative to that of the fast-frequency component provide a direct measure of modulation depth. This relationship is central to the proposed framework because it connects the temporal property of \ac{PAC} to a compact spectral signature that can be recovered from an identified dynamical model.
\begin{figure*}
  \centering
  \includegraphics[width=\textwidth]{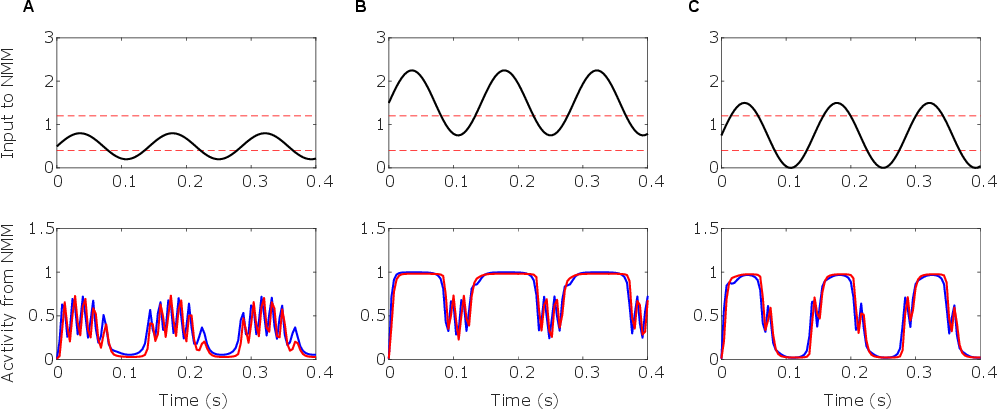} 
  \caption{\textbf{Different forms of PAC induced by a 7Hz theta input to the excitatory population of an \ac{NMM}.} The following parameters were used for equations \eqref{eq:NMM_E}-\eqref{eq:NMM_sig} as given in \cite{Onslow2014}; $\tau_E = \tau_I = 0.0032$, $w_{EE} = 2.4$, $w_{IE} = w_{EI} = 2$, $b = 4$. The inhibitory input $x_I(t)=0$. The excitatory input is the 7 Hz theta sinusoid $x_E(t)=A_E\cos(2\pi \times 7 \times t)+\overline{x}_E$ (black), while the red and blue curves show inhibitory and excitatory activity, respectively. This \ac{NMM} produces intrinsic gamma activity when $0.4 \leq x_E(t) \leq 1.2$ (red dotted lines in the top panels). \textbf{A} $A_E=0.3$ and $\overline{x}_E=0.5$: gamma bursts align with the theta peak. \textbf{B} $A_E=0.75$ and $\overline{x}_E=1.5$: gamma bursts align with the theta trough. \textbf{C} $A_E=0.75$ and $\overline{x}_E=0.75$: gamma bursts occur on both the rising and falling phases, because $x_E(t)$ crosses $0.4 \leq x_E(t) \leq 1.2$ twice per theta cycle.}
  \label{fig:NMM_sine}
\end{figure*}

Non-sinusoidal amplitude modulation (Fig. \ref{fig:simple_gen_model}D) can generate additional intermodulation components, whereas a non-sinusoidal slow oscillation (Fig. \ref{fig:simple_gen_model_vdp}) generates harmonics that may, in turn, produce further intermodulations (see $y(t)$ in Fig. \ref{fig:simple_gen_model_vdp}A--B). These additional spectral components contribute to the detailed waveform and spectral structure of the observed PAC signal (Fig. \ref{fig:simple_gen_model}C--D and Fig. \ref{fig:simple_gen_model_vdp}). However, as discussed later in Section \ref{sec:Dynm_sys_PAC} in the context of nonlinear dynamics, they are not necessary for \ac{PAC} generation. Moreover, harmonic and additional intermodulation components may themselves contribute to spurious \ac{PAC} detection, as discussed in Section \ref{sec:spuriousPAC}.

\ac{PAC} can be broadly classified as monophasic or biphasic \cite{Tort2010}. In monophasic \ac{PAC}, the amplitude of the high-frequency oscillation is modulated by a single phase of the low-frequency oscillation, producing one burst of high-frequency activity per low-frequency cycle (Fig. \ref{fig:simple_gen_model_biphasic}A). In biphasic \ac{PAC}, modulation occurs at two phases approximately separated by $180^\circ$ ($\pi$ radians), such as the peak and trough of the low-frequency oscillation, producing two bursts per cycle (Fig. \ref{fig:simple_gen_model_biphasic}B-C). Spectrally, biphasic \ac{PAC} is characterised by sidebands whose magnitudes can approach or exceed that of the high-frequency component (Fig. \ref{fig:simple_gen_model_biphasic}B--C).

While the analytical models discussed above provide insight into the basic temporal and spectral mechanisms underlying \ac{PAC} generation, they do not capture the emergence of \ac{PAC} arising from dynamic interactions between distinct neural populations. An \ac{NMM} provides a physiological description of such interactions by representing the collective dynamics of neuronal populations through coupled differential equations \cite{Onslow2014}.
%
\subsection{Neural mass models for generating PAC}\label{sec:NMM}
An \ac{NMM} simulates \ac{PAC} as an emergent population-level neural phenomenon rather than as an externally imposed modulation. To illustrate this, we consider the simplified Wilson--Cowan \ac{NMM} proposed in \cite{Onslow2014}, in which \ac{PAC} arises through interactions between excitatory and inhibitory neural populations. The model represents the mean activity of an excitatory population, $E$, and an inhibitory population, $I$, as
\begin{equation}\label{eq:NMM_E}
    \tau_E \dot{E}(t) = -E(t) + f\Big(x_E(t) + w_{EE} E(t) - w_{IE} I(t)\Big),
\end{equation}
\begin{equation}\label{eq:NMM_I}
    \tau_I \dot{I}(t) = -I(t) + f\Big(x_I(t) + w_{EI} E(t)\Big).
\end{equation}
In both equations \eqref{eq:NMM_E} and \eqref{eq:NMM_I},  a sigmoid function, 
\begin{equation}\label{eq:NMM_sig}
    f(z) = \frac{1}{1 + e^{-b(z-1)}},
\end{equation}
is used as an activation function for both $E$ and $I$ neural populations.

The reciprocal coupling terms $w_{IE} I(t)$ and $w_{EI} E(t)$, together with the excitatory self-feedback term $w_{EE} E(t)$ and the external inputs $x_E(t)$ and $x_I(t)$, enable intrinsic fast oscillations that can become modulated by the slow input rhythm. Varying the amplitude and mean level of the slow input can shift the preferred phase of the coupling and produce monophasic or biphasic \ac{PAC} (Fig. \ref{fig:NMM_sine}) \cite{Onslow2014}. Importantly, this population-level \ac{NMM} can be linked explicitly to the simpler analytical \ac{PAC} model introduced in equation \eqref{eq:PAC_math}. As shown in Appendix \ref{appndx:taylor_approx_NMM}, a Taylor expansion of the sigmoid activation function $f(z)$ (equation \eqref{eq:NMM_sig}) in equation \eqref{eq:NMM_E} yields a form in which the \ac{NMM} output can be interpreted as an internally generated fast dynamics, $h(t)=w_{EE}E(t)-w_{IE}I(t)$, whose amplitude is modulated by the slow input $x_E(t)$. The resulting activity can therefore be decomposed into a slow component, $x(t)$, and an amplitude-modulated fast component, $y(t)$, consistent with the general PAC decomposition in Figure \ref{fig:pac_decom}. Although analytical PAC models and NMMs differ in their physiological detail, both can be reduced to a similar dynamical form, in which a slow oscillation interacts nonlinearly with a faster oscillatory process to produce amplitude modulation. This shared structure motivates the canonical representation of PAC developed in Section III.
%
\subsection{Common metrics used for measuring PAC}\label{sec:Crrnt_mthds_lmts}
\begin{figure}[!h]
	\centerline{\includegraphics[scale=0.44]{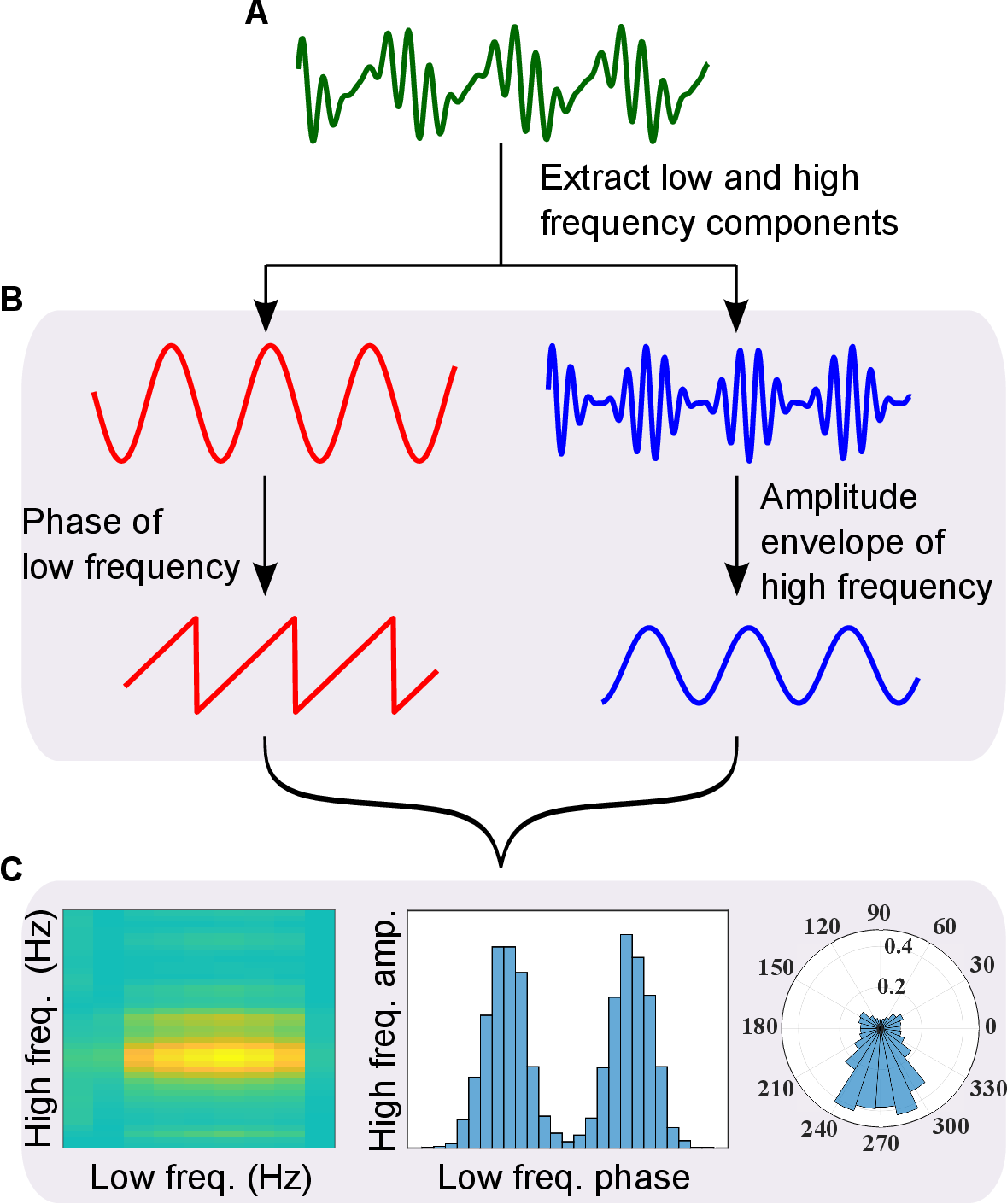}}
	\caption{\textbf{General procedures in standard methods for detecting \ac{PAC}}. The \ac{PAC} signal (\textbf{A}) is decomposed or bandpass-filtered to obtain the low-frequency and high-frequency oscillations (\textbf{B}). As shown in \textbf{B}, the phase of the low-frequency oscillation is estimated, and the amplitude envelope of the high-frequency oscillation is extracted. Depending on the metric used, one then assesses whether the variation of the amplitude envelope is synchronised with the phase of the low-frequency oscillation. Regardless of the metric used, this information is usually visualised in a comodulogram (first panel in \textbf{C}). The second panel in \textbf{C} shows how the fast oscillation’s amplitude varies with the slow oscillation’s phase, as used in the \ac{MI} metric \cite{Tort2010}. The third panel illustrates the vector representation used for the \ac{MVL} metric \cite{Canolty2006}, where each vector’s angle is given by the slow oscillation phase, and its length is the corresponding fast oscillation amplitude at a certain time instance.}
	\label{fig:PAC_stnd_procd}
\end{figure}

Most commonly used \ac{PAC} metrics quantify the relationship between the instantaneous phase of a low-frequency oscillation and the amplitude envelope of a higher-frequency oscillation. The standard workflow involves bandpass filtering the signal, estimating the low-frequency instantaneous phase and the high-frequency amplitude envelope, often through the Hilbert transform, and then evaluating the dependency between these two quantities (Fig. \ref{fig:PAC_stnd_procd}) \cite{Tort2010, Hulsemann2019}. Widely used measures include the \ac{MI}, \ac{PLV}, \ac{MVL}, and \ac{GLM-CFC} \cite{Hulsemann2019}.

Although these methods all aim to quantify \ac{PAC}, they differ in formulation and in their sensitivity to factors such as coupling type, noise, data length, and sampling rate \cite{Tort2010, ONSLOW2011, Hulsemann2019}. In general, \ac{MI} and \ac{GLM-CFC} have been reported to be more robust to noise and to perform well for biphasic coupling, whereas \ac{MVL} can be more sensitive under higher \ac{SNR} conditions, and \ac{PLV} is less robust in some settings \cite{Hulsemann2019}. No single metric is therefore optimal across all conditions, and the choice of method should be guided by the characteristics of the data and the specific analysis objective \cite{CANOLTY2010, ONSLOW2011, Hulsemann2019}. As discussed in the next section, these techniques can also falsely suggest the presence of \ac{PAC} \cite{KRAMER2008, DVORAK2014}. This motivates both careful interpretation of conventional PAC measures and methods that more directly characterise the nonlinear dynamics underlying the observed coupling.
%
\subsection{Spurious PAC}\label{sec:spuriousPAC}

\begin{figure*}
  \centering
  \includegraphics[width=\textwidth]{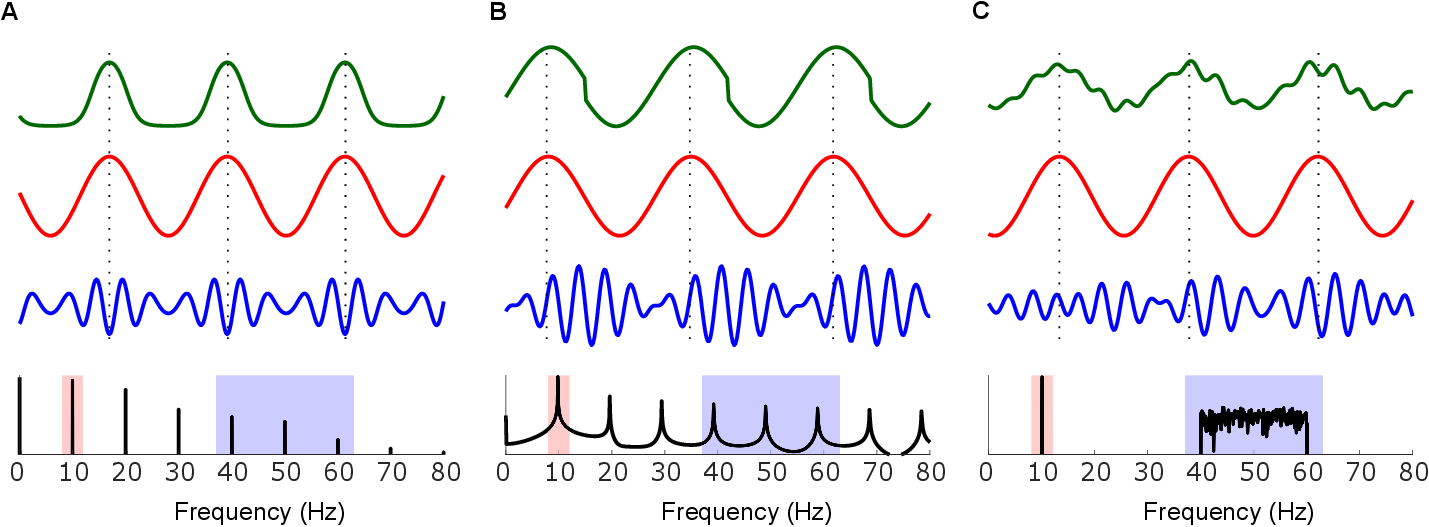} 
  \caption{\textbf{Signals that can lead to spurious detection of \ac{PAC} in neural recordings}. The three types of signals that may produce spurious \ac{PAC} are shown in green, and the corresponding spectrum is shown in black. The bandpass-filtered low-frequency component is shown in red (with the corresponding passband indicated by the red shaded region). Similarly, the corresponding high-frequency component is depicted in blue (with its passband indicated by the blue shaded region in the spectrum). \textbf{A} and \textbf{B} show non-sinusoidal waveforms whose harmonics can be misinterpreted as cross-frequency coupling. \textbf{C} is a sinusoidal signal with noise that may show an apparent phase-locking between the high-frequency activity and the low-frequency signal within certain time windows. Such signal characteristics (green) are known to give rise to spurious detection of \ac{PAC}.}
  \label{fig:spurious_pac}
\end{figure*}

Standard \ac{PAC} procedures are particularly susceptible to spurious coupling when the signal contains sharp edges, non-sinusoidal waveforms, or periodic spiking (Fig. \ref{fig:spurious_pac}) \cite{KRAMER2008, ARU2015, Gerber2016, Jensen2016, GIEHL2021}. In such cases, phase-locked harmonics generated by a base (a single non-sinusoidal) rhythm can be misinterpreted as genuine interactions between distinct slow- and fast-frequency processes.

The choice of filter bandwidths and analysis window lengths can also influence PAC estimates \cite{DVORAK2014}. Wider high-frequency filter bands can encompass sub-bands that reflect distinct processes and phase relationships \cite{Colgin2009}, while overly narrow low-frequency filter bands can distort non-sinusoidal rhythms \cite{Belluscio2012, COLE2017} and bias phase estimation. Similarly, shorter analysis windows risk inflating \ac{PAC} due to noise, whereas longer windows may introduce non-stationarities. These limitations motivate methods that go beyond instantaneous phase and amplitude-envelope extraction. Although inspection of unfiltered data, spectra, amplitude envelopes, and bicoherence can help identify or reduce spurious detections \cite{KRAMER2008, ARU2015, GIEHL2021}, robust detection and verification of genuine \ac{PAC} ultimately requires modelling the nonlinear interactions that generate it \cite{KRAMER2008, VELARDE2019, Dellavale2020}. Higher-order spectral analyses can capture some of these nonlinear dynamics, but on their their performance can be limited for short or noisy recordings \cite{Elgar1988,Siu2008,KOVACH2018}. These considerations motivate the model-based nonlinear dynamical systems approach developed in the following section.
%
\section{Proposed method} \label{sec:Method}
This section develops the proposed system identification framework for detecting and for characterising \ac{PAC}. We first introduce a nonlinear-systems perspective on \ac{PAC} and identify QPC as its key dynamical signature (Section \ref{sec:Dynm_sys_PAC}). We then show that a \ac{PAC} signal generated by more complex dynamics can be transformed into a canonical form that can be expressed as a nonlinear dynamical system (Section \ref{sec:canonnical_pac}). Together, Sections \ref{sec:Dynm_sys_PAC} and \ref{sec:canonnical_pac} provide the dynamical basis for the canonical representation. Section \ref{sec:canon_apprx_sysid} then describes how this canonical form can be estimated using nonlinear system identification, a process referred to here as canonical approximation. Finally, Sections \ref{sec:pract_impl} and \ref{sec:procedures} describe the implementation and main procedural steps of the proposed method.

%
\subsection{A nonlinear systems perspective on PAC}\label{sec:Dynm_sys_PAC}
The proposed method is based on the quadratic phase-coupled relationship between a slow oscillation, a fast oscillation, and the immediate intermodulation components (sidebands) generated by their interaction. The aim of this subsection is to identify the minimal second-order dynamical structure sufficient to captures this relationship (i.e. \ac{QPC}), thereby providing a concrete target for model-based \ac{PAC} detection.

In nonlinear systems, interactions between oscillatory components can generate new frequencies that are absent from the input spectrum, including harmonics and intermodulations \cite{Lang1996, Lang1997, Lang2005}. Harmonics occur at integer multiples of an input frequency (e.g. $\omega_1 \rightarrow 3\omega_1$), whereas intermodulation components arise from combinations of two or more input frequencies (e.g. $\{\omega_1, \omega_2\} \rightarrow \omega_1 + \omega_2$). These dynamical nonlinear phenomena are associated, respectively, with resonant neural processing (harmonics) \cite{Herrmann2001, TOBIMATSU1999, Herrmann2016, COLE2017} and functional integration across neural circuits (intermodulations) \cite{ChenX_2013, GIANI2012, GORDON2019, LUFF2024}. More generally, an $n$\textsuperscript{th}-order nonlinearity ($n \geq 2$, as $n=1$ is the linear case) determines how many input frequency components interact (are coupled) within that order to generate $n$\textsuperscript{th}-order harmonics and intermodulations \cite{Lang1996, Lang1997}.

\ac{PAC} can be expressed in this framework as a second-order (quadratic) coupling between slow- and fast-frequency oscillations \cite{Sigl1994, WITTE2000, hasselmann1963, Hyafil2015b, KOVACH2018}. This can be illustrated using the basic \ac{PAC} model defined by equations \eqref{eq:PAC_math} and \eqref{eq:simple_PAC}. Let $x(t) = \cos(\omega_{S}t + \phi_{S})$ denote the slow oscillation and $h(t) = \cos(\omega_{F}t + \phi_{F})$ the fast oscillation. Substituting these into equation \eqref{eq:simple_PAC}, gives the simplest form of amplitude modulation $y(t) = \left[ 1 + m \cos( \omega_{S}t + \phi_{S} ) \right] \cos( \omega_{F}t + \phi_{F})$ \cite{Hyafil2015b, KOVACH2018}, which can be expanded using trigonometric product identities as
\begin{multline}\label{eq:amp_modl_sin}
    y(t) = \frac{m}{2} \cos\Big( (\omega_F - \omega_S)t \ + \ (\phi_{F}-\phi_{s}) 
           \Big) + \\
           \ \cos( \omega_Ft + \phi_{F}) + 
           \ \ \ \ \ \ \ \ \ \ \ \ \ \ \ \ \ \ \ \ \ \ \ \ \ \ \ \\
           \ \frac{m}{2} \cos\Big( (\omega_F + \omega_S)t \ + \ (\phi_{F}+\phi_{s}) \Big).
           \ \ \ \ \ \
\end{multline}
In equation \eqref{eq:simple_PAC}, the nonlinear effect arises from the product $x(t) \times h(t)$. As equation \eqref{eq:amp_modl_sin} shows, this quadratic interaction generates the second-order intermodulations at $\omega_F \pm \omega_S$. Thus, the defining dynamical operation underlying the basic \ac{PAC} model is a quadratic interaction between slow and fast oscillations.

Introducing a time shift $t = t + \delta t$ (in equation \eqref{eq:amp_modl_sin}) in the amplitude-modulated signal, $y(t+\delta t)$, and substituting it into equation \eqref{eq:PAC_math} gives a more general basic \ac{PAC} signal:

\noindent
\begin{multline}\label{eq:PAC_sin_phase}
    z(t) = \cos( \omega_St + \phi_{S}) + 
           \frac{m}{2} \cos\Big( \omega_{\Delta} t \ + \ (\phi_{\Delta}+\omega_{\Delta} \delta t) \Big) + \\
           \cos\Big( \omega_Ft \ + \ (\phi_{F} + \omega_F\delta t) \Big) + 
           \ \ \ \ \ \ \ \ \ \ \ \ \\
           \frac{m}{2} \cos\Big( \omega_{\Sigma} t \ + \  (\phi_{\Sigma}+\omega_{\Sigma} \delta t) \Big),
           \ \ \ \ \ \ \ \ \ \ \ \ \ 
\end{multline}
where $\omega_{\Delta} = \omega_F - \omega_S$, $\phi_{\Delta} = \phi_{F} - \phi_{S}$, $\omega_{\Sigma} = \omega_F + \omega_S$, and $\phi_{\Sigma} = \phi_{F} + \phi_{S}$.

Equation \eqref{eq:PAC_sin_phase} shows that the slow and fast oscillations, $\omega_S$ and $\omega_F$, have a fixed phase relationship with their intermodulation frequencies $\omega_{\Delta}$ and $\omega_{\Sigma}$. This second-order intermodulation forms the basis of \ac{PAC} and is referred to as \ac{QPC} \cite{hasselmann1963, Sigl1994, WITTE2000, Hyafil2015b, Yang2016, KOVACH2018}. More generally, when two frequencies generate a third frequency through a second-order nonlinearity, all three frequencies are quadratically phase-coupled (either as a second-order harmonic or as a second-order intermodulation). Formally, \ac{QPC} between $\omega_1$, $\omega_2$, and $\omega_3 = \omega_1 + \omega_2 $ is defined as
\begin{equation}\label{eq:qpc_def}
    \left| \Delta\varphi \right| = \left| \varphi( \omega_{1} ) + \varphi( \omega_{2} ) - \varphi( \omega_{3} ) \right| = C ,
\end{equation}
where $\varphi(\omega_i)$ denotes the phase of frequency $\omega_i$, and $C$ is a constant. If $\left| \Delta\varphi \right|$ remains constant, then $\omega_1$ and $\omega_2$ are said to be quadratically phase-coupled with $\omega_3$. For equation \eqref{eq:PAC_sin_phase}, taking $\omega_1=\omega_S$, $\omega_2=\omega_F$, and $\omega_3=\omega_{\Sigma}$ gives $\varphi(\omega_S)=\omega_S t+\phi_S$, $\varphi(\omega_F)=\omega_F t+(\phi_F+\omega_F\delta t)$, and $\varphi(\omega_{\Sigma})=\omega_{\Sigma} t+(\phi_{\Sigma}+\omega_{\Sigma}\delta t)$, from which $\left| \Delta\varphi \right|=\omega_S\delta t$. By contrast, if the phases of the frequency components in $z(t)$ are independent and random, this phase combination $\left| \Delta\varphi \right|$ is not constant, and the components $z(t)$ reduces to a superposition of signals at those frequencies, rather than QPC. Accurately characterising this dynamic relationship therefore provides robust \ac{PAC} detection and quantification that does not rely on instantaneous phase and amplitude envelopes from bandpass-filtered signals.

Equation \eqref{eq:PAC_sin_phase} therefore identifies a minimal spectral-dynamical signature of \ac{PAC}: comprising a low-frequency component $\omega_S$, a high-frequency component $\omega_F$, and the two symmetric second-order intermodulation components at $\omega_F \pm \omega_S$, linked through QPC (equation \eqref{eq:qpc_def}). This provides the dynamical basis for the canonical representation of \ac{PAC} introduced in the next subsection and defines the dynamical target that the proposed system-identification method seeks to recover. 
%
\subsection{Canonical form of \ac{PAC} signals}\label{sec:canonnical_pac}
Realistic \ac{PAC} signals often contain frequency components beyond those present in the basic representation of equation \eqref{eq:PAC_sin_phase}. Non-sinusoidal slow oscillations introduce harmonics, while stronger or more complex nonlinearities can generate additional or higher-order intermodulations (e.g. Fig. \ref{fig:simple_gen_model}D and Fig. \ref{fig:simple_gen_model_vdp}). Although these components may contribute to the detailed waveform and spectrum, they are not essential for representing the defining quadratic interaction underlying \ac{PAC}. For detection purposes, it is therefore useful to reduce a \ac{PAC} signal to a minimal representation that preserves this second-order interaction. Here, preservation does not imply exact retention of all higher-order spectral details, but rather retention of the core quadratic phase-coupled structure relevant to \ac{PAC} detection and characterisation (Section \ref{sec:Dynm_sys_PAC}).

Although \ac{PAC} can arise through different mechanisms, the observed \ac{PAC} signal can still exhibit a common reduced dynamical structure as described in Section \ref{sec:Dynm_sys_PAC}. In the simple analytical models (Section \ref{sec:PAC_models}), a slow oscillation explicitly modulates a fast oscillation through a nonlinear interaction, whereas in the \ac{NMM} (Section \ref{sec:NMM}), the amplitude-modulated fast oscillation emerges from internal excitatory--inhibitory dynamics. As shown in Appendix \ref{appndx:taylor_approx}, Taylor approximations of these more complex \ac{PAC}-generating mechanisms recover terms with an analogous interaction structure to that of the basic \ac{PAC} model. This motivates representing the observed PAC dynamics by an effective two-input, single-output second-order nonlinear system, in which the interaction between a slow oscillation and a fast oscillation generates the second-order intermodulations. This reduced representation should therefore be viewed as a minimal dynamical approximation for \ac{PAC} detection and characterisation, rather than as a complete model of the observed signal or its spectrum.

Under this approximation, any \ac{PAC} signal can be reduced to a sum of four sinusoids with frequencies $\{\omega_S, \omega_F, \omega_F + \omega_S, \omega_F - \omega_S\}$, as given in equation \eqref{eq:PAC_sin_phase}. Their relative phases specify \ac{QPC}, while the magnitudes of the intermodulation components ($\omega_F \pm \omega_S$), relative to the fast oscillation at $\omega_F$, provide a measure of modulation strength. We refer to this reduced representation as the \textit{canonical form} of \ac{PAC}. It provides a compact dynamical target for model-based \ac{PAC} detection. The following subsection describes how this canonical form can be approximated from observed data using nonlinear system identification.
%
\subsection{Canonical approximation of \ac{PAC} using system identification}\label{sec:canon_apprx_sysid}
In this study, we propose using nonlinear system identification, specifically polynomial \ac{NARX} modelling, to detect and characterise \ac{PAC}. Polynomial \ac{NARX} models are well suited to this task because they represent system outputs using linear and nonlinear combinations of delayed inputs and outputs. Their explicit polynomial structure also makes the identified dynamics interpretable, enabling second-order intermodulation components to be identified and separated from the linear dynamics \cite{LJUNG2010, Chen1989a}. Here, we show that the canonical form of \ac{PAC} can be expressed using a second-order input-only \ac{NARX} model, and that this representation provides the basis for \ac{PAC} detection and characterisation. We then define a \ac{PAC} modulation index from the simulated output of the identified model and show how the identified model terms can be grouped into clusters for \ac{PAC} detection and interpretation.

Polynomial \ac{NARX} identification has been widely applied to complex dynamical systems \cite{Chiras2002, WANG2024, ZAINOL2022, RITZBERGER2017, Gao2023, HE2016, HE2021, LIU2024}. The implementation used here is based on the open-source MATLAB toolbox \textit{NonSysID} \cite{Gunawardena2024, nonsysid}; a brief overview of the \ac{NARX} model structure and identification procedure is provided in Appendix \ref{appndx:Brief_sysID_FRA}.

\begin{figure*}
	\centerline{\includegraphics[width=\textwidth]{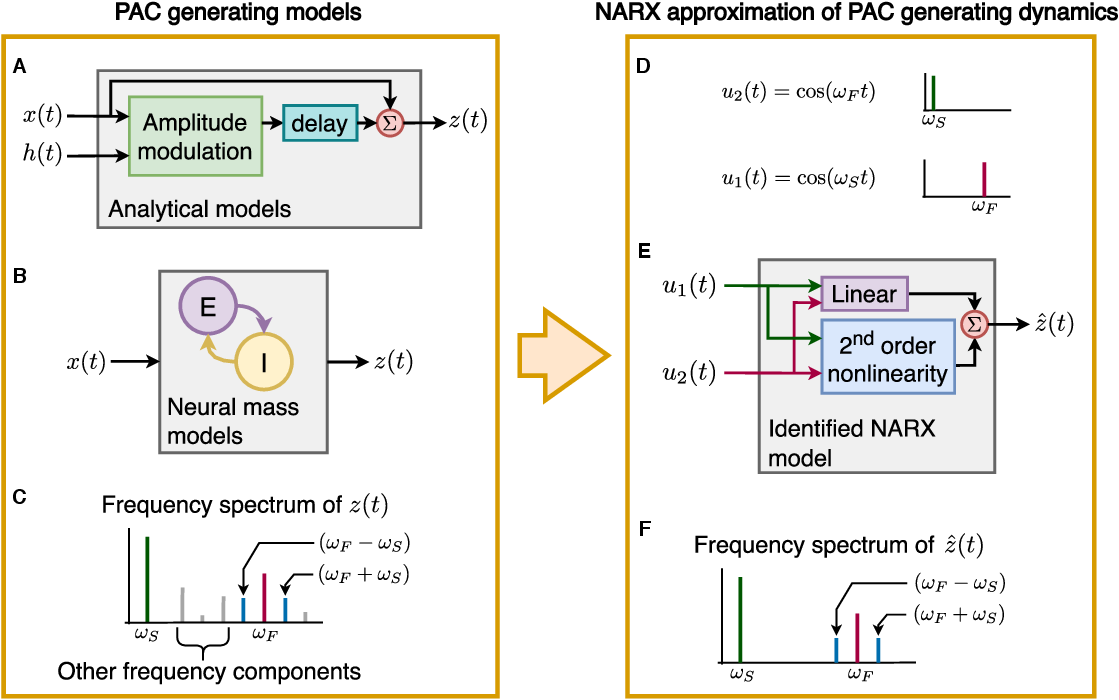}}
	\caption{\textbf{A two-input single-output \ac{NARX} model can be used to approximate \ac{PAC}-generating dynamics}. \textbf{A} shows simple analytical \ac{PAC} models in which a nonlinear interaction between a slow oscillation $x(t)$ and a fast oscillation $h(t)$ produces $z(t)$. \textbf{B} shows an \ac{NMM} that models the emergence of \ac{PAC} from interacting neural populations. \textbf{C} illustrates the typical spectrum of the resulting \ac{PAC} signal, where non-sinusoidal slow oscillations and non-sinusoidal amplitude modulations can introduce additional frequency components. \textbf{D} shows that two sinusoidal inputs under a second-order nonlinearity can provide a canonical approximation of \ac{PAC} dynamics. \textbf{E} and \textbf{F} illustrate the proposed system identification approach: a two-input single-output \ac{NARX} model is identified to capture underlying \ac{PAC} dynamics (\ac{QPC}), and its simulated spectrum provides the canonical approximation of the actual \ac{PAC} signal.}
	\label{fig:PAC_approx}
\end{figure*}

We first show analytically how the canonical \ac{PAC} form arises naturally within the \ac{NARX} structure. Essentially, \ac{QPC} between two frequencies is captured by identifying \ac{NARX} models using \textit{only past input values}. Let $u_1(t) = \cos(\omega_S t)$ and $u_2(t) = \cos(\omega_F t)$, where $\omega_S < \omega_F$. Introducing arbitrary time delays $\kappa$ and $\lambda$, we have $u_1(t-\kappa) = \cos(\omega_S t - \omega_S\kappa)$ and $u_2(t-\lambda) = \cos(\omega_F t - \omega_F\lambda)$. The product $u_1(t-\kappa) \times u_2(t-\lambda)$, when expanded via the double-angle trigonometric identities, yields two sinusoidal components at $\omega_\Delta = \omega_F - \omega_S$ and $\omega_\Sigma = \omega_F + \omega_S$, with their respective phase angles determined by the phases of $u_1(t-\kappa)$ and $u_2(t-\lambda)$.

It is a well-established result that the sum of two sinusoids of identical frequency produces a single sinusoid of the same frequency, with amplitude and phase determined by the original components \cite{LathiGreen2017} (also called the phasor addition rule). Accordingly, by employing appropriate combinations of delayed terms, the following decompositions hold:
\begin{equation}\label{eq:low_freq_compsn}
    A_1\cos(\omega_S t + \phi_1) = \sum_{i} a^{i}_{1} \times u_1(t-\kappa_i) \ \ \ \ \ \ \ \ \ \ \ \ \ \ \ \
\end{equation}
\begin{equation}\label{eq:high_freq_compsn}
    A_2\cos(\omega_F t + \phi_2) = \sum_{j} a^{j}_{2} \times u_2(t-\lambda_j) \ \ \ \ \ \ \ \ \ \ \ \ \ \ \ \
\end{equation}
\begin{multline}\label{eq:intrmod_freq_compsn}
    \frac{A_{3}}{2} \bigg( \cos(\omega_\Delta t + \phi_\Delta) +
    \cos(\omega_\Sigma t + \phi_\Sigma) \bigg) = \\ \sum_{i,j} a^{i,j}_{3} \times u_1(t-\kappa_i)u_2(t-\lambda_j),
\end{multline}
where $a^{i}_{1}$, $a^{j}_{2}$ and $a^{i,j}_{3}$ (for the respective $i$ and $j$) are real-valued constants. These decompositions show that the canonical \ac{PAC} structure can be expressed directly in terms of linear and second-order nonlinear delayed input terms. Accordingly, in analogy with equation \eqref{eq:PAC_sin_phase}, the basic \ac{PAC} signal can be estimated as 
\begin{multline}\label{eq:narx_pac_approx}
    \hat{z}(t) = \sum_{i} a^{i}_{1} u_1(t-\kappa_i) + \sum_{j} a^{j}_{2} u_2(t-\lambda_j) \\ 
    + \sum_{i,j} a^{i,j}_{3} u_1(t-\kappa_i)u_2(t-\lambda_j).
\end{multline}

Equation \eqref{eq:narx_pac_approx} is a two-input, single-output, 2\textsuperscript{nd}-order input-only \ac{NARX} model, with $u_1(t) = \cos(\omega_S t)$ and $u_2(t) = \cos(\omega_F t)$ representing the slow- and fast-frequency inputs, respectively. By selecting appropriate delays and coefficients, this model can reproduce the magnitudes and phases of the constituent terms in equation \eqref{eq:PAC_sin_phase}. Identifying such a model therefore yields a canonical approximation of the \ac{PAC} signal (Fig. \ref{fig:PAC_approx}).
\begin{figure*}
  \centering
  \includegraphics[width=\textwidth]{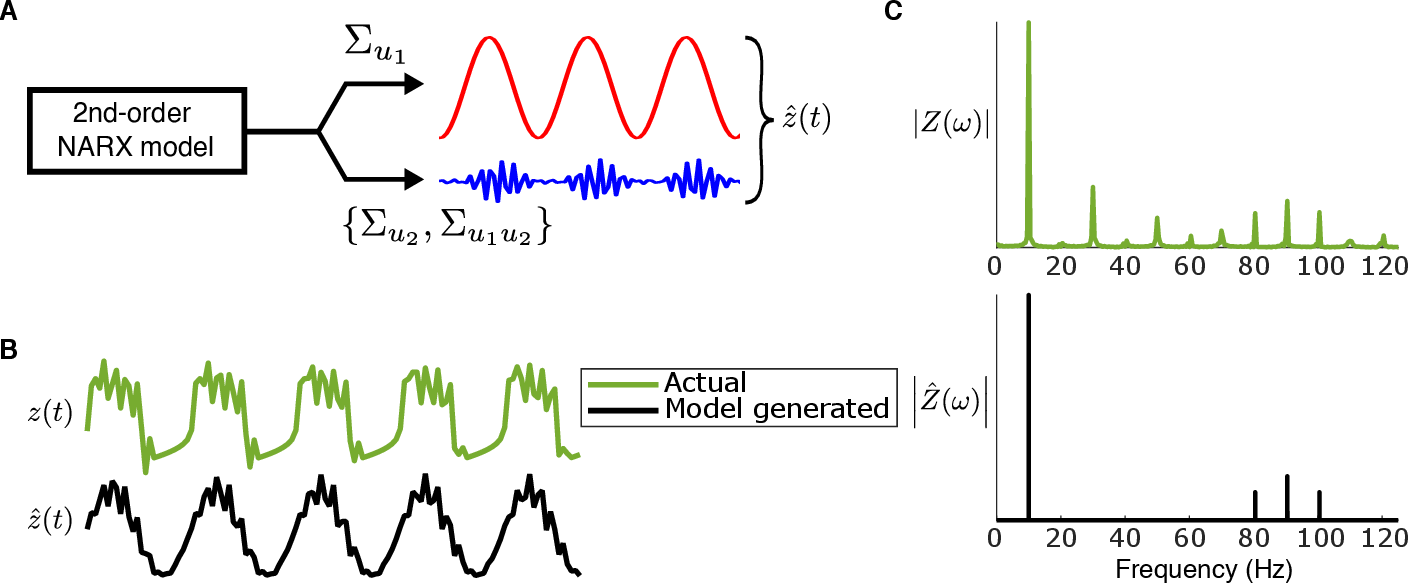} 
  \caption{\textbf{Canonical approximation of a \ac{PAC} signal using an identified 2\textsuperscript{nd}-order NARX model}. The \ac{PAC} signal is generated using the model given in equations \eqref{eq:PAC_math} and \eqref{eq:nonsine_PAC} with a non-sinusoidal slow oscillation defined by equation \eqref{eq:van_der_pol}. As shown in \textbf{A}, for the identified 2\textsuperscript{nd}-order \ac{NARX} model, the response from the term cluster $\Sigma_{u_{1}}$ gives the low-frequency component, while the response from the term clusters $\left\{ \Sigma_{u_{2}} , \Sigma_{u_{1}u_{2}} \right\}$ gives the high-frequency amplitude modulation component. Only the response from the clusters $\left\{ \Sigma_{u_{1}} , \Sigma_{u_{2}} , \Sigma_{u_{1}u_{2}} \right\}$ is considered as the canonical approximation, $\hat{z}(t)$, of the actual \ac{PAC} signal, $z(t)$. \textbf{B} shows the actual \ac{PAC} signal, $z(t)$, and its canonical approximation generated by an identified 2\textsuperscript{nd}-order \ac{NARX} model. The respective magnitude spectra of the actual \ac{PAC} signal, $\left|Z(\omega)\right|$, and its canonical approximation from the identified \ac{NARX} model, $\left|\hat{Z}(\omega)\right|$ are shown in \textbf{C}.}
  \label{fig:mthd_exmpl}
\end{figure*}

We quantify the effective modulation (coupling) strength, i.e. the modulation depth or modulation index, using the simulated output spectrum of the identified \ac{NARX} model (canonical approximation), $\hat{Z}(\omega)$ (similar to Figs. \ref{fig:simple_gen_model}–\ref{fig:simple_gen_model_biphasic}). Specifically, we define it as the mean magnitude ratio of the intermodulation components ($\omega_\Sigma=\omega_F + \omega_S$ and $\omega_\Delta=\omega_F - \omega_S$) to the magnitude of the high-frequency component ($\omega_F$). Accordingly, the \ac{PAC} modulation index in the proposed \ac{NARX}-based approach is
\begin{equation}\label{eq:NARX_PAC_MI}
    \text{MI}=\frac{\left|\hat{Z}(\omega_\Sigma)\right|+\left|\hat{Z}(\omega_\Delta)\right|}{2\left|\hat{Z}(\omega_F)\right|}.
\end{equation}
The spectrum $\hat{Z}(\omega)$ is computed via a normalised \ac{FFT} of the simulated output $\hat{z}(t)$:
\begin{equation}\label{eq:NARX_PAC_FFT}
    \hat{Z}(\omega)=\frac{T_s}{N} \ \text{FFT}\big\{\hat{z}(t)\big\},
\end{equation}
where $T_s$ is the sampling time ($T_s=1/F_s$, with $F_s$ the sampling frequency in Hz), and $N$ is the number of samples in $\hat{z}(t)$. Similar to conventional methods, the proposed modulation index $\text{MI}$ can be evaluated over a grid of slow- and fast-frequency pairs ($\omega_1$ and $\omega_2$) and visualised using a map known as the comodulogram, $\mathcal{C}(\omega_1, \omega_2)$.    

Similar to Figure \ref{fig:simple_gen_model_biphasic}, for monophasic \ac{PAC}, the intermodulation components at $\omega_F \pm \omega_S$ in the canonical approximation have magnitudes smaller than that of the high-frequency component at $\omega_F$, whereas the reverse holds for biphasic \ac{PAC}. Consequently, in equation \eqref{eq:NARX_PAC_MI}, $\text{MI}<1$ for monophasic coupling and $\text{MI}>1$ for biphasic coupling, providing a straightforward interpretation of the \ac{PAC} type from this metric.

Essentially, when $\text{MI}<1$, the value reflects the effective modulation depth of monophasic coupling (Fig. \ref{fig:simple_gen_model_biphasic}A), whereas $\text{MI}>1$ indicates the strength of biphasic coupling (Fig. \ref{fig:simple_gen_model_biphasic}B--C). However, it should be noted that in practice, the transition from monophasic to biphasic occurs as $\text{MI}$ approaches $1$, i.e. at around $0.7$.     

For practical identification of the canonical form in equation \eqref{eq:narx_pac_approx}, suitable maximum delays must be specified for the two input terms, i.e. the maximum values of $\kappa_i$ and $\lambda_j$ for $u_1(t)$ and $u_2(t)$, respectively. These determine the maximum number of past input values considered \cite{mendes1998a}. This selection is analogous to defining the embedding window in time-delay embedding (embedding time window) \cite{mendes1998a}. In periodic dynamics, it is generally recommended to include delays covering up to one full period (or quasi-period) of the frequency of interest to adequately reconstruct the underlying dynamics \cite{KUGIUMTZIS1996, Tan2023}. Based on extensive empirical evaluation, we found that a maximum delay of one-quarter period for the slow oscillation ( $\text{t}_S/4$ ) and one full period for the fast oscillation ($\text{t}_F$) provides an effective and robust configuration. Here, $\text{t}_S$ and $\text{t}_F$ denote the periods, in seconds, of the slow and fast oscillations, respectively. Next, we show how the identified \ac{NARX} model can be structurally interpreted and decomposed to analyse the captured \ac{PAC} dynamics.

The identified \ac{NARX} model terms can be systematically grouped into \textit{term clusters} according to their monomial structure \cite{AGUIRRE1995, AGUIRRE1996}. Each cluster represents a subset of terms with similar algebraic forms. For example, $u(t-1)u(t-3)$, $u(t-1)^2$, and $u(t-5)^2$ all belong to the cluster $\Sigma_{uu}$, which contains products of two delayed instances of the input $u(t)$. In a two-input, single-output \ac{NARX} model, clusters such as $\Sigma_{u_{1}^{2}u_{2}}$ or $\Sigma_{u_{1}u_{1}u_{2}}$ represent terms composed of two delayed instances of $u_1(t)$ and one delayed instance of $u_2(t)$. Importantly, different clusters can be associated with distinct underlying dynamics \cite{AGUIRRE1995, AGUIRRE1996, Aguirre1997}. Accordingly, by relating equation \eqref{eq:PAC_sin_phase} to equation \eqref{eq:narx_pac_approx}, the terms in equation \eqref{eq:narx_pac_approx} can be grouped into the following term clusters:
\begin{equation}
    \begin{aligned}
        a^{i}_{1} u_1(t-\kappa_i) &\in \Sigma_{u_{1}} \\
        a^{j}_{2} u_2(t-\lambda_j) &\in \Sigma_{u_{2}} \\
        a^{i,j}_{3} u_1(t-\kappa_i)u_2(t-\lambda_j) &\in \Sigma_{u_{1}u_{2}}.
    \end{aligned}
\end{equation}
Therefore, in a second-order \ac{NARX} model, the canonical \ac{PAC} representation is associated with the three term clusters $\{\Sigma_{u_{1}}, \Sigma_{u_{2}}, \Sigma_{u_{1}u_{2}}\}$. This term-cluster representation provides the structural basis for interpreting the identified \ac{NARX} model and for detecting \ac{PAC} in unknown signals.

The cluster $\Sigma_{u_{1}}$ represents the low-frequency component of the canonical approximation, analogous to $x(t)$ in equation \eqref{eq:PAC_math}. The clusters $\{\Sigma_{u_{2}}, \Sigma_{u_{1}u_{2}}\}$ encode the amplitude-modulated high-frequency component, analogous to $y(t+\delta t)$, with $\Sigma_{u_{1}u_{2}}$ capturing the second-order dynamics underlying \ac{PAC}, that is, \ac{QPC}. This allows the identified dynamics to be decomposed explicitly into slow and amplitude-modulated fast components, as illustrated in Figure \ref{fig:mthd_exmpl}. Building on this formulation, \ac{PAC} in an unknown signal is detected by identifying second-order \ac{NARX} models across a grid of low– and high-frequency pairs. A frequency pair is considered a candidate \ac{PAC} interaction, if the resulting \ac{NARX} model includes the clusters $\{\Sigma_{u_{1}}, \Sigma_{u_{2}}, \Sigma_{u_{1}u_{2}}\}$. That is, all three clusters must be present in the identified model. Here, a cluster is considered present if the identified model contains at least one selected term belonging to that cluster. 

Although the identified term clusters $\{\Sigma_{u_1}, \Sigma_{u_2}, \Sigma_{u_1u_2}\}$ indicate candidate \ac{PAC}, spurious detections can still occur. In particular, because $\Sigma_{u_1u_2}$ captures quadratic interaction associated with QPC, it may also capture harmonic-related QPC, leading to false positives. Moreover, when the immediate intermodulations are interpreted as high-frequency components, they can produce spurious couplings that appear as additional clusters in the comodulogram above and below the genuine \ac{PAC}. Therefore, after candidate \ac{PAC} pairs are identified from the canonical approximation, two post-detection checks are applied: an existing harmonic-related check and a new intermodulation-related check.
\FloatBarrier
\subsection{Rejection of harmonic-related spurious coupling}\label{sec:avoid_hSC}
Harmonic-related spurious \ac{PAC}, discussed earlier in Section \ref{sec:spuriousPAC}, arises from non-sinusoidal waveforms such as those shown in Figure \ref{fig:spurious_pac}A--B. This type of spurious coupling can be addressed relatively simply in the proposed \ac{NARX}-based \ac{PAC} method by following \cite{Fabus2022} and using the concept of instantaneous frequency, defined as time derivative of the instantaneous phase.

Fabus \textit{et al.} \cite{Fabus2022} used the observation that a sum of harmonic oscillations will behave like a single non-sinusoidal waveform, rather than as a superposition of individual oscillations. In this case, the instantaneous frequency is consistently positive. For a low-frequency $\omega_S$ and high-frequency $\omega_F$, a general expression for a simple \ac{PAC} between $\{\omega_S, \omega_F\}$ is given in $\hat{z}(t)$. Therefore, \ac{PAC} detected between the frequencies $\{ \omega_S, \omega_F\}$ is considered spurious and attributable to non-sinusoidal waveforms if
\begin{equation}
    \frac{\partial \varphi\big( \hat{z}(t)  \big)}{\partial t} > 0, \forall t, 
\end{equation}
where $\hat{z}(t)$ is the simulated output from the clusters $\left\{ \Sigma_{u_{1}} , \Sigma_{u_{2}} , \Sigma_{u_{1}u_{2}} \right\}$ within the identified input-only \ac{NARX} model (Fig. \ref{fig:mthd_exmpl}). Therefore, this condition can be applied directly to the simulated model output, avoiding ambiguity introduced by measurement noise.
\FloatBarrier
\subsection{Discrimination of intermodulation-related spurious coupling}\label{sec:avoid_iSC}
A genuine \ac{PAC} interaction generates sidebands, or intermodulation frequency components, around the high-frequency oscillation (Fig. \ref{fig:simple_gen_model}). In many \ac{PAC} analysis methods, these components may be treated as independent high-frequency oscillations, causing a single interaction to appear as multiple \ac{PAC} peaks or as a smeared region in the comodulogram \cite{Caiola2019, ZANDVOORT2021}. 

\begin{figure}
    \centering
    \includegraphics[width=0.77\linewidth]{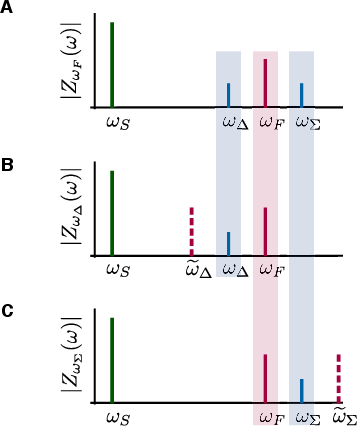}
    \caption{\textbf{Intermodulation-related spurious \ac{PAC}.} \textbf{A} illustrates the spectrum of a genuine \ac{PAC} between the low-frequency $\omega_S$ and the high-frequency $\omega_F$, producing the immediate intermodulation (sidebands) at $\omega_\Delta$ and $\omega_\Sigma$. \textbf{B} shows the corresponding apparent \ac{PAC} when the lower intermodulation $\omega_\Delta$ is treated as a candidate high-frequency component. In this case, $\omega_F$ becomes the corresponding intermodulation and an additional frequency component $\widetilde{\omega}_\Delta$ is implied (dashed red). \textbf{C} shows the symmetric case when the upper intermodulation $\omega_\Sigma$ is treated as a candidate high-frequency component. Again, $\omega_F$ becomes the corresponding intermodulation and an additional frequency component $\widetilde{\omega}_\Sigma$ is implied (dashed red). In filtering-based \ac{PAC} methods, the three cases correspond to centring the wide-band high-frequency filter on the true high-frequency component $\omega_F$ (A) or on one of its immediate intermodulations, $\omega_\Delta$ (B) or $\omega_\Sigma$ (C).}
    \label{fig:intrm_sc}
\end{figure}

From a dynamical-systems perspective, using equation \eqref{eq:PAC_sin_phase} as an example, the high-frequency component $\omega_F$ and its intermodulation components $\omega_{\Delta}$ and $\omega_{\Sigma}$ are all quadratically phase-coupled with the low-frequency oscillation $\omega_S$ (equation \eqref{eq:qpc_def}). Consequently, if $\omega_{\Delta}$ and $\omega_{\Sigma}$ are treated as high-frequency oscillators, the same underlying interaction may be detected as multiple \ac{PAC} peaks. This is illustrated in Figure \ref{fig:intrm_sc}. Such detections are spurious and are referred to as intermodulation-related spurious \ac{PAC}. They should therefore be excluded when identifying the genuine coupling pair. This section presents a method for distinguishing these intermodulation-related spurious detections from genuine \ac{PAC}.

Let the slow oscillation $x(t) = A\cos( \omega_St + \phi_{S})$ and the fast oscillation $h(t) = B\cos( \omega_Ft + \phi_{F})$. Substituting these into equations \eqref{eq:PAC_math} and \eqref{eq:simple_PAC} will result in
\begin{multline}\label{eq:PAC_mag_gen}
    z_{\omega_F}(t) = A\cos\Big( \varphi(\omega_S) \Big) +
           \frac{mAB}{2} \cos\Big( \varphi(\omega_{\Delta}) \Big) + \\
           B\cos\Big( \varphi(\omega_F) \Big) +
           \frac{mAB}{2} \cos\Big( \varphi(\omega_{\Sigma})  \Big),
\end{multline}
where $\varphi(\omega_S)=\omega_S t+\phi_S$, $\varphi(\omega_F)=\omega_F t+(\phi_F+\omega_F\delta t)$, $\varphi(\omega_{\Delta})=\omega_{\Delta} t \ + \ (\phi_{\Delta}+\omega_{\Delta} \delta t)$ and $\varphi(\omega_{\Sigma})=\omega_{\Sigma} t+(\phi_{\Sigma}+\omega_{\Sigma}\delta t)$ are the phases of the respective frequency components. Equation \eqref{eq:PAC_mag_gen} describes the genuine \ac{PAC} shown in Figure \ref{fig:intrm_sc}A. The cases in Figure \ref{fig:intrm_sc}B and C arise when one of the immediate intermodulations, $\omega_{\Delta}$ or $\omega_{\Sigma}$, is treated as an apparent high-frequency component. For $\omega_1 < \omega_2$, let $\mathcal{C}(\omega_1,\omega_2)$ denote the corresponding modulation strength of \ac{PAC} in the comodulogram. Let $\mathcal{H}(\omega_1,\omega_2)$ denote the magnitude of the high-frequency component in the high-frequency magnitude map. From Figure \ref{fig:intrm_sc} and equation \eqref{eq:PAC_mag_gen}, $\left| Z_{\omega_F}(  \omega_F) \right| = \left| Z_{\omega_\Delta}(\omega_F) \right| = \left| Z_{\omega_\Sigma}(\omega_F) \right| = B$, $\left| Z_{\omega_F}(\omega_\Delta) \right| = \left| Z_{\omega_\Delta}(\omega_\Delta) \right| = mAB/2$ and $\left| Z_{\omega_F}(\omega_\Sigma) \right| = \left| Z_{\omega_\Sigma}(\omega_\Sigma) \right| = mAB/2$. Consequently,
\begin{equation}\label{eq:intrmd_SpuC_rel}
    \begin{aligned}
        \mathcal{C}( \omega_S , \omega_F ) &= \frac{mA}{2} \\
        \mathcal{C}( \omega_S , \omega_\Delta ) = \mathcal{C}( \omega_S , \omega_\Sigma ) &= \frac{2}{mA} \\
        \mathcal{H}( \omega_S , \omega_F ) &= B \\
        \mathcal{H}( \omega_S , \omega_\Delta ) = \mathcal{H}( \omega_S , \omega_\Sigma ) &= \frac{mAB}{2}. \\
    \end{aligned}
\end{equation}
As shown by equation \eqref{eq:intrmd_SpuC_rel}, for $\omega_2 \in \{ \omega_{\Delta}, \omega_F, \omega_{\Sigma} \}$, $\mathcal{C}(\omega_S,\omega_2)$ tends to vary inversely with $\mathcal{H}(\omega_S,\omega_2)$. This relationship is used below to identify intermodulation-related spurious coupling.

The two maps $\mathcal{C}(\omega_1,\omega_2)$ and $\mathcal{H}(\omega_1,\omega_2)$ are normalised across the high-frequency axis for each fixed low frequency and compared through the discriminator $\mathcal{D}(\omega_1,\omega_2)$, defined as the sign of their difference. For a low-frequency $\omega_S$,
\begin{equation}\label{eq:commod_diff}
\mathcal{D}(\omega_S , \omega_2) =
    \begin{cases}
        +1 \ \ \text{if} \ \ 0 < \overline{\mathcal{H}}(\omega_S , \omega_2)
            - \overline{\mathcal{C}}(\omega_S , \omega_2)  \\[5pt]
       -1 \ \ \text{if} \ \ 0 > \overline{\mathcal{H}}(\omega_S , \omega_2)
            - \overline{\mathcal{C}}(\omega_S , \omega_2)
    \end{cases}
\end{equation}
where $\overline{\mathcal{H}}$ and $\overline{\mathcal{C}}$ denote the high-frequency magnitude map and the comodulogram after normalisation across the high-frequency axis with respect to individual low frequencies;

\noindent
\begin{equation}\label{eq:intrmd_norm_maps}
\overline{\mathcal{C}}(\omega_S , \omega_2) = \frac{\mathcal{C}(\omega_S,\omega_2)}{\sigma_{\mathcal{C}}(\omega_S)},
\qquad
\overline{\mathcal{H}}(\omega_S , \omega_2) = \frac{\mathcal{H}(\omega_S , \omega_2)}{\sigma_{\mathcal{H}}(\omega_S)},
\end{equation}
where $\sigma_{\mathcal{C}}(\omega_S)$ and $\sigma_{\mathcal{H}}(\omega_S)$ are the standard deviations of $\mathcal{C}(\omega_S , \omega_2)$ and $\mathcal{H}(\omega_S , \omega_2)$, respectively, across all high-frequencies ($\forall\omega_2$) under the same low-frequency $\omega_S$. Therefore, the discriminator $\mathcal{D}=+1$ indicates that the high-frequency magnitude is relatively larger than the modulation strength, that is, $\mathcal{H}(\omega_S,\omega_2) > \mathcal{C}(\omega_S,\omega_2)$. In contrast, the discriminator $\mathcal{D}=-1$ indicates that the modulation strength is relatively larger than the high-frequency magnitude, that is, $\mathcal{C}(\omega_S,\omega_2) > \mathcal{H}(\omega_S,\omega_2)$.

For a low-frequency component at $\omega_S$ and a high-frequency component at $\omega_F$, let $\omega_{\Delta}=\omega_F-\omega_S$ and $\omega_{\Sigma}=\omega_F+\omega_S$ denote the corresponding immediate intermodulations. For genuine \ac{PAC} at $\{ \omega_S, \omega_F \}$, the sign of the discriminator at $\omega_F$ should then be opposite to those at $\omega_{\Delta}$ and $\omega_{\Sigma}$:
\begin{equation}\label{eq:intrmd_sign_condition}
    \begin{aligned}
        \operatorname{sgn}\Big( \mathcal{D}(\omega_S , \omega_F) \Big)
    &= -\operatorname{sgn}\Big( \mathcal{D}(\omega_S , \omega_{\Delta}) \Big) \\
    &= -\operatorname{sgn}\Big( \mathcal{D}(\omega_S , \omega_{\Sigma}) \Big).
    \end{aligned}
\end{equation}
Equation \eqref{eq:intrmd_sign_condition} provides strong evidence that the \ac{PAC} detected at $\{ \omega_S, \omega_F \}$ is genuine, and the detections at $\{ \omega_S, \omega_{\Delta} \}$ and $\{ \omega_S, \omega_{\Sigma} \}$ are intermodulation-related spurious \ac{PAC}. 

In practice, equation \eqref{eq:intrmd_sign_condition} can be implemented algorithmically or assessed by inspecting the \ac{NARX}-based $\mathcal{D}$ map for a local alternating-sign pattern between a genuine \ac{PAC} region and the adjacent intermodulation-related spurious \ac{PAC} regions, as demonstrated throughout Section \ref{sec:Results} (Fig. \ref{fig:D_map}). If this pattern is absent, there is no evidence of intermodulation-related spurious \ac{PAC}.

A simple and effective condition, analogous to that used in the previous section to avoid harmonics, would be desirable. However, deriving such a condition would require further investigation. The development so far has established the proposed method for idealised sinusoidal oscillations. In the next section, the proposed framework is adapted to practical neural recordings, where \ac{PAC}-generating oscillations are non-stationary and embedded in noise. 

It should be noted that, in the preceding mathematical derivations and model definitions, frequencies are expressed using angular-frequency notation (when $\omega$ is used, in rad/s). For ease of interpretation, however, frequency values and \ac{PAC} ranges are reported in Hz throughout the remainder of the paper.
\FloatBarrier
\subsection{Practical implementation}\label{sec:pract_impl}
In neural recordings, \ac{PAC}-related oscillations are often non-stationary, with time-varying amplitude, frequency, and phase \cite{Axmacher2010, Canolty2006, Bragin1995}, and are embedded in concurrent electrophysiological activity often approximated as pink noise \cite{Miskovic2019, BiyuHE2014, Pettersen2014, Ao2025}. Under these conditions, a system-identification approach based solely on idealised sinusoidal inputs $u_1(t)$ and $u_2(t)$ is insufficient to capture pronounced non-stationarities and low-\ac{SNR} cases.

To handle non-stationarity and noise, the signal is first bandpass-filtered around each candidate low- and high-frequency component, and the filtered components are used as inputs $u_1(t)$ and $u_2(t)$. Filtering is implemented using the \ac{FFT} and inverse \ac{FFT}, allowing efficient application even to short signal segments. To reduce ringing associated with sharp filter windows (brick-wall filtering), the \ac{FFT}-based filter used here employs a smooth cosine roll-off with an empirically chosen $10\%$ transition band. In practice, a fixed 1 Hz bandwidth is used for the fast oscillation, while a 2 Hz bandwidth is generally adequate for the slow oscillation, although accuracy improves when this bandwidth matches the true slow-oscillation bandwidth. For purely sinusoidal oscillations, or cases where the relevant frequency components are discrete and point-based, a narrower slow-oscillation bandwidth of 0.5 or 1 Hz is more appropriate.

To enable consistent evaluation of the canonical \ac{PAC} spectrum $\hat{Z}(\omega)$ and $\text{MI}$ in equation \eqref{eq:NARX_PAC_MI} while avoiding the effects of non-stationary, the identified \ac{NARX} model is simulated using stationary sinusoidal inputs at the centre frequencies of the filtered slow and fast oscillations: $u_1(t)=A_S\cos(\omega_S t)$ and $u_2(t)=A_F\cos(\omega_F t)$. To preserve the relative amplitude scale of the identified dynamics, the amplitudes are chosen to match the standard deviations of the corresponding filtered components, such that $A_S=\sigma_S\sqrt{2}$ and $A_F=\sigma_F\sqrt{2}$, where $\sigma_S$ and $\sigma_F$ denote the standard deviations of the filtered slow and fast oscillations, respectively. To mitigate detections arising from noise, a frequency pair is considered phase-amplitude coupled only when $0.2 < \text{MI} < 5$ and $|\hat{Z}(\omega_F)|/|\hat{Z}(\omega_S)| \leq 0.7$. The former bounds ensure that the intermodulation and high-frequency components are not excessively small relative to one another, while the latter helps exclude detections in which the low-frequency component is negligible relative to the high-frequency component, particularly at small variances. 

For intermodulation-related spurious detections in broadband cases, the discriminator $\mathcal{D}$ (equation \eqref{eq:commod_diff}) is first segmented into locally sign-consistent regions along the high-frequency axis before applying the alternating-sign relationship in equation \eqref{eq:intrmd_sign_condition}, reducing sensitivity to isolated sign variations. 

The next section summarises these implementation steps in pseudocode for \ac{PAC} detection and quantification.
\FloatBarrier
\subsection{Procedures}\label{sec:procedures}
This section summarises the operational workflow for the proposed method. Let $\varsigma(t)$ denote the signal to be analysed, and $\wp_{S}$ and $\wp_{F}$ the low- and high-frequency query points, respectively. $\wp_{S}$ and $\wp_{F}$ define the frequency grid over which \ac{PAC} is to be detected. The method returns a comodulogram $\mathcal{C}$ and, for frequency pairs identified as exhibiting \ac{PAC}, the corresponding identified models $\mathcal{M}$. $2\omega^{R}_{1}$ and $2\omega^{R}_{2}$ denote the bandwidths of the slow- and fast-oscillation filters, respectively.

The procedure has two stages. First, an initial scan is performed over all frequency pairs $\{\omega_1,\omega_2\}$ such that $\omega_1 \in \wp_{S}$, $\omega_2 \in \wp_{F}$, and $\omega_1 < \omega_2$. This stage uses linear \ac{ARX} model identification, which is substantially faster than fitting \ac{NARX} models, to retain only those pairs for which the signal contains stable slow and fast oscillatory components. These pairs are recorded in $\overline{\mathcal{C}}$ such that $\overline{\mathcal{C}}(\omega_1,\omega_2)=1$ (and $0$ otherwise). This initial \ac{ARX}-based stage rapidly prunes the frequency grid, so the more computationally expensive 2\textsuperscript{nd}-order \ac{NARX} identification is applied only to the shortlisted pairs.

%
\begin{algorithm}[H]
\caption{\ac{NARX}-based \ac{PAC}: Initial scan}\label{alg:ARX_scan}
\begin{algorithmic}[1]
\State \textbf{Input:} $\varsigma(t)$, $\wp_{S}$, $\wp_{F}$, $\omega^{R}_{1}$
\State \textbf{Initialise:} $\overline{\mathcal{C}}$
\LComment{Bandwidth ($2\omega^{R}_{2}$) for extracting the fast oscillation}
\State $\omega^{R}_{2} \gets 0.5$
\For{$\omega_1 \in \wp_{S}$}
    \For{$\omega_2 \in \wp_{F}$}
        \If{$\omega_1 < \omega_2$}
            \State $u_1(t) \gets \mathcal{F}\big( \varsigma(t), \omega_1 - \omega^{R}_{1}, \omega_1 + \omega^{R}_{1} \big)$
            \State $u_2(t) \gets \mathcal{F}\big( \varsigma(t), \omega_2 - \omega^{R}_{2}, \omega_2 + \omega^{R}_{2} \big)$
            \State $n_{b_1} \gets \text{t}_S/4$
            \State $n_{b_2} \gets \text{t}_F$
            \State $\overline{ \eta } \gets \text{SysIdL}( \varsigma(t), u_1(t) , u_2(t) , n_{b_1} , n_{b_2})$
            \If{$\overline{ \eta }$ contains $\left\{ \Sigma_{u_{1}} , \Sigma_{u_{2}} \right\}$}
                \LComment{$\omega_1$ and $\omega_2$ are significant}
                \State $\overline{\mathcal{C}}(\omega_1 , \omega_2) \gets 1$
            \Else
                \LComment{$\omega_1$ and/or $\omega_2$ are not significant}
                \State $\overline{\mathcal{C}}(\omega_1 , \omega_2) \gets 0$
            \EndIf
        \EndIf
    \EndFor
\EndFor
\State \textbf{Output:} $\overline{\mathcal{C}}$
\end{algorithmic}
\end{algorithm}
%

After the initial scan, only the shortlisted pairs are examined using a 2\textsuperscript{nd}-order \ac{NARX} model identification. The identified model is denoted by $\eta$. A pair $\{\omega_1,\omega_2\}$ is accepted as exhibiting \ac{PAC} only if $\eta$ contains the term clusters $\{\Sigma_{u_{1}}, \Sigma_{u_{2}}, \Sigma_{u_{1}u_{2}} \}$. That is, all three required clusters must be present in $\eta$. For accepted pairs, the coupling strength, $\text{MI}$ (equation \eqref{eq:NARX_PAC_MI}), is stored in $\mathcal{C}(\omega_1,\omega_2)$ and the identified model is stored in $\mathcal{M}(\omega_1,\omega_2)=\eta$. These models can then be used to quantify and interpret the characteristics of the detected \ac{PAC} (see Fig. \ref{fig:mthd_exmpl}).

Algorithm \ref{alg:ARX_scan} summarises the initial scan, and Algorithm \ref{alg:NARX_PAC} summarises the full detection and quantification procedure. In both algorithms, $\mathcal{F}\big(\varsigma(t), \omega_a, \omega_b\big)$ denotes bandpass filtering of $\varsigma(t)$ within the frequency range $\left[\omega_a, \omega_b\right]$ (see Section \ref{sec:pract_impl}). In Algorithm \ref{alg:NARX_PAC}, $\text{SysIdN}(\varsigma(t),u_1(t),u_2(t),n_{b_1},n_{b_2})$ denotes identification of a two-input single-output 2\textsuperscript{nd}-order \ac{NARX} model, where $n_{b_1}$ and $n_{b_2}$ are the maximum numbers of past instances (embedding time window) of $u_1(t)$ and $u_2(t)$ considered (see Appendix \ref{appndx:Brief_sysID_FRA}). Similarly, $\text{SysIdL}(\varsigma(t),u_1(t),u_2(t),n_{b_1},n_{b_2})$ denotes identification of a two-input single-output linear \ac{ARX} model.

%
\begin{algorithm}[H]
\caption{\ac{NARX}-based \ac{PAC}: Detection and quantification}\label{alg:NARX_PAC}
\begin{algorithmic}[1]
\State \textbf{Input:} $\varsigma(t)$, $\wp_{S}$, $\wp_{F}$, $\omega^{R}_{1}$ 
\State \textbf{Initialise:} $\mathcal{C}$, $\mathcal{M}$
\State $\overline{\mathcal{C}} \gets$ Run initial scan, Algorithm \ref{alg:ARX_scan}
\
\LComment{Bandwidth ($2\omega^{R}_{2}$) for extracting the fast oscillation}
\State $\omega^{R}_{2} \gets 0.5$ 
\For{$\omega_1 \in \wp_{S}$}
    \For{$\omega_2 \in \wp_{F}$}
        \If{$\omega_1 < \omega_2$}
            \LComment{Check $\{\omega_1 , \omega_2 \}$ has passed the initial scan}
            \If{$\overline{\mathcal{C}}(\omega_1 , \omega_2) = 1$}
                \State $u_1(t) \gets \mathcal{F}\big( \varsigma(t), \omega_1 - \omega^{R}_{1}, \omega_1 + \omega^{R}_{1} \big)$
                \State $u_2(t) \gets \mathcal{F}\big( \varsigma(t), \omega_2 - \omega^{R}_{2}, \omega_2 + \omega^{R}_{2} \big)$
                \State $n_{b_1} \gets \text{t}_S/4$
                \State $n_{b_2} \gets \text{t}_F$
                \State $\eta \gets \text{SysIdN}( \varsigma(t), u_1(t) , u_2(t) , n_{b_1} , n_{b_2})$
                \If{$\eta$ contains $\left\{ \Sigma_{u_{1}} , \Sigma_{u_{2}} , \Sigma_{u_{1}u_{2}} \right\}$}
                    \LComment{$\omega_1$ and $\omega_2$ are coupled}
                    \State $\mathcal{C}(\omega_1 , \omega_2) \gets \text{MI}$
                    \State $\mathcal{M}(\omega_1 , \omega_2) \gets \eta$
                \Else
                    \LComment{$\omega_1$ and $\omega_2$ are uncoupled}
                    \State $\mathcal{C}(\omega_1 , \omega_2) \gets 0$
                    \State $\mathcal{M}(\omega_1 , \omega_2) \gets 0$
                \EndIf
            \Else
                \LComment{$\omega_1$ and $\omega_2$ are uncoupled}
                \State $\mathcal{C}(\omega_1 , \omega_2) \gets 0$
                \State $\mathcal{M}(\omega_1 , \omega_2) \gets 0$
            \EndIf    
        \EndIf
    \EndFor
\EndFor
\State \textbf{Output:} $\mathcal{C}$, $\mathcal{M}$
\end{algorithmic}
\end{algorithm}
%

Statistical significance can be assessed by replacing $\varsigma(t)$ with a surrogate signal and applying Algorithm \ref{alg:NARX_PAC} to each surrogate realisation. To reduce computational cost, testing may be restricted to frequency pairs that have already been detected to be in \ac{PAC}. The proposed method estimates the canonical \ac{PAC} form by detecting \ac{QPC} through a second-order input-only \ac{NARX} model (Fig. \ref{fig:mthd_exmpl}). A phase-randomised surrogate technique is therefore appropriate for testing whether the observed nonlinear phase relationships are genuine \cite{Siu2008, LANCASTER2018}.

Following the detection of frequency pairs exhibiting \ac{PAC}, together with the corresponding comodulogram $\mathcal{C}$ and identified models $\mathcal{M}$, the detections are first assessed for intermodulation-related ambiguity using the discriminator-based procedure in Section \ref{sec:avoid_iSC}. Harmonic-related spurious \ac{PAC} is then rejected among the remaining detections using the criterion in Section \ref{sec:avoid_hSC}.

Polynomial input-only \ac{NARX} model identification in this study was carried out using the open-source MATLAB package \textit{NonSysID} \cite{Gunawardena2024, nonsysid}, which includes built-in stability checks for model simulation. Furthermore, the \ac{NARX} models identified using \textit{NonSysID} are cross-validated using the leave-one-out method (see Appendix \ref{appndx:SysID_alg}).

\FloatBarrier
\section{Experimental Results} \label{sec:Results}
This section presents a comparative evaluation of the proposed \ac{NARX}-based framework for \ac{PAC} detection and characterisation against several established benchmark methods. The evaluation uses both synthetic and real electrophysiological datasets. Synthetic datasets are generated according to the models described in Section \ref{sec:PAC_models}, while benchmark comparisons include filtering-based \ac{PAC} metrics introduced by Ozkurt \textit{et al.} \cite{OZKURT2011}, Canolty \textit{et al.} \cite{Canolty2006}, and Penny \textit{et al.} \cite{PENNY2008}.
\begin{figure}[ht]
  \centering
  \includegraphics[scale=0.65]{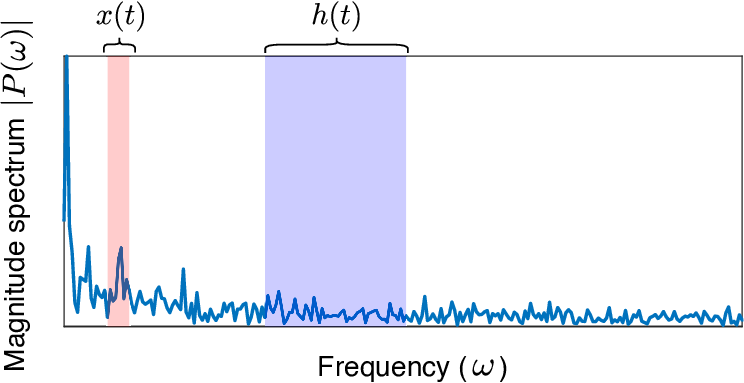} 
  \caption{\textbf{Generating non-stationary slow and fast oscillations.} To obtain non-stationary slow and fast oscillations $x(t)$ and $h(t)$, respectively, a randomly generated pink noise sample $p(t)$ is bandpass-filtered in the frequency ranges of interest. This is illustrated using the magnitude spectrum $\left| P(\omega) \right|$ of a pink noise sample $p(t)$. The red and blue shaded regions indicate the frequency ranges of interest for the non-stationary slow, $x(t)$, and fast, $h(t)$, oscillations.}
  \label{fig:pink_pac}
\end{figure}

Non-stationary slow- and fast-frequency oscillations were generated by bandpass filtering pink noise over the frequency ranges of interest (Fig. \ref{fig:pink_pac}). The pink noise was produced by shaping white noise to obtain a power spectral density that decreased inversely with frequency, approximating the spectral profile of many \ac{EEG} and \ac{MEG} recordings \cite{Miskovic2019, BiyuHE2014, Pettersen2014, Ao2025}. This procedure was used to generate all synthetic datasets in this section. Each synthetic signal was 10 seconds long, which is generally considered the minimum duration required for robust \ac{PAC} analysis using filtering-based methods \cite{DVORAK2014, Caiola2019}. To improve filtering accuracy, signals analysed using filtering-based methods were sampled at 1000 Hz. For the proposed \ac{NARX}-based method, the signals were down-sampled to 250 Hz to enable accurate and computationally efficient system identification \cite{SPINELLI2005, ANDERSON2007}. Finally, an independent pink-noise realisation was added to obtain an \ac{SNR} of $3$ (defined as a ratio of signal to noise standard deviation), corresponding to $9.54\,\mathrm{dB}$. The following three synthetic datasets were generated using the selected analytical model:
\begin{enumerate}
    \item A basic sinusoidal case with coupling between a 7 Hz slow oscillation and a 63 Hz fast oscillation.
    \item A case with a 6–7 Hz slow oscillation coupled to a fast oscillation in the 55–60 Hz range. 
    \item A more complex case in which a 9-10 Hz slow oscillation is simultaneously coupled to two distinct high-frequency bands: 35–40 Hz and 70–80 Hz.
\end{enumerate}
Experiments 2 and 3 are particularly useful for assessing the method’s ability to detect \ac{PAC} with broadband oscillatory components under non-stationary conditions. Furthermore, to demonstrate the repeatability and specificity of our approach, we average comodulograms obtained across multiple realisations of the same \ac{PAC} signal (cases 2 and 3 above), each generated using a different pink-noise sample (i.e. different random-number-generator seeds). Detailed descriptions for each setup and the corresponding results are provided in the following sections.
\FloatBarrier
\subsection{Single frequency oscillations with non-sinusoidal amplitude modulation}

\begin{figure*}
  \centering
  \includegraphics[width=\textwidth]{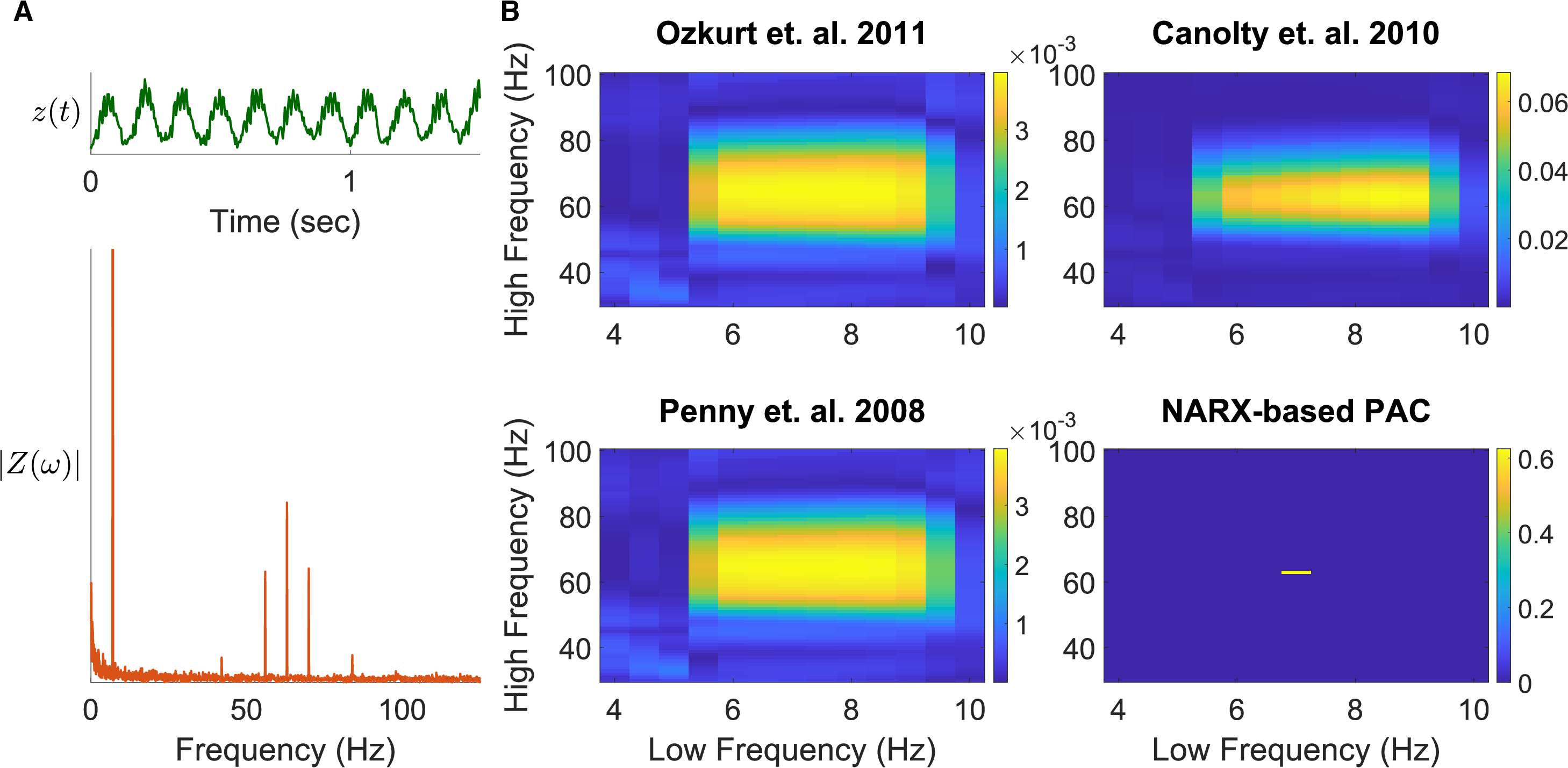} 
  \caption{\textbf{Sinusoidal slow and fast oscillations coupled through non-sinusoidal amplitude modulation.} A 7 Hz slow oscillation is coupled to a 63 Hz fast oscillation to obtain the 10-second \ac{PAC} signal $z(t)$. Panel \textbf{A} displays a segment of the signal $z(t)$ alongside its magnitude spectrum $\left| Z(\omega) \right|$. Panel \textbf{B} compares the proposed \ac{NARX}-\ac{PAC} method with alternative methods, using the comodulogram obtained from each method. The comodulogram from \ac{NARX}-based \ac{PAC} is the post-processed result with intermodulation-related spurious \ac{PAC} suppressed (Section \ref{sec:avoid_iSC}).}
  \label{fig:sngl_freq_cmd}
\end{figure*}

The first simulated \ac{PAC} example, shown in Figure \ref{fig:sngl_freq_cmd}, involves a 7 Hz slow oscillation coupled to a 63 Hz high-frequency sinusoidal oscillation. Since the slow and fast oscillations in this example are discrete sinusoidal components, filtering bandwidths of 0.5 Hz and 1 Hz were used for the slow and fast oscillations, respectively. Figure \ref{fig:sngl_freq_cmd}A displays a short segment of the simulated signal $z(t)$ alongside its magnitude spectrum $\left| Z(\omega) \right|$. As expected, the spectrum contains intermodulation components at 56 Hz and 70 Hz, as well as several other phase-locked components. Many of these are integer multiples of the 7 Hz base frequency, raising the potential for spurious couplings. Notably, the 63 Hz component itself is also such a multiple. Nevertheless, the experiment shows that the proposed method accurately identifies genuine \ac{PAC} while robustly rejecting spurious couplings with high specificity. Figure \ref{fig:D_map}A shows the corresponding discriminator map. At the genuine 7--63 Hz \ac{PAC} detection, $\mathcal{D}=+1$, whereas the intermodulation-related spurious \ac{PAC} detections at 7--56 Hz and 7--70 Hz have $\mathcal{D}=-1$. This alternating-sign pattern satisfies equation \eqref{eq:intrmd_sign_condition}, supporting the 7--63 Hz interaction as genuine and identifying the adjacent detections as intermodulation-related spurious \ac{PAC}.

\begin{figure*}
  \centering
  \includegraphics[width=\textwidth]{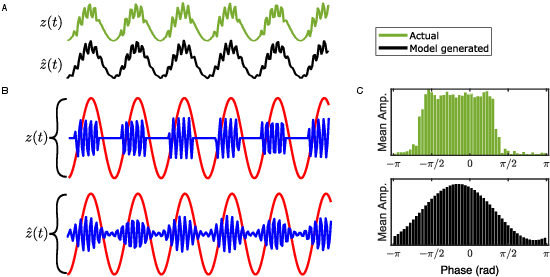} 
  \caption{\textbf{\ac{PAC} analysis using the proposed method}. This figure demonstrates the level of detail achievable in \ac{PAC} analysis with the proposed method. The analysis corresponds to the synthetic experiment presented in Figure \ref{fig:sngl_freq_cmd}, where sinusoidal oscillations of 7 Hz and 63 Hz are coupled. \textbf{A} compares the true \ac{PAC} signal $z(t)$ with the canonical approximation produced by the \ac{NARX} model, $\hat{z}(t)$. \textbf{B} shows comparisons between the low-frequency (red) and the high-frequency amplitude modulation (blue) components of $z(t)$ and $\hat{z}(t)$. In \textbf{C}, the histograms illustrate how, for both $z(t)$ and $\hat{z}(t)$, the amplitude envelope of the modulated fast oscillation varies with the phase of the slow oscillation.}
  \label{fig:sngl_freq_ana}
\end{figure*}

A more detailed examination of the identified couplings can be carried out, as shown in Figure \ref{fig:sngl_freq_ana}. Figure \ref{fig:sngl_freq_ana}B compares the original \ac{PAC} signal with the canonical approximation generated by the \ac{NARX} model (see Sections \ref{sec:canonnical_pac} and \ref{sec:canon_apprx_sysid}). The \ac{NARX} model for each detected coupling can also be decomposed into their constituent components—the slow oscillation and the amplitude-modulated fast oscillation (as illustrated in Fig. \ref{fig:mthd_exmpl}).

Figure \ref{fig:sngl_freq_ana}B shows both the actual and the model-generated slow and fast oscillatory components. Since the canonical approximation consists of purely sinusoidal components (see Section \ref{sec:canonnical_pac}), the Hilbert transform can be applied directly—without additional narrowband filtering—to extract instantaneous phase information. The relationship between the phase of the slow oscillation and the amplitude envelope of the fast oscillation is visualised using a phase–amplitude histogram \cite{Tort2010}, where the slow-phase cycle is divided into bins, and the mean fast-amplitude within each bin is computed. As shown in Figure \ref{fig:sngl_freq_ana}C, for both the actual signal and the canonical approximation, this procedure reveals the slow phase at which the fast oscillation reaches its peak amplitude. Taken together, Figures \ref{fig:sngl_freq_ana}B–C demonstrate that the \ac{NARX}-based method accurately identifies the phase of fast oscillatory bursts, confirming its effectiveness for \ac{PAC} characterisation.
%
%
\subsection{Non-stationary oscillations and pink noise}
\begin{figure*}
  \centering
  \includegraphics[width=\textwidth]{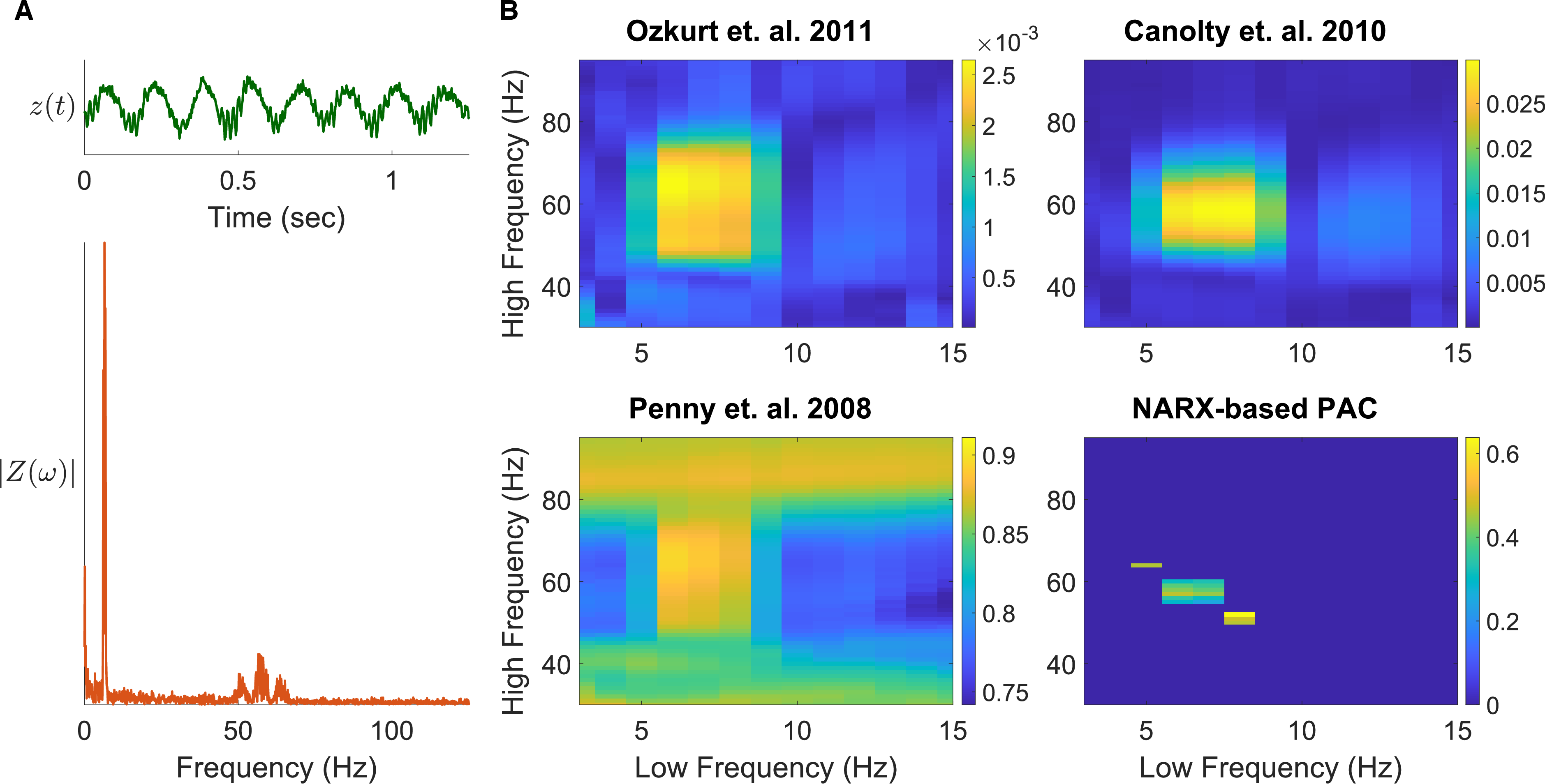} 
  \caption{\textbf{Non-stationary slow oscillation coupled to non-stationary fast oscillation using a non-sinusoidal amplitude modulation.} A 6--7 Hz slow oscillation is coupled with 55--60 Hz non-stationary fast oscillations to obtain the 10-second \ac{PAC} signal $z(t)$. \textbf{A} shows a snippet of the signal $z(t)$ and its magnitude spectrum $\left| Z(\omega) \right|$. Comparisons between the proposed \ac{NARX}-\ac{PAC} method and other methods are shown in \textbf{B} using the comodulogram evaluated from each method. The comodulogram from \ac{NARX}-based \ac{PAC} is the post-processed result with intermodulation-related spurious \ac{PAC} suppressed (Section \ref{sec:avoid_iSC}).}
  \label{fig:rng_PinkNoise_cmd}
\end{figure*}
\begin{figure*}
  \centering
  \includegraphics[width=\textwidth]{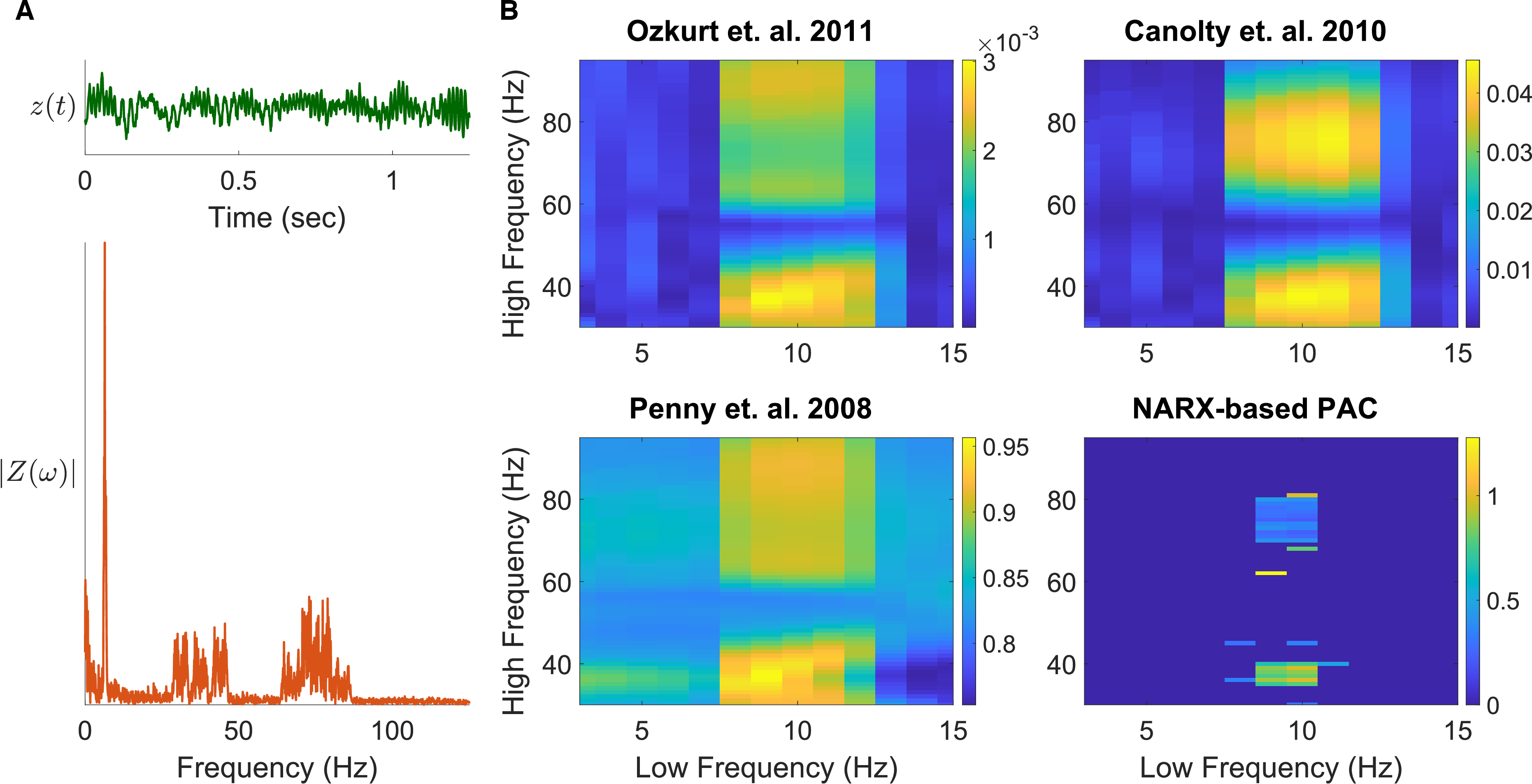} 
  \caption{\textbf{Non-stationary slow oscillation coupled to two distinct non-stationary fast oscillations using a non-sinusoidal amplitude modulation.} A 9--10 Hz slow oscillation is coupled with 35--40 Hz and 70--80 Hz non-stationary fast oscillations to obtain the 10-second \ac{PAC} signal $z(t)$. \textbf{A} shows a snippet of the signal $z(t)$ and its magnitude spectrum $\left| Z(\omega) \right|$. Comparisons between the proposed \ac{NARX}-\ac{PAC} method and other methods are shown in \textbf{B} using the comodulogram evaluated from each method. The comodulogram from \ac{NARX}-based \ac{PAC} is the post-processed result with intermodulation-related spurious \ac{PAC} suppressed (Section \ref{sec:avoid_iSC}).}
  \label{fig:rng_PinkNoise_multpl_cmd}
\end{figure*}

A slow oscillation, $x_1(t)$, in the 6–7 Hz range and a fast oscillation, $h_1(t)$, in the 55–60 Hz range were extracted via bandpass filtering of pink noise (Fig. \ref{fig:pink_pac}). These components were phase–amplitude coupled using the analytical model with non-sinusoidal amplitude modulation (equations \eqref{eq:PAC_math} and \eqref{eq:nonsine_PAC}), with monophasic coupling expressed as:
\begin{equation}
    \begin{cases}
        z_1(t)    = x_1(t) + k_1 y_1(t + \delta t_1),	\\[5pt]
        y_1(t) = \left( 1  - \frac{1}{ 1 + \exp\Big( -\alpha\Big( x_1(t) - c \Big) \Big)} \right) h_1(t),
\end{cases}
\end{equation}
where $k_1 = 0.5$ controls the variance of the amplitude modulation applied to the slow oscillation. $\delta t_1 = 0.016\text{sec.}$, $\alpha = 200$ and $c = 1\times10^{-6}$. A sample of pink noise was added to the generated \ac{PAC} signal, yielding an \ac{SNR} of $3$ ($9.54\,\mathrm{dB}$). 

\begin{figure*}
  \centering
  \includegraphics[width=\textwidth]{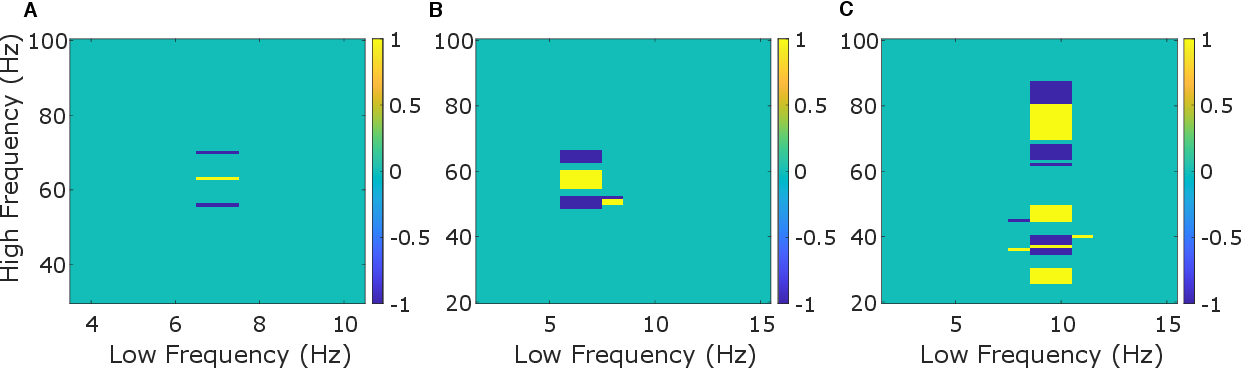} 
  \caption{\textbf{Discriminator ($\mathcal{D}$) maps for distinguishing genuine \ac{PAC} from intermodulation-related spurious \ac{PAC}.} The discriminator $\mathcal{D}$ is evaluated according to equation \eqref{eq:commod_diff}; yellow and blue indicate $\mathcal{D}=+1$ and $\mathcal{D}=-1$, respectively. A candidate \ac{PAC} region is supported as genuine when its discriminator sign is opposite to the signs of the corresponding intermodulation-related spurious \ac{PAC} regions at the lower and upper intermodulation frequencies, as specified by equation \eqref{eq:intrmd_sign_condition}. \textbf{A}, \textbf{B}, and \textbf{C} correspond to the experiments presented in Figures \ref{fig:sngl_freq_cmd}, \ref{fig:rng_PinkNoise_cmd}, and \ref{fig:rng_PinkNoise_multpl_cmd}, respectively. In each panel, the local alternating-sign pattern distinguishes the genuine \ac{PAC} region from its adjacent intermodulation-related spurious \ac{PAC} regions.}
  \label{fig:D_map}
\end{figure*}

Representative comodulograms, comparing the proposed method with conventional approaches, are shown in Figure \ref{fig:rng_PinkNoise_cmd}. For the conventional filtering-based methods, a sampling frequency of $1000$ Hz was used, as filtering is more accurate at higher sampling rates. For the proposed \ac{NARX}-based \ac{PAC} method, the synthetic signal was downsampled to a sampling frequency of $250$ Hz to reduce the computational cost of system identification process. Furthermore, the slow oscillation filter bandwidth for the \ac{NARX}-based \ac{PAC} method was set to $2$ Hz (corresponding to $\omega^{R}_{1}=1$ Hz in Algorithms \ref{alg:ARX_scan} and \ref{alg:NARX_PAC}). The corresponding discriminator map in Figure \ref{fig:D_map}B shows that the genuine \ac{PAC} region between 6--7 Hz and 55--60 Hz has the opposite discriminator sign to the adjacent intermodulation-related spurious \ac{PAC} regions. The resulting alternating-sign pattern satisfies equation \eqref{eq:intrmd_sign_condition} and supports the detected interaction between 6--7 Hz and 55--60 Hz as genuine.

A more complex scenario, using the same sampling rates as above, is illustrated in Figure \ref{fig:rng_PinkNoise_multpl_cmd}, where a slow 9–10 Hz oscillation, $x_2(t)$, is simultaneously coupled to two distinct high-frequency bands: 35–40 Hz, $h_2(t)$, and 70–80 Hz, $h_3(t)$. $x_2(t)$ was phase-amplitude coupled to $h_2(t)$ and $h_3(t)$ using the basic \ac{PAC} model (equations \eqref{eq:PAC_math} and \eqref{eq:nonsine_PAC}) such that $\{x_2, h_2 \}$ coupling is biphasic and $\{x_2, h_3 \}$ is monophasic. This \ac{PAC} signal can be expressed as   
\begin{equation}
    \begin{cases}
        z_2(t) = x_2(t) + k_2 y_2(t + \delta t_2) + k_3 y_3(t + \delta t_3),	\\[5pt]
        y_2(t) = \Big( 1 + m_2 \times x_2(t) \Big) h_2(t) ,	\\[5pt]
        y_3(t) = \Big( 1 + m_3 \times x_2(t) \Big) h_3(t),
    \end{cases}
\end{equation}
where $k_2 = 0.5$ and $k_3 = 0.25$ control the variances of the corresponding amplitude modulations embedded on the slow oscillation. $\delta t_2 = 0.09\text{s}$, $\delta t_3 = 0.016\text{s}$, $m_2 = 150$ and $m_3 = 50$.

As shown in Figure \ref{fig:rng_PinkNoise_multpl_cmd}, the proposed method not only localises the two interactions more sharply than conventional methods but also provides a direct interpretation of their coupling types (see Section \ref{sec:canon_apprx_sysid} and equation \eqref{eq:NARX_PAC_MI}). For the coupling at 35-40 Hz, the \ac{NARX}-based \ac{PAC} $\text{MI} \gtrapprox 0.7$ correctly indicates biphasic coupling. By contrast, the coupling at 70-80 Hz is correctly identified as monophasic, with $\text{MI} \lessapprox 0.7$. Figure \ref{fig:D_map}C shows that each genuine \ac{PAC} region at 35--40 Hz and 70--80 Hz is flanked by intermodulation-related spurious \ac{PAC} regions with the opposite discriminator sign. Although the discriminator values at the two genuine \ac{PAC} regions have opposite signs, each genuine region has a sign opposite to that of its adjacent intermodulation-related spurious \ac{PAC} regions, thereby satisfies equation \eqref{eq:intrmd_sign_condition}, supporting both interactions as genuine.

Taken together, Figure \ref{fig:D_map} demonstrates that the discriminator depends on the local sign reversal between a genuine \ac{PAC} region and its adjacent intermodulation-related spurious \ac{PAC} regions, rather than on the absolute sign of $\mathcal{D}$. This enables genuine \ac{PAC} regions to be distinguished from intermodulation-related spurious \ac{PAC} regions in both single-frequency and broadband, non-stationary cases.

The supplementary Figures \ref{fig:rng_PinkNoise_cmd_stat} and \ref{fig:rng_PinkNoise_multpl_cmd_stat} (see Appendix \ref{appndx:extra_plts}) illustrate the reproducibility and specificity of the proposed \ac{NARX}-based method. Figure \ref{fig:rng_PinkNoise_cmd_stat} shows the median comodulograms obtained from 100 realisations of the 6–7 Hz coupled with 55–60 Hz case, while Figure \ref{fig:rng_PinkNoise_multpl_cmd_stat} shows the corresponding median comodulograms for the two-band monophasic and biphasic case in Figure \ref{fig:rng_PinkNoise_multpl_cmd}. Supplementary Figure \ref{fig:rng_PinkNoise_multpl_noise_cmd_stat} further shows that the proposed method remains robust under higher noise levels, including at an \ac{SNR} of 2 ($6.02\,\mathrm{dB}$), and fairly robust even at an \ac{SNR} of 1 ($0\,\mathrm{dB}$). Supplementary Figure \ref{fig:rng_PinkNoise_multpl_cmd__TmWndw_stat} shows that the method also performs well for short signal lengths, including 5-second windows and, to a lesser extent, even 2–3 second windows. From all these examples (Figs. \ref{fig:rng_PinkNoise_cmd}, \ref{fig:rng_PinkNoise_multpl_cmd}, \ref{fig:rng_PinkNoise_cmd_stat}, \ref{fig:rng_PinkNoise_multpl_cmd_stat}, \ref{fig:rng_PinkNoise_multpl_noise_cmd_stat}, and \ref{fig:rng_PinkNoise_multpl_cmd__TmWndw_stat}), the proposed method accurately identifies the coupled frequency bands and localises them to the regions of the comodulogram where genuine phase–amplitude interactions occur. In contrast, conventional filtering-based methods yield broader, less precise coupling profiles, characterised by increased frequency smearing and greater susceptibility to false positive detections.
\subsection{Evaluating performance on signals prone to spurious PAC}\label{sec:exp_rslt_spurious_PAC}
Signals prone to spurious \ac{PAC} were discussed in section \ref{sec:spuriousPAC}. As discussed, non-sinusoidal waveforms, waveforms with sharp edges, and periodic spike-train signals are known to generate false-positive \ac{PAC} detections in conventional filtering-based methods because they contain phase-locked harmonic structure or sharp periodic transients \citep{KRAMER2008, Gerber2016}.

\begin{figure*}
  \centering
  \includegraphics[width=\textwidth]{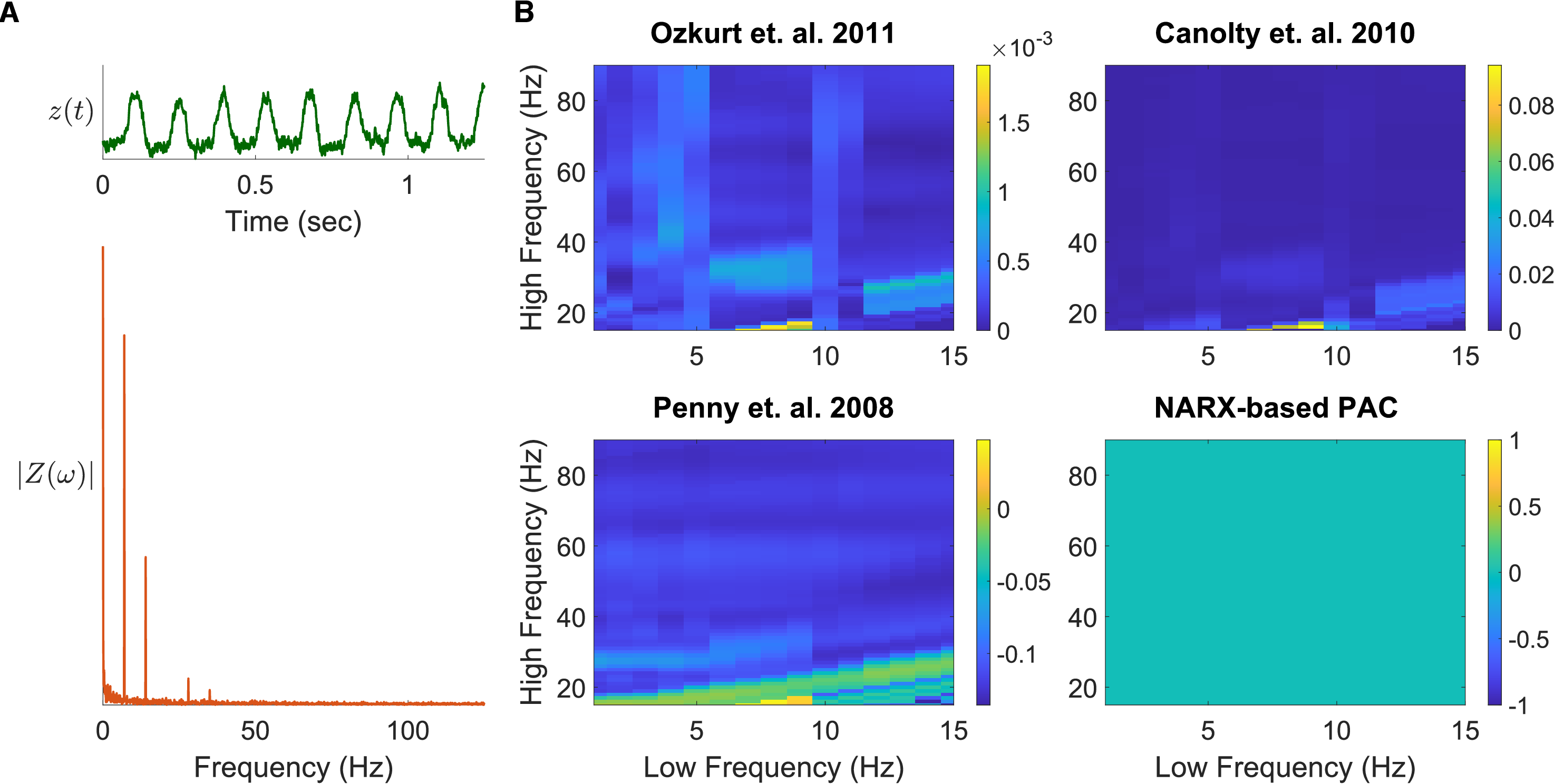} 
  \caption{\textbf{Comparison using a simulated non-sinusoidal waveform that produces spurious \ac{PAC}}. This is the same signal shown in Figure \ref{fig:spurious_pac}A. A 20-second non-sinusoidal signal $z(t)$ with added pink noise is used here to compare how the proposed method can reject spurious couplings due to harmonics. \textbf{A} shows a snippet of the signal $z(t)$ and its magnitude spectrum $\left| Z(\omega) \right|$. Comparisons between the proposed \ac{NARX}-\ac{PAC} (after post-processing for harmonics) method and other methods are shown in \textbf{B} using the comodulogram evaluated from each method. Note that the comodulogram from the method by Penny \textit{et al.} is shown on a natural logarithmic scale. The comodulogram from \ac{NARX}-based \ac{PAC} is the post-processed result with harmonic-related spurious \ac{PAC} suppressed (Section \ref{sec:avoid_hSC}).}
  \label{fig:nonsine_spuc_cmd}
\end{figure*}

\begin{figure*}
  \centering
  \includegraphics[width=\textwidth]{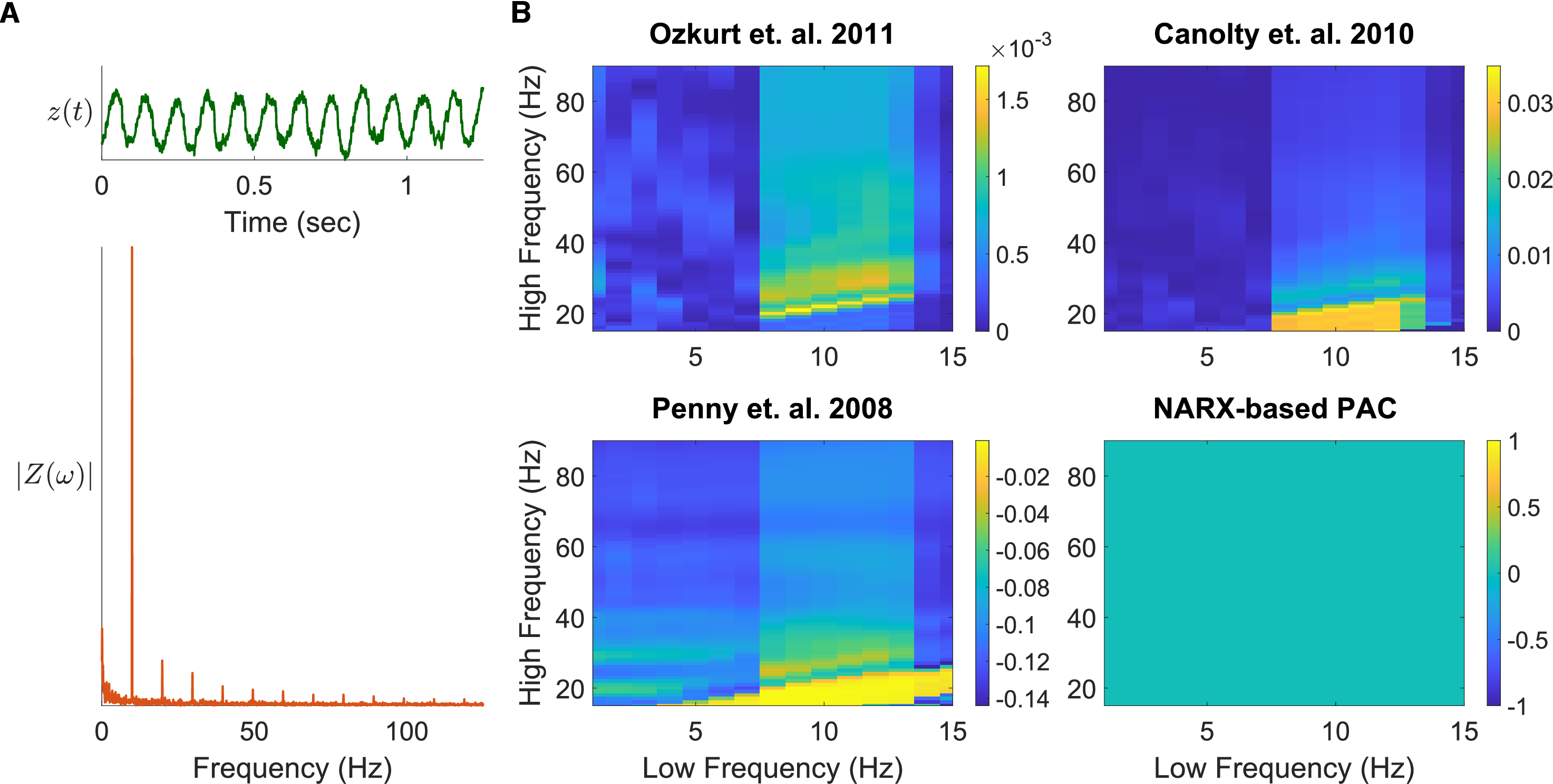} 
  \caption{\textbf{Comparison using a simulated non-sinusoidal signal with a sharp edge that produces spurious \ac{PAC}}. This is the same signal shown in Figure \ref{fig:spurious_pac}B. A 20-second non-sinusoidal signal with a sharp edge $z(t)$ and added pink noise is used here to compare how the proposed method can reject spurious couplings due to harmonics. \textbf{A} shows a snippet of the signal $z(t)$ and its magnitude spectrum $\left| Z(\omega) \right|$. Comparisons between the proposed \ac{NARX}-\ac{PAC} (after post-processing for harmonics) method and other methods are shown in \textbf{B} using the comodulogram evaluated from each method. Note that the comodulogram from the method by Penny \textit{et al.} is shown on a natural logarithmic scale. The comodulogram from \ac{NARX}-based \ac{PAC} is the post-processed result with harmonic-related spurious \ac{PAC} suppressed (Section \ref{sec:avoid_hSC}).}
  \label{fig:shrp_edg_spuc_cmd}
\end{figure*}

Figures \ref{fig:nonsine_spuc_cmd} and \ref{fig:shrp_edg_spuc_cmd} show the results for the non-sinusoidal waveform and sharp-edge cases, respectively. In both examples, the proposed \ac{NARX}-\ac{PAC} method, after post-processing based on the harmonic-rejection criterion described in Section \ref{sec:avoid_hSC}, suppresses spurious couplings arising from harmonics. By contrast, the conventional methods continue to show broader and less specific coupling patterns in the comodulogram. These results demonstrate that the proposed method can systematically reject spurious \ac{PAC} arising from non-sinusoidal waveforms and sharp transients.

Sharp periodic spikes, or spike trains, contain phase-locked harmonics and are frequently observed in \ac{ECoG}, particularly in sensorimotor regions \cite{Gerber2016}. Figure \ref{fig:spike_train_cmd} compares the proposed method with standard techniques using a simulated periodic spike-train signal generated according to \cite{Gerber2016}. All methods show apparent \ac{PAC}; however, the \ac{NARX}-based comodulogram reveals the characteristic spike-related pattern \cite{Gerber2016} more clearly. This clearer representation supports the diagnostic identification of spike-related spurious \ac{PAC}, although a formal rejection rule for this type of spurious \ac{PAC} is not introduced here.
%
\subsection{Actual LFP data from the rat hippocampus}
Scheffer-Teixeira et al. \cite{Teixeira2011} analysed \ac{LFP}s recorded from the CA1 region of the rat hippocampus during different sleep–wake states, including REM sleep and active wakefulness. Signals were sampled at 1000 Hz, and \ac{PAC} was assessed using the modulation index proposed by Tort et al. \cite{Tort2008, Tort2010}. The study found that the low-frequency theta phase modulated two distinct high-frequency oscillations: (1) high-gamma (HG) oscillations around 80 Hz, predominant in deep CA1 layers (stratum lacunosum–moleculare), and (2) high-frequency oscillations (HFOs) around 140 Hz, which were dominant in superficial CA1 layers (stratum oriens–alveus). In the original study \cite{Teixeira2011}, HG and HFO were used specifically to distinguish these frequency bands. Here, HFO refers specifically to the higher gamma-band activity and should not be confused with the broader use of “high-frequency oscillation” elsewhere in this paper.
\begin{figure*}
  \centering
  \includegraphics[width=\textwidth]{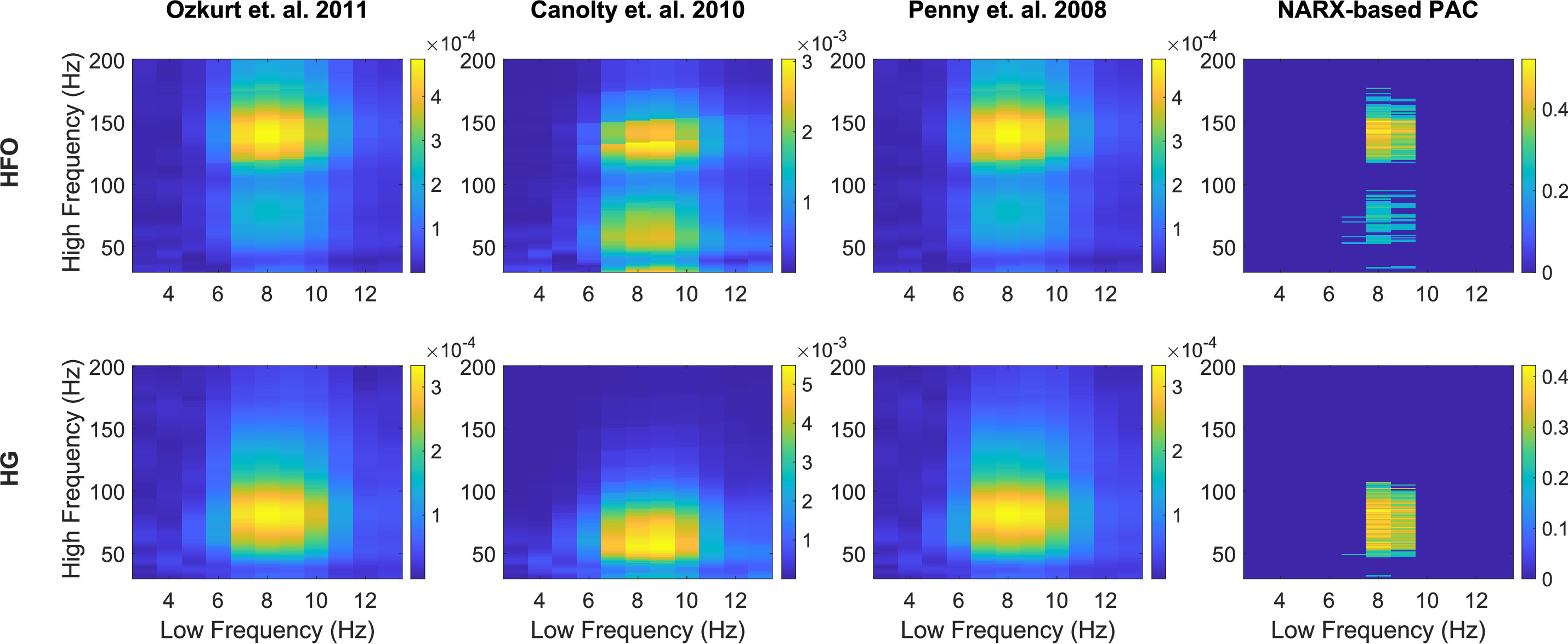} 
  \caption{\textbf{Comparison of \ac{PAC} results between commonly used methods and the proposed \ac{NARX}-\ac{PAC} method on the \ac{LFP} data from \cite{Teixeira2011, Tortlab}}. \ac{PAC} results are shown for the two LFPs that were provided in \cite{Tortlab}. One of the \ac{LFP} samples contains dominant frequencies around the HFO gamma frequency range (superficial CA1 layers, stratum oriens–alveus) and the other contains frequencies around the HG frequency range (deeper CA1 layers, stratum lacunosum–moleculare).}
  \label{fig:LFP_HFO_HG}
\end{figure*}
\begin{figure*}
  \centering
  \includegraphics[width=\textwidth]{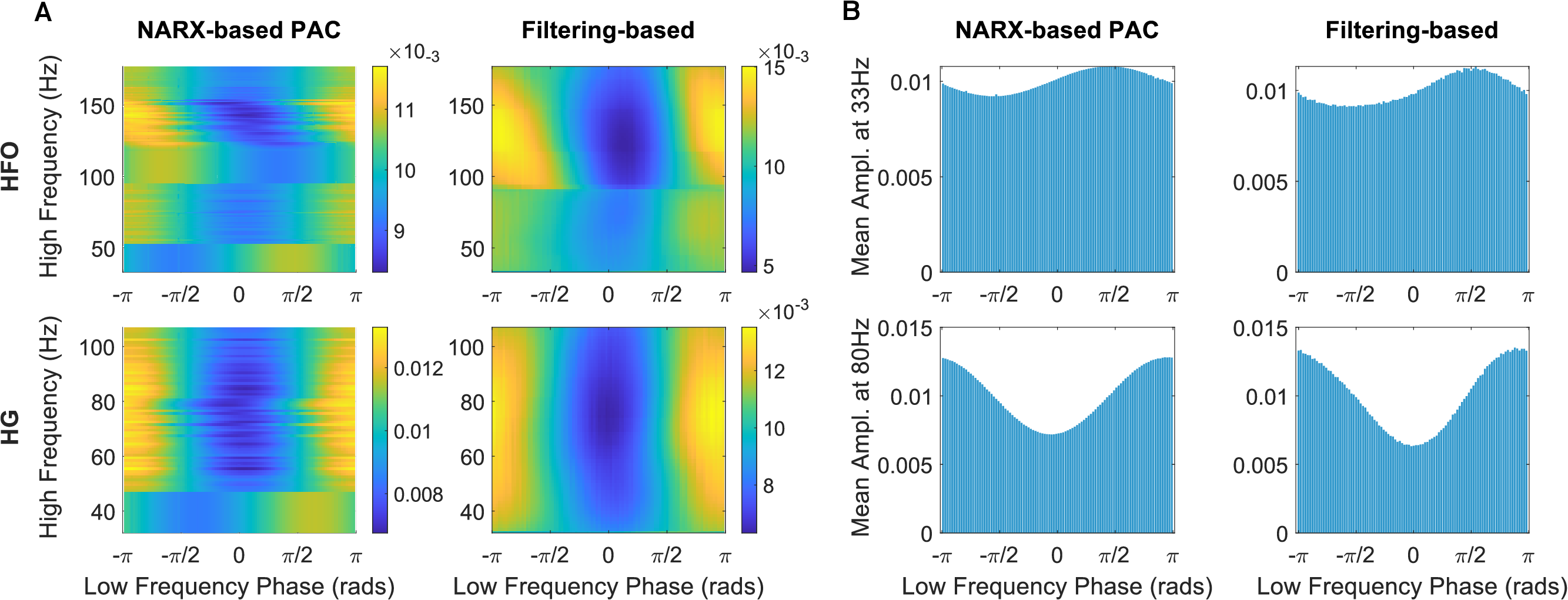} 
  \caption{\textbf{Comparison of phase information for the detected \ac{PAC}s involving the 8 Hz low-frequency component}. From the detected \ac{PAC}s shown in Figure \ref{fig:LFP_HFO_HG}, this figure compares the phase information of all \ac{PAC}s involving the 8 Hz low-frequency component between the proposed \ac{NARX}-based \ac{PAC} method and the filtering-based \ac{PAC} detection methods. Panel \textbf{A} shows comparisons of phase information for all couplings involving the 8 Hz low-frequency component. In contrast, panel \textbf{B} shows the comparison for a particular high-frequency coupling involving the 8 Hz low-frequency component.}
  \label{fig:LFP_phs_HFO_HG}
\end{figure*}

Two sample \ac{LFP} recordings from the REM sleep state, originally reported in \cite{Teixeira2011}, are available via the GitHub repository \cite{Tortlab}. One recording originates from deep CA1 layers (stratum lacunosum–moleculare), where high-gamma activity is prominent, and the other from superficial CA1 layers (stratum oriens–alveus), where HFOs dominate. These recordings were analysed using the proposed \ac{NARX}-based method and were compared with conventional filtering-based approaches. For conventional methods, a sampling rate of $1000$ Hz was used. For the \ac{NARX}-based method, a sampling rate of $500$ Hz was used. Furthermore, the slow oscillation filter bandwidth for the \ac{NARX}-based method was set to $2$ Hz (corresponding to $\omega^{R}_{1}=1$ Hz in Algorithms \ref{alg:ARX_scan} and \ref{alg:NARX_PAC}).

Figure \ref{fig:LFP_HFO_HG} shows the detected \ac{PAC} frequency bands, and Figure \ref{fig:LFP_phs_HFO_HG} depicts the corresponding phase–coupling relationships associated with the 8 Hz low-frequency component. From the comodulograms in Figure \ref{fig:LFP_HFO_HG}, the \ac{NARX}-based method provides sharper and more precise localisation of coupling. Figure \ref{fig:LFP_phs_HFO_HG} compares the phase information associated with the coupling; specifically, the phase of the low-frequency oscillation at which the high-frequency amplitude envelope peaks. As shown in Figure \ref{fig:LFP_phs_HFO_HG}B, the \ac{NARX}-based method can accurately identify the preferred phase of the coupling. 

%
\section{Discussion and Conclusion} \label{sec:Dissc}
\ac{PAC}, a form of \ac{CFC}, is thought to support cognitive functions and large-scale neural communication, making accurate detection and characterisation important for understanding brain function. In this study, we introduced a \ac{QPC}-grounded dynamical-systems framework in which \ac{PAC} is modelled as a nonlinear interaction between slow and fast oscillations and identified directly from data using nonlinear system identification. Unlike methods based primarily on filtered phase and amplitude-envelope fluctuations, the proposed approach identifies a generative model of the coupling and reproduces its canonical form. This provides a dynamical representation of \ac{PAC}, enabling direct estimation of modulation strength, preferred phase, and \ac{PAC} type, while providing a principled basis for distinguishing genuine and spurious detections. Overall, the method should be viewed as a dynamically grounded framework for \ac{PAC} detection and characterisation, within which dynamically motivated procedures for rejecting spurious \ac{PAC} can be incorporated, refined, and extended as the underlying spurious dynamics become better understood.
%
\subsection{Discussion of the results}
A key advantage of the proposed method is that it identifies a generative model of \ac{PAC}, enabling noise-free simulation of the learned dynamics and post-hoc analysis of coupling structure. This supports estimation of modulation strength and preferred phase, improves robustness to variations in slow-frequency power, and provides a dynamical basis for mitigating spurious coupling.

The main advantages of the proposed framework can be grouped into three closely related aspects: its ability to identify a generative canonical model of \ac{PAC}, its systematic dynamical basis for mitigating spurious coupling, and its practical robustness to noise, non-stationarity, and short-data lengths.

A major challenge in \ac{PAC} analysis is the risk of spurious coupling detection \cite{KRAMER2008, Gerber2016, Soldevilla2016, AMIRI2016, Tour2017, Jurkiewicz2021}. The dynamical-systems formulation proposed here provides a systematic framework for addressing this problem, although the approach is at different stages of development for different types of spurious \ac{PAC}. For harmonic-related spurious \ac{PAC}, the method provides an explicit dynamically derived rejection criterion based on instantaneous frequency (Section \ref{sec:avoid_hSC}). The robustness of the proposed method against such artefacts was demonstrated in Section \ref{sec:exp_rslt_spurious_PAC} and Figures \ref{fig:nonsine_spuc_cmd} and \ref{fig:shrp_edg_spuc_cmd}, where the \ac{NARX}-\ac{PAC} method systematically avoided spurious \ac{PAC} arising from harmonics associated with non-sinusoidal waveforms and sharp transients.

For periodic spike-related artefacts, the present method does not yet provide a formal rejection rule. However, with the \ac{NARX}-\ac{PAC} method, the characteristic phase-locked spike pattern is more clearly identifiable than with the other methods considered here, making this type of spurious coupling easier to diagnose and reject. This is important because periodic spikes are known to produce false-positive \ac{PAC} detections across a range of methods \cite{Jurkiewicz2021, Gerber2016, Tour2017}.

By contrast, spurious coupling due to intermodulations has received much less explicit attention in the literature. As far as we are aware, this study is the first to explicitly discuss and examine intermodulation-related spurious \ac{PAC}, and the first to propose a structured solution to this problem. In the present framework, the discriminator-based analysis provides a systematic way to distinguish genuine \ac{PAC} from intermodulation-related spurious \ac{PAC} detections by exploiting the dynamical relationships between the true high-frequency component and its intermodulations. However, a more complete analytical condition remains a subject for future work.

These advantages yield sharper and more frequency-specific localisation than commonly used filtering-based \ac{PAC} measures (supplementary Figs. \ref{fig:rng_PinkNoise_cmd_stat}, \ref{fig:rng_PinkNoise_multpl_cmd_stat}, and \ref{fig:rng_PinkNoise_multpl_noise_cmd_stat}). The method remains robust at an \ac{SNR} of 2 ($6.02\,\mathrm{dB}$) and reasonably robust at an \ac{SNR} of 1 ($0\,\mathrm{dB}$), and performs well for 5-second windows while remaining usable with shorter 2--3 second windows (supplementary Fig. \ref{fig:rng_PinkNoise_multpl_cmd__TmWndw_stat}). 

The \ac{NARX}-\ac{PAC} method specifically captures \ac{QPC} through the second-order nonlinear terms (the term cluster $\Sigma_{u_1u_2}$, Fig. \ref{fig:mthd_exmpl}). Although the bispectrum can be used to detect \ac{QPC} and provides a useful higher-order spectral description of \ac{PAC}-related interactions \cite{KRAMER2008,DVORAK2014,Hyafil2015b,KOVACH2018}, it remains primarily a descriptive statistic and often requires relatively long, high-quality recordings for reliable estimation \cite{Elgar1988,Siu2008,KOVACH2018}. By contrast, the proposed \ac{NARX}-\ac{PAC} method identifies an explicit nonlinear dynamical model of the coupling, allowing the observed interaction to be reduced to a canonical \ac{PAC} form and simulated without noise. This provides several practical advantages: sharper localisation of genuine coupling, direct estimation of modulation strength and preferred phase, and improved robustness to short analysis windows, noise, and spurious \ac{PAC} arising from harmonics or intermodulation-related effects. Thus, while the bispectrum offers a broader view of higher-order frequency interactions, \ac{NARX}-\ac{PAC} provides a more interpretable and \ac{PAC}-specific framework for detection and characterisation.

It should be noted that the methodology presented by Tour \textit{et al.} \cite{Tour2017} shares conceptual similarities with the approach presented here. Their framework employs nonlinear autoregressive models to capture \ac{QPC}, but it focuses exclusively on modelling the amplitude modulation component of a \ac{PAC} signal. By contrast, the \ac{NARX}-based method introduced here models the full \ac{PAC} signal, consistent with the analytical formulations described in Section \ref{sec:PAC_models}.
%
\subsection{Limitations and future work} \label{sec:Concl}
This study demonstrates that a dynamical systems approach to \ac{PAC} detection and characterisation via nonlinear system identification is both advantageous and practically feasible. This opens opportunities for developing further methods based on the same principles. Nonetheless, the \ac{NARX}-based approach presented here does have certain limitations.

The most evident limitation of the proposed method is its computational cost. A separate two-input, single-output \ac{NARX} model must be identified for each low- and high-frequency pair in the comodulogram. A potential solution is to use a multi-input, single-output \ac{NARX} model that incorporates multiple low- and high-frequency inputs simultaneously. This would allow a single model to evaluate multiple frequency pairs, thereby greatly reducing the computational overhead.

The present method uses a second-order input-only \ac{NARX} model with term clusters $\{\Sigma_{u_{1}}, \Sigma_{u_{2}}, \Sigma_{u_{1}u_{2}}\}$ (Section \ref{sec:canon_apprx_sysid}). Extending the model to higher-order clusters, such as $\{\Sigma_{u_{1}^{2}u_{2}}, \Sigma_{u_{1}^{3}u_{2}}, \Sigma_{u_{1}^{2}u_{2}^{3}}\}$, could capture higher-order intermodulations and non-sinusoidal amplitude modulations, which may carry physiological information \cite{COLE2017}. Such extensions may also improve robustness to higher-order phase-locked artefacts. However, higher-order cross-terms could also capture phase-coupling relationships beyond the specific \ac{QPC} structure considered here, so they should be introduced cautiously.

Performance may also improve through data-driven selection of the slow-oscillation bandwidth, such as the adaptive procedure proposed in \cite{Tour2017}. In addition, the optimal sampling rate for \ac{NARX}-based \ac{PAC} identification should be investigated, since it may depend on the fast-oscillation frequency under consideration.

The methodology for mitigating spurious \ac{PAC} due to intermodulations (Section \ref{sec:avoid_iSC}) is not yet complete. Ideally, a simple and effective dynamically derived condition, similar to the one derived in Section \ref{sec:avoid_hSC}, should be established to identify and reject these types of spurious \ac{PAC}. However, deriving such a condition requires a separate study and will be addressed in future work.
%
\section*{Data and Code Availability}
The MATLAB implementation of the proposed \ac{NARX}-based \ac{PAC} method, together with the experimental scripts, saved data, and plotting routines used in this study, is available at \url{https://github.com/raj-gun/NARX-based-PAC}. The repository also includes a standalone version of the \textit{NonSysID-i} routines required by the proposed method and scripts implementing the filtering-based methods used for comparison. The scripts implementing the filtering-based methods were adapted from the MATLAB code accompanying \cite{OZKURT2011} and from the MATLAB routines available in the TortLab GitHub repository \cite{Tortlab}. In these adapted scripts, the high-frequency filter bandwidths were set adaptively according to the low frequency under investigation, following \cite{Berman2012}.
%
\section*{Acknowledgements} 

RG and FH were supported by EPSRC grant [EP/X020193/1].

\FloatBarrier
\bibliographystyle{ieeetr}
\bibliography{citations}

@ARTICLE{Fabus2022,
author={Fabus, Marco S. and Woolrich, Mark W. and Warnaby, Catherine W. and Quinn, Andrew J.},
journal={IEEE Open Journal of Signal Processing}, 
title={Understanding Harmonic Structures Through Instantaneous Frequency}, 
year={2022},
volume={3},
number={},
pages={320-334},
doi={10.1109/OJSP.2022.3198012}
}

@article{CANOLTY2010,
title = {The functional role of cross-frequency coupling},
journal = {Trends in Cognitive Sciences},
volume = {14},
number = {11},
pages = {506-515},
year = {2010},
issn = {1364-6613},
doi = {10.1016/j.tics.2010.09.001},
author = {Ryan T. Canolty and Robert T. Knight}
}

@article{Voloh2015,
author = {Benjamin Voloh  and Taufik A. Valiante  and Stefan Everling  and Thilo Womelsdorf },
title = {Theta–gamma coordination between anterior cingulate and prefrontal cortex indexes correct attention shifts},
journal = {Proceedings of the National Academy of Sciences},
volume = {112},
number = {27},
pages = {8457-8462},
year = {2015},
doi = {10.1073/pnas.1500438112}
}

@article {Bragin1995,
author = {A Bragin and G Jando and Z Nadasdy and J Hetke and K Wise and G Buzsaki},
title = {Gamma (40-100 Hz) oscillation in the hippocampus of the behaving rat},
volume = {15},
number = {1},
pages = {47--60},
year = {1995},
doi = {10.1523/JNEUROSCI.15-01-00047.1995},
publisher = {Society for Neuroscience},
issn = {0270-6474},
journal = {Journal of Neuroscience}
}

@article{Axmacher2010,
author = {Nikolai Axmacher  and Melanie M. Henseler  and Ole Jensen  and Ilona Weinreich  and Christian E. Elger  and Juergen Fell },
title = {Cross-frequency coupling supports multi-item working memory in the human hippocampus},
journal = {Proceedings of the National Academy of Sciences},
volume = {107},
number = {7},
pages = {3228-3233},
year = {2010},
doi = {10.1073/pnas.0911531107},
}

@article {Yanagisawa2012,
author = {Takufumi Yanagisawa and Okito Yamashita and Masayuki Hirata and Haruhiko Kishima and Youichi Saitoh and Tetsu Goto and Toshiki Yoshimine and Yukiyasu Kamitani},
title = {Regulation of Motor Representation by Phase{\textendash}Amplitude Coupling in the Sensorimotor Cortex},
volume = {32},
number = {44},
pages = {15467--15475},
year = {2012},
doi = {10.1523/JNEUROSCI.2929-12.2012},
publisher = {Society for Neuroscience},
issn = {0270-6474},
journal = {Journal of Neuroscience}
}

@article{Roux2014,
title = {Working memory and neural oscillations: alpha–gamma versus theta–gamma codes for distinct WM information?},
journal = {Trends in Cognitive Sciences},
volume = {18},
number = {1},
pages = {16-25},
year = {2014},
issn = {1364-6613},
doi = {10.1016/j.tics.2013.10.010},
author = {Frédéric Roux and Peter J. Uhlhaas},
}

@article{Aurimas2025,
title = {Phase-amplitude coupling during auditory steady-state stimulation: a methodological review},
author = {Aurimas Mockevičius and Inga Griškova-Bulanova},
pages = {577--586},
volume = {36},
number = {6},
journal = {Reviews in the Neurosciences},
doi = {10.1515/revneuro-2024-0165},
year = {2025}
}

@article{ESGHAEI2022,
title = {Dynamic coupling of oscillatory neural activity and its roles in visual attention},
journal = {Trends in Neurosciences},
volume = {45},
number = {4},
pages = {323-335},
year = {2022},
issn = {0166-2236},
doi = {10.1016/j.tins.2022.01.003},
author = {Moein Esghaei and Stefan Treue and Trichur R. Vidyasagar}
}

@article{Lakatos2008,
author = {Peter Lakatos  and George Karmos  and Ashesh D. Mehta  and Istvan Ulbert  and Charles E. Schroeder },
title = {Entrainment of Neuronal Oscillations as a Mechanism of Attentional Selection},
journal = {Science},
volume = {320},
number = {5872},
pages = {110-113},
year = {2008},
doi = {10.1126/science.1154735},
}

@article{Tort2009,
author = {Adriano B. L. Tort  and Robert W. Komorowski  and Joseph R. Manns  and Nancy J. Kopell  and Howard Eichenbaum },
title = {Theta–gamma coupling increases during the learning of item–context associations},
journal = {Proceedings of the National Academy of Sciences},
volume = {106},
number = {49},
pages = {20942-20947},
year = {2009},
doi = {10.1073/pnas.0911331106},
}

@article{Jensen2007,
title = {Cross-frequency coupling between neuronal oscillations},
journal = {Trends in Cognitive Sciences},
volume = {11},
number = {7},
pages = {267-269},
year = {2007},
doi = {10.1016/j.tics.2007.05.003},
author = {Ole Jensen and Laura L. Colgin}
}

@article{Deco2008,
author = {Deco, Gustavo AND Jirsa, Viktor K. AND Robinson, Peter A. AND Breakspear, Michael AND Friston, Karl},
journal = {PLOS Computational Biology},
publisher = {Public Library of Science},
title = {The Dynamic Brain: From Spiking Neurons to Neural Masses and Cortical Fields},
year = {2008},
month = {08},
volume = {4},
pages = {1-35},
number = {8},
doi = {10.1371/journal.pcbi.1000092}
}

@article{Ghosh2008,
author={Ghosh, A. and Rho, Y. and McIntosh, A. R. and K{\"o}tter, R. and Jirsa, V. K.},
title={Cortical network dynamics with time delays reveals functional connectivity in the resting brain},
journal={Cognitive Neurodynamics},
year={2008},
month={06},
day={01},
volume={2},
number={2},
pages={115-120},
doi={10.1007/s11571-008-9044-2}
}

@book{Buzsaki2006,
author = {Buzsáki, György},
title = "{Rhythms of the Brain}",
publisher = {Oxford University Press},
year = {2006},
month = {10},
isbn = {9780195301069},
doi = {10.1093/acprof:oso/9780195301069.001.0001}
}

@article{Chehelcheraghi2017,
doi = {10.1371/journal.pone.0173776},
author = {Chehelcheraghi, Mojtaba AND van Leeuwen, Cees AND Steur, Erik AND Nakatani, Chie},
journal = {PLOS ONE},
publisher = {Public Library of Science},
title = {A neural mass model of cross-frequency coupling},
year = {2017},
month = {04},
volume = {12},
pages = {1-19},
number = {4}
}

@article{Cohen2008,
title = {Assessing transient cross-frequency coupling in EEG data},
journal = {Journal of Neuroscience Methods},
volume = {168},
number = {2},
pages = {494-499},
year = {2008},
doi = {10.1016/j.jneumeth.2007.10.012},
author = {Michael X Cohen}
}

@article{Florin2015,
title = {The brain's resting-state activity is shaped by synchronized cross-frequency coupling of neural oscillations},
journal = {NeuroImage},
volume = {111},
pages = {26-35},
year = {2015},
doi = {10.1016/j.neuroimage.2015.01.054},
author = {Esther Florin and Sylvain Baillet}
}

@ARTICLE{Florin2018, 
AUTHOR={Florin, Esther  and Baillet, Sylvain },         
TITLE={Commentary: Evaluation of Phase-Amplitude Coupling in Resting State Magnetoencephalographic Signals: Effect of Surrogates and Evaluation Approach},        
JOURNAL={Frontiers in Computational Neuroscience},
VOLUME={Volume 12 - 2018},
YEAR={2018},
DOI={10.3389/fncom.2018.00026},
ISSN={1662-5188}
}

@article{Sadaghiani2022,
title = {Connectomics of human electrophysiology},
journal = {NeuroImage},
volume = {247},
pages = {118788},
year = {2022},
doi = {10.1016/j.neuroimage.2021.118788},
author = {Sepideh Sadaghiani and Matthew J Brookes and Sylvain Baillet}
}

@article{Friese2013,
title = {Successful memory encoding is associated with increased cross-frequency coupling between frontal theta and posterior gamma oscillations in human scalp-recorded EEG},
journal = {NeuroImage},
volume = {66},
pages = {642-647},
year = {2013},
doi = {10.1016/j.neuroimage.2012.11.002},
author = {Uwe Friese and Moritz Köster and Uwe Hassler and Ulla Martens and Nelson Trujillo-Barreto and Thomas Gruber}
}

@article{Wang2021,
author = {Wang, Xiaoyue and Cui, Xiaohong and Ding, Congli and Li, Dandan and Cheng, Chen and Wang, Bin and Xiang, Jie},
title = {Deficit of Cross-Frequency Integration in Mild Cognitive Impairment and Alzheimer's Disease: A Multilayer Network Approach},
journal = {Journal of Magnetic Resonance Imaging},
volume = {53},
number = {5},
pages = {1387-1398},
doi = {10.1002/jmri.27453},
year = {2021}
}

@article{Fraga2013,
	doi = {10.1371/journal.pone.0072240},
	author = {Fraga, Francisco and Falk, Tiago and Kanda, Paulo and Anghinah, Renato},
	year = {2013},
	month = {08},
	pages = {e72240},
	title = {Characterizing {A}lzheimer’s Disease Severity via Resting-Awake {EEG} Amplitude Modulation Analysis},
	volume = {8},
	journal = {PloS one},
}

@article {Jackson2019,
author = {Jackson, Nicko and Cole, Scott R. and Voytek, Bradley and Swann, Nicole C.},
title = {Characteristics of Waveform Shape in Parkinson's Disease Detected with Scalp Electroencephalography},
volume = {6},
number = {3},
year = {2019},
doi = {10.1523/ENEURO.0151-19.2019},
publisher = {Society for Neuroscience},
journal = {eNeuro}
}

@article{YAKUBOV2022,
title = {Cross-frequency coupling in psychiatric disorders: A systematic review},
journal = {Neuroscience \& Biobehavioral Reviews},
volume = {138},
pages = {104690},
year = {2022},
issn = {0149-7634},
doi = {10.1016/j.neubiorev.2022.104690},
author = {Boris Yakubov and Sushmit Das and Reza Zomorrodi and Daniel M. Blumberger and Peter G. Enticott and Melissa Kirkovski and Tarek K. Rajji and Pushpal Desarkar}
}

@INPROCEEDINGS{Efstratia2023,
author={Ganiti-Roumeliotou, Efstratia and Ziogas, Ioannis and Lamprou, Charalampos and Alhussein, Ghada and Alfalahi, Hessa and Shehhi, Aamna Al and Dias, Sofia and Jelinek, Herbert F. and Stouraitis, Thanos and Hadjileontiadis, Leontios J.},
booktitle={2023 45th Annual International Conference of the IEEE Engineering in Medicine \& Biology Society (EMBC)}, 
title={Classification of children with ADHD through task-related EEG recordings via Swarm-Decomposition-based Phase Locking Value}, 
year={2023},
volume={},
number={},
pages={1-5},
doi={10.1109/EMBC40787.2023.10340329}
}

@ARTICLE{ChenXi2023,  
AUTHOR={Chen, Xi  and Li, Yingjie  and Li, Renren  and Yuan, Xiao  and Liu, Meng  and Zhang, Wei  and Li, Yunxia },         
TITLE={Multiple cross-frequency coupling analysis of resting-state EEG in patients with mild cognitive impairment and Alzheimer’s disease},        
JOURNAL={Frontiers in Aging Neuroscience},       
VOLUME={Volume 15 - 2023},
YEAR={2023},
DOI={10.3389/fnagi.2023.1142085},
ISSN={1663-4365}
}

@article{TANG2025,
title = {Brain functional differences among ADHD subtypes in children revealed by phase-amplitude coupling analysis of resting-state EEG},
journal = {International Journal of Psychophysiology},
volume = {215},
pages = {113222},
year = {2025},
issn = {0167-8760},
doi = {10.1016/j.ijpsycho.2025.113222},
author = {Wanting Tang and Jiuchuan Jiang and Haixian Wang}
}

@ARTICLE{Murphy2020,  
AUTHOR={Murphy, Nicholas  and Ramakrishnan, Nithya  and Walker, Christopher P.  and Polizzotto, Nicola R.  and Cho, Raymond Y. },         
TITLE={Intact Auditory Cortical Cross-Frequency Coupling in Early and Chronic Schizophrenia},        
JOURNAL={Frontiers in Psychiatry},        
VOLUME={Volume 11 - 2020},
YEAR={2020},
DOI={10.3389/fpsyt.2020.00507},
ISSN={1664-0640}
}

@Article{Mariscal2021,
author={Mariscal, Michael. G.
and Berry-Kravis, Elizabeth
and Buxbaum, Joseph D.
and Ethridge, Lauren E.
and Filip-Dhima, Rajna
and Foss-Feig, Jennifer H.
and Kolevzon, Alexander
and Modi, Meera. E.
and Mosconi, Matthew W.
and Nelson, Charles A.
and Powell, Craig M.
and Siper, Paige M.
and Soorya, Latha
and Thaliath, Andrew
and Thurm, Audrey
and Zhang, Bo
and Sahin, Mustafa
and Levin, April R.
and the Developmental Synaptopathies Consortium},
title={Shifted phase of EEG cross-frequency coupling in individuals with Phelan-McDermid syndrome},
journal={Molecular Autism},
year={2021},
month={04},
day={28},
volume={12},
number={1},
pages={29},
issn={2040-2392},
doi={10.1186/s13229-020-00411-9}
}

@Article{Hirano2018,
author={Hirano, Shogo
and Nakhnikian, Alexander
and Hirano, Yoji
and Oribe, Naoya
and Kanba, Shigenobu
and Onitsuka, Toshiaki
and Levin, Margaret
and Spencer, Kevin M.},
title={Phase-Amplitude Coupling of the Electroencephalogram in the Auditory Cortex in Schizophrenia},
journal={Biological Psychiatry: Cognitive Neuroscience and Neuroimaging},
year={2018},
month={01},
day={01},
publisher={Elsevier},
volume={3},
number={1},
pages={69-76},
issn={2451-9022},
doi={10.1016/j.bpsc.2017.09.001}
}

@Article{DePieri2025,
author={De Pieri, Marco
and Sabe, Michel
and Rochas, Vincent
and Poglia, Greta
and Bartolomei, Javier
and Kirschner, Matthias
and Kaiser, Stefan},
title={Resting-state EEG and MEG gamma frequencies in schizophrenia: a systematic review and exploratory power-spectrum meta-analysis},
journal={Schizophrenia},
year={2025},
month={03},
day={21},
volume={11},
number={1},
pages={48},
issn={2754-6993},
doi={10.1038/s41537-025-00596-z}
}

@ARTICLE{Sacks2021,  
AUTHOR={Sacks, Dashiell D.  and Schwenn, Paul E.  and McLoughlin,  Larisa T.  and Lagopoulos, Jim  and Hermens, Daniel F. },         
TITLE={Phase–Amplitude Coupling, Mental Health and Cognition: Implications for Adolescence},        
JOURNAL={Frontiers in Human Neuroscience},       
VOLUME={Volume 15 - 2021},
YEAR={2021},
DOI={10.3389/fnhum.2021.622313},
ISSN={1662-5161}
}

@ARTICLE{Wang2025,  
AUTHOR={Wang, Hongli  and Shi, Xiaoning  and Wang, Chenyang  and Qi, Yongsheng  and Gao, Michel  and Zhao, Yingying },         
TITLE={Cross-frequency coupling between low frequency and gamma oscillations altered in cognitive biotype of depression},        
JOURNAL={Frontiers in Psychiatry},      
VOLUME={Volume 16 - 2025},
YEAR={2025},
DOI={10.3389/fpsyt.2025.1596191},
ISSN={1664-0640}
}

@article{KLEPL2023,
title = {Cross-Frequency Multilayer Network Analysis with Bispectrum-based Functional Connectivity: A Study of Alzheimer’s Disease},
journal = {Neuroscience},
volume = {521},
pages = {77-88},
year = {2023},
issn = {0306-4522},
doi = {10.1016/j.neuroscience.2023.04.008},
author = {Dominik Klepl and Fei He and Min Wu and Daniel J. Blackburn and Ptolemaios G. Sarrigiannis}
}

@article{Muthuraman2020,
author = {Muthuraman, Muthuraman and Bange, Manuel and Koirala, Nabin and Ciolac, Dumitru and Pintea, Bogdan and Glaser, Martin and Tinkhauser, Gerd and Brown, Peter and Deuschl, Günther and Groppa, Sergiu},
title = {Cross-frequency coupling between gamma oscillations and deep brain stimulation frequency in Parkinson’s disease},
journal = {Brain},
volume = {143},
number = {11},
pages = {3393-3407},
year = {2020},
month = {11},
issn = {0006-8950},
doi = {10.1093/brain/awaa297}
}

@article{Tort2008,
author = {Adriano B. L. Tort  and Mark A. Kramer  and Catherine Thorn  and Daniel J. Gibson  and Yasuo Kubota  and Ann M. Graybiel  and Nancy J. Kopell },
title = {Dynamic cross-frequency couplings of local field potential oscillations in rat striatum and hippocampus during performance of a T-maze task},
journal = {Proceedings of the National Academy of Sciences},
volume = {105},
number = {51},
pages = {20517-20522},
year = {2008},
doi = {10.1073/pnas.0810524105}
}

@article{FIEBELKORN2019,
title = {A Rhythmic Theory of Attention},
journal = {Trends in Cognitive Sciences},
volume = {23},
number = {2},
pages = {87-101},
year = {2019},
issn = {1364-6613},
doi = {10.1016/j.tics.2018.11.009},
author = {Ian C. Fiebelkorn and Sabine Kastner}
}

@article{KRAMER2008,
title = {Sharp edge artifacts and spurious coupling in EEG frequency comodulation measures},
journal = {Journal of Neuroscience Methods},
volume = {170},
number = {2},
pages = {352-357},
year = {2008},
doi = {10.1016/j.jneumeth.2008.01.020},
author = {Mark A. Kramer and Adriano B.L. Tort and Nancy J. Kopell}
}

@article{DVORAK2014,
title = {Toward a proper estimation of phase–amplitude coupling in neural oscillations},
journal = {Journal of Neuroscience Methods},
volume = {225},
pages = {42-56},
year = {2014},
issn = {0165-0270},
doi = {10.1016/j.jneumeth.2014.01.002},
author = {Dino Dvorak and André A. Fenton},
}

@article {Jensen2016,
author = {Ole Jensen and Eelke Spaak and Hyojin Park},
title = {Discriminating Valid from Spurious Indices of Phase-Amplitude Coupling},
volume = {3},
number = {6},
elocation-id = {ENEURO.0334-16.2016},
year = {2016},
doi = {10.1523/ENEURO.0334-16.2016},
publisher = {Society for Neuroscience},
journal = {eNeuro}
}

@article{Dellavale2020,
title = {Complex interplay between spectral harmonicity and different types of cross-frequency couplings in nonlinear oscillators and biologically plausible neural network models},
author = {Dellavale, Dami\'an and Velarde, Osvaldo Mat\'{\i}as and Mato, Germ\'an and Urdapilleta, Eugenio},
journal = {Phys. Rev. E},
volume = {102},
issue = {6},
pages = {062401},
numpages = {43},
year = {2020},
month = {12},
publisher = {American Physical Society},
doi = {10.1103/PhysRevE.102.062401}
}

@article{ZANDVOORT2021,
title = {Defining the filter parameters for phase-amplitude coupling from a bispectral point of view},
journal = {Journal of Neuroscience Methods},
volume = {350},
pages = {109032},
year = {2021},
issn = {0165-0270},
doi = {10.1016/j.jneumeth.2020.109032},
author = {Coen S. Zandvoort and Guido Nolte}
}

@article{GIEHL2021,
title = {Dissociating harmonic and non-harmonic phase-amplitude coupling in the human brain},
journal = {NeuroImage},
volume = {227},
pages = {117648},
year = {2021},
issn = {1053-8119},
doi = {10.1016/j.neuroimage.2020.117648},
author = {Janet Giehl and Nima Noury and Markus Siegel}
}

@article{VELARDE2019,
title = {Bifurcation structure determines different phase-amplitude coupling patterns in the activity of biologically plausible neural networks},
journal = {NeuroImage},
volume = {202},
pages = {116031},
year = {2019},
issn = {1053-8119},
doi = {10.1016/j.neuroimage.2019.116031},
author = {Osvaldo Matías Velarde and Eugenio Urdapilleta and Germán Mato and Damián Dellavale}
}

@article{Colgin2009,
author={Colgin, Laura Lee and Denninger, Tobias and Fyhn, Marianne and Hafting, Torkel and Bonnevie, Tora and Jensen, Ole and Moser, May-Britt and Moser, Edvard I.},
title={Frequency of gamma oscillations routes flow of information in the hippocampus},
journal={Nature},
year={2009},
month={11},
day={01},
volume={462},
number={7271},
pages={353-357},
issn={1476-4687},
doi={10.1038/nature08573}
}

@article {Belluscio2012,
author = {Mariano A. Belluscio and Kenji Mizuseki and Robert Schmidt and Richard Kempter and Gy{\"o}rgy Buzs{\'a}ki},
title = {Cross-Frequency Phase{\textendash}Phase Coupling between Theta and Gamma Oscillations in the Hippocampus},
volume = {32},
number = {2},
pages = {423--435},
year = {2012},
doi = {10.1523/JNEUROSCI.4122-11.2012},
publisher = {Society for Neuroscience},
issn = {0270-6474},
journal = {Journal of Neuroscience}
}

@article{COLE2017,
title = {Brain Oscillations and the Importance of Waveform Shape},
journal = {Trends in Cognitive Sciences},
volume = {21},
number = {2},
pages = {137-149},
year = {2017},
issn = {1364-6613},
doi = {10.1016/j.tics.2016.12.008},
author = {Scott R. Cole and Bradley Voytek}
}

@article{ARU2015,
title = {Untangling cross-frequency coupling in neuroscience},
journal = {Current Opinion in Neurobiology},
volume = {31},
pages = {51-61},
year = {2015},
note = {SI: Brain rhythms and dynamic coordination},
issn = {0959-4388},
doi = {10.1016/j.conb.2014.08.002},
author = {Juhan Aru and Jaan Aru and Viola Priesemann and Michael Wibral and Luiz Lana and Gordon Pipa and Wolf Singer and Raul Vicente}
}

@article{JIANG2015,
title = {Measuring directionality between neuronal oscillations of different frequencies},
journal = {NeuroImage},
volume = {118},
pages = {359-367},
year = {2015},
issn = {1053-8119},
doi = {10.1016/j.neuroimage.2015.05.044},
author = {Haiteng Jiang and Ali Bahramisharif and Marcel A.J. {van Gerven} and Ole Jensen}
}

@ARTICLE{Jirsa2013,
AUTHOR={Jirsa, Viktor and Müller, Viktor},   
TITLE={Cross-frequency coupling in real and virtual brain networks},      
JOURNAL={Frontiers in Computational Neuroscience},      
VOLUME={7},           	
YEAR={2013},      
DOI={10.3389/fncom.2013.00078}
}

@Article{Nandi2019,
author={Nandi, Bijurika
and Swiatek, Peter
and Kocsis, Bernat
and Ding, Mingzhou},
title={Inferring the direction of rhythmic neural transmission via inter-regional phase-amplitude coupling (ir-PAC)},
journal={Scientific Reports},
year={2019},
month={05},
day={06},
volume={9},
number={1},
pages={6933},
issn={2045-2322},
doi={10.1038/s41598-019-43272-w}
}

@article{Onslow2014,
doi = {10.1371/journal.pone.0102591},
author = {Onslow, Angela C. E. AND Jones, Matthew W. AND Bogacz, Rafal},
journal = {PLOS ONE},
publisher = {Public Library of Science},
title = {A Canonical Circuit for Generating Phase-Amplitude Coupling},
year = {2014},
month = {08},
volume = {9},
pages = {1-15},
number = {8}
}

@article{CHACKO2018,
title = {Distinct phase-amplitude couplings distinguish cognitive processes in human attention},
journal = {NeuroImage},
volume = {175},
pages = {111-121},
year = {2018},
issn = {1053-8119},
doi = {10.1016/j.neuroimage.2018.03.003},
author = {Ravi V. Chacko and Byungchan Kim and Suh Woo Jung and Amy L. Daitch and Jarod L. Roland and Nicholas V. Metcalf and Maurizio Corbetta and Gordon L. Shulman and Eric C. Leuthardt}
}

@article{Dong2022,
title = {Intrinsic phase–amplitude coupling on multiple spatial scales during the loss and recovery of consciousness},
journal = {Computers in Biology and Medicine},
volume = {147},
pages = {105687},
year = {2022},
issn = {0010-4825},
doi = {10.1016/j.compbiomed.2022.105687},
author = {Kangli Dong and Delin Zhang and Qishun Wei and Guozheng Wang and Fan Huang and Xing Chen and Kanhar G. Muhammad and Yu Sun and Jun Liu},
}

@ARTICLE{Soldevilla2016,  
AUTHOR={Lozano-Soldevilla, Diego  and ter Huurne, Niels  and Oostenveld, Robert },         
TITLE={Neuronal Oscillations with Non-sinusoidal Morphology Produce Spurious Phase-to-Amplitude Coupling and Directionality},        
JOURNAL={Frontiers in Computational Neuroscience},
VOLUME={Volume 10 - 2016},
YEAR={2016},
DOI={10.3389/fncom.2016.00087},
ISSN={1662-5188}
}

@article{AMIRI2016,
title = {High Frequency Oscillations and spikes: Separating real HFOs from false oscillations},
journal = {Clinical Neurophysiology},
volume = {127},
number = {1},
pages = {187-196},
year = {2016},
issn = {1388-2457},
doi = {10.1016/j.clinph.2015.04.290},
author = {Mina Amiri and Jean-Marc Lina and Francesca Pizzo and Jean Gotman}
}

@article{Friston2000,
author = {Friston, Karl J. },
title = {The labile brain. I. Neuronal transients and nonlinear coupling},
journal = {Philosophical Transactions of the Royal Society of London. Series B: Biological Sciences},
volume = {355},
number = {1394},
pages = {215-236},
year = {2000},
doi = {10.1098/rstb.2000.0560}
}

@article{Hyafil2015,
title = {Neural Cross-Frequency Coupling: Connecting Architectures, Mechanisms, and Functions},
journal = {Trends in Neurosciences},
volume = {38},
number = {11},
pages = {725-740},
year = {2015},
author = {Alexandre Hyafil and Anne-Lise Giraud and Lorenzo Fontolan and Boris Gutkin},
doi = {10.1016/j.tins.2015.09.001}
}

@article{Stone1974,
ISSN = {00359246},
author = {M. Stone},
journal = {Journal of the Royal Statistical Society. Series B (Methodological)},
number = {2},
pages = {111--147},
publisher = {[Royal Statistical Society, Oxford University Press]},
title = {Cross-Validatory Choice and Assessment of Statistical Predictions},
urldate = {2024-11-01},
volume = {36},
doi = {10.1111/j.2517-6161.1974.tb00994.x},
year = {1974}
}

@article{Little2017,
author = {Little, Max A and Varoquaux, Gael and Saeb, Sohrab and Lonini, Luca and Jayaraman, Arun and Mohr, David C and Kording, Konrad P},
title = "{Using and understanding cross-validation strategies. Perspectives on Saeb et. al.}",
journal = {GigaScience},
volume = {6},
number = {5},
pages = {gix020},
year = {2017},
month = {03},
issn = {2047-217X},
doi = {10.1093/gigascience/gix020}
}

@inbook{LathiGreen2017,
title        = {Linear Systems and Signals},
author       = {Lathi, Bhagwandas Pannalal and Green, Roger A.},
edition      = {3},
year         = {2017},
publisher    = {Oxford University Press},
address      = {Oxford, UK},
pages        = {18--20},
chapter      = {B.2},
isbn         = {978-0-19-020017-6}
}

@article{Gunawardena2024,
title = {NonSysId: A nonlinear system identification package with improved model term selection for NARMAX models}, 
author = {Rajintha Gunawardena and Zi-Qiang Lang and Fei He},
journal = {Journal of Open Source Software},
year = {2025},
doi  = {10.21105/joss.08028},
}

@misc{nonsysid,
author = {Rajintha Gunawardena and Zi-Qiang Lang and Fei He},
title = {{G}it{H}ub - {N}on{S}ys{I}{D}: {A} {M}at{L}ab package for {S}ystem {I}dentification using linear and nonlinear auto-regresive models--({N}){AR}, ({N}){ARX} and ({N}){ARMAX} models},
year = {2024},
publisher = {GitHub},
journal = {GitHub repository},
howpublished = {\url{https://github.com/raj-gun/NonSysID}},
}

@article{Wei2004,
title={Term and variable selection for non-linear system identification},
author={Wei, Hua-Liang and Billings, Steve A and Liu, Jianhua},
journal={International Journal of Control},
volume={77},
number={1},
pages={86--110},
year={2004},
publisher={Taylor \& Francis},
doi = {10.1080/00207170310001639640}
}

@Article{Herrmann2001,
author={Herrmann, Christoph S.},
title={Human {EEG} responses to 1--100Hz flicker: resonance phenomena in visual cortex and their potential correlation to cognitive phenomena},
journal={Experimental Brain Research},
year={2001},
month={04},
day={01},
volume={137},
number={3},
pages={346-353},
issn={1432-1106},
doi={10.1007/s002210100682}
}

@article {Herrmann2016,
author = {Herrmann, Christoph S. and Murray, Micah M. and Ionta, Silvio and Hutt, Axel and Lefebvre, J{\'e}r{\'e}mie},
title = {Shaping Intrinsic Neural Oscillations with Periodic Stimulation},
volume = {36},
number = {19},
pages = {5328--5337},
year = {2016},
doi = {10.1523/JNEUROSCI.0236-16.2016},
publisher = {Society for Neuroscience},
issn = {0270-6474},
journal = {Journal of Neuroscience}
}

@article{TOBIMATSU1999,
title = {Steady-state vibration somatosensory evoked potentials: physiological characteristics and tuning function},
journal = {Clinical Neurophysiology},
volume = {110},
number = {11},
pages = {1953-1958},
year = {1999},
issn = {1388-2457},
doi = {10.1016/S1388-2457(99)00146-7},
author = {Shozo Tobimatsu and You Min Zhang and Motohiro Kato}
}

@article{ChenX_2013,
doi = {10.1088/1741-2560/10/6/066009},
year = {2013},
month = {10},
publisher = {IOP Publishing},
volume = {10},
number = {6},
pages = {066009},
author = {Xiaogang Chen and Zhikai Chen and Shangkai Gao and Xiaorong Gao},
title = {Brain–computer interface based on intermodulation frequency},
journal = {Journal of Neural Engineering}
}

@article{LUFF2024,
title = {The neuron mixer and its impact on human brain dynamics},
journal = {Cell Reports},
volume = {43},
number = {6},
pages = {114274},
year = {2024},
issn = {2211-1247},
doi = {10.1016/j.celrep.2024.114274},
author = {Charlotte E. Luff and Robert Peach and Emma-Jane Mallas and Edward Rhodes and Felix Laumann and Edward S. Boyden and David J. Sharp and Mauricio Barahona and Nir Grossman}
}

@article{GIANI2012,
title = {Steady-state responses in MEG demonstrate information integration within but not across the auditory and visual senses},
journal = {NeuroImage},
volume = {60},
number = {2},
pages = {1478-1489},
year = {2012},
issn = {1053-8119},
doi = {10.1016/j.neuroimage.2012.01.114},
author = {Anette S. Giani and Erick Ortiz and Paolo Belardinelli and Mario Kleiner and Hubert Preissl and Uta Noppeney},
}

@article{GORDON2019,
title = {From intermodulation components to visual perception and cognition-a review},
journal = {NeuroImage},
volume = {199},
pages = {480-494},
year = {2019},
issn = {1053-8119},
doi = {10.1016/j.neuroimage.2019.06.008},
author = {Noam Gordon and Jakob Hohwy and Matthew James Davidson and Jeroen J.A. {van Boxtel} and Naotsugu Tsuchiya}
}

@ARTICLE{Elgar1988,
author={Elgar, S. and Guza, R.T.},
journal={IEEE Transactions on Acoustics, Speech, and Signal Processing}, 
title={Statistics of bicoherence}, 
year={1988},
volume={36},
number={10},
pages={1667-1668},
doi={10.1109/29.7555}
}

@ARTICLE{Siu2008,
author={Siu, Kin L. and Ahn, Jae M. and Ju, Kihwan and Lee, Myoungho and Shin, Kunsoo and Chon, Ki H.},
journal={IEEE Transactions on Biomedical Engineering}, 
title={Statistical Approach to Quantify the Presence of Phase Coupling Using the Bispectrum}, 
year={2008},
volume={55},
number={5},
pages={1512-1520},
doi={10.1109/TBME.2007.913418}
}

@article{LANCASTER2018,
title = {Surrogate data for hypothesis testing of physical systems},
journal = {Physics Reports},
volume = {748},
pages = {1-60},
year = {2018},
note = {Surrogate data for hypothesis testing of physical systems},
issn = {0370-1573},
doi = {10.1016/j.physrep.2018.06.001},
author = {Gemma Lancaster and Dmytro Iatsenko and Aleksandra Pidde and Valentina Ticcinelli and Aneta Stefanovska}
}

@ARTICLE{Hyafil2015b,
AUTHOR={Hyafil, Alexandre },
TITLE={Misidentifications of specific forms of cross-frequency coupling: three warnings},
JOURNAL={Frontiers in Neuroscience},
VOLUME={9},
YEAR={2015},
DOI={10.3389/fnins.2015.00370},
ISSN={1662-453X}
}

@article{KOVACH2018,
title = {The bispectrum and its relationship to phase-amplitude coupling},
journal = {NeuroImage},
volume = {173},
pages = {518-539},
year = {2018},
issn = {1053-8119},
doi = {10.1016/j.neuroimage.2018.02.033},
author = {Christopher K. Kovach and Hiroyuki Oya and Hiroto Kawasaki}
}

@article{WITTE2000,
title = {Quantification of transient quadratic phase couplings within EEG burst patterns in sedated patients during electroencephalic burst-suppression period},
journal = {Journal of Physiology-Paris},
volume = {94},
number = {5},
pages = {427-434},
year = {2000},
issn = {0928-4257},
doi = {10.1016/S0928-4257(00)01086-X},
author = {Herbert Witte and Bärbel Schack and Marko Helbig and Peter Putsche and Christoph Schelenz and Karin Schmidt and Martin Specht}
}

@Article{Sigl1994,
author={Sigl, Jeffrey C.
and Chamoun, Nassib G.},
title={An introduction to bispectral analysis for the electroencephalogram},
journal={Journal of Clinical Monitoring},
year={1994},
month={11},
day={01},
volume={10},
number={6},
pages={392-404},
issn={1573-2614},
doi={10.1007/BF01618421}
}

@article{Yang2016,
author = {Yang, Yuan and Solis-Escalante, Teodoro and Yao, Jun and Daffertshofer, Andreas and Schouten, Alfred C. and van der Helm, Frans C. T.},
title = {A General Approach for Quantifying Nonlinear Connectivity in the Nervous System Based on Phase Coupling},
journal = {International Journal of Neural Systems},
volume = {26},
number = {01},
pages = {1550031},
year = {2016},
doi = {10.1142/S0129065715500318}
}

@inproceedings{hasselmann1963,
title={Bispectra of ocean waves},
author={Hasselmann, Klaus and Munk, Walter and MacDonald, Gordon JF},
booktitle={Symposium on time series analysis},
pages={125--139},
year={1963},
organization={Brown University},
editor={Rodenblatt, Murray},
publisher={John Wiley, New York}
}

@article{Prendergast2010, 
author = {Prendergast, Garreth and Johnson, Sam R. and Green, Gary G. R.},
title = {Temporal dynamics of sinusoidal and non-sinusoidal amplitude modulation},
journal = {European Journal of Neuroscience},
volume = {32},
number = {9},
pages = {1599-1607},
doi = {10.1111/j.1460-9568.2010.07423.x},
year = {2010}
}

@article{Gerber2016,
doi = {10.1371/journal.pone.0167351},
author = {Gerber, Edden M. AND Sadeh, Boaz AND Ward, Andrew AND Knight, Robert T. AND Deouell, Leon Y.},
journal = {PLOS ONE},
publisher = {Public Library of Science},
title = {Non-Sinusoidal Activity Can Produce Cross-Frequency Coupling in Cortical Signals in the Absence of Functional Interaction between Neural Sources},
year = {2016},
month = {12},
volume = {11},
url = {https://doi.org/10.1371/journal.pone.0167351},
pages = {1-19},
}

@ARTICLE{Seymour2017,
AUTHOR={Seymour, Robert A.  and Rippon, Gina  and Kessler, Klaus },    
TITLE={The Detection of Phase Amplitude Coupling during Sensory Processing},    
JOURNAL={Frontiers in Neuroscience},    
VOLUME={Volume 11 - 2017},
YEAR={2017},
DOI={10.3389/fnins.2017.00487},
ISSN={1662-453X},
}

@article{Teixeira2011,
author = {Scheffer-Teixeira, Robson and Belchior, Hindiael and Caixeta, Fábio V. and Souza, Bryan C. and Ribeiro, Sidarta and Tort, Adriano B. L.},
title = {Theta Phase Modulates Multiple Layer-Specific Oscillations in the CA1 Region},
journal = {Cerebral Cortex},
volume = {22},
number = {10},
pages = {2404-2414},
year = {2011},
month = {11},
issn = {1047-3211},
doi = {10.1093/cercor/bhr319}
}

@misc{Tortlab,
author = {Tortlab},
title = {{G}it{H}ub - tortlab/phase-amplitude-coupling: {M}atlab routines for computing the {M}odulation {I}ndex and {C}omodulogram, as described in {T}ort et al., {J} {N}europhysiol 2010 --- github.com},
howpublished = {\url{https://github.com/tortlab/phase-amplitude-coupling}},
year = {2018},
note = {Accessed 06-08-2025},
}

@article{Tort2010,
author = {Tort, Adriano B. L. and Komorowski, Robert and Eichenbaum, Howard and Kopell, Nancy},
title = {Measuring Phase-Amplitude Coupling Between Neuronal Oscillations of Different Frequencies},
journal = {Journal of Neurophysiology},
volume = {104},
number = {2},
pages = {1195-1210},
year = {2010},
doi = {10.1152/jn.00106.2010}
}

@article{OZKURT2011,
title = {A critical note on the definition of phase–amplitude cross-frequency coupling},
journal = {Journal of Neuroscience Methods},
volume = {201},
number = {2},
pages = {438-443},
year = {2011},
issn = {0165-0270},
doi = {10.1016/j.jneumeth.2011.08.014},
author = {Tolga Esat Özkurt and Alfons Schnitzler},
}

@Article{Jurkiewicz2021,
author={Jurkiewicz, Gabriela J.
and Hunt, Mark J.
and {\.{Z}}ygierewicz, Jaros{\l}aw},
title={Addressing Pitfalls in Phase-Amplitude Coupling Analysis with an Extended Modulation Index Toolbox},
journal={Neuroinformatics},
year={2021},
month={04},
day={01},
volume={19},
number={2},
pages={319-345},
issn={1559-0089},
doi={10.1007/s12021-020-09487-3}
}

@article{Berman2012,
author = {Jeffrey I. Berman and Jonathan McDaniel and Song Liu and Lauren Cornew and William Gaetz and Timothy P.L. Roberts and J. Christopher Edgar},
title ={Variable Bandwidth Filtering for Improved Sensitivity of Cross-Frequency Coupling Metrics},
journal = {Brain Connectivity},
volume = {2},
number = {3},
pages = {155-163},
year = {2012},
doi = {10.1089/brain.2012.0085},
}

@article{Tour2017,
doi = {10.1371/journal.pcbi.1005893},
author = {Dupré la Tour, Tom AND Tallot, Lucille AND Grabot, Laetitia AND Doyère, Valérie AND van Wassenhove, Virginie AND Grenier, Yves AND Gramfort, Alexandre},
journal = {PLOS Computational Biology},
publisher = {Public Library of Science},
title = {Non-linear auto-regressive models for cross-frequency coupling in neural time series},
year = {2017},
month = {12},
volume = {13},
pages = {1-32},
number = {12}
}

@article{VANWIJK2015,
title = {Parametric estimation of cross-frequency coupling},
journal = {Journal of Neuroscience Methods},
volume = {243},
pages = {94-102},
year = {2015},
issn = {0165-0270},
doi = {10.1016/j.jneumeth.2015.01.032},
author = {B.C.M. {van Wijk} and A. Jha and W. Penny and V. Litvak}
}

@article{Canolty2006,
author = {R. T. Canolty  and E. Edwards  and S. S. Dalal  and M. Soltani  and S. S. Nagarajan  and H. E. Kirsch  and M. S. Berger  and N. M. Barbaro  and R. T. Knight },
title = {High Gamma Power Is Phase-Locked to Theta Oscillations in Human Neocortex},
journal = {Science},
volume = {313},
number = {5793},
pages = {1626-1628},
year = {2006},
doi = {10.1126/science.1128115}
}

@article{PENNY2008,
title = {Testing for nested oscillation},
journal = {Journal of Neuroscience Methods},
volume = {174},
number = {1},
pages = {50-61},
year = {2008},
issn = {0165-0270},
doi = {10.1016/j.jneumeth.2008.06.035},
author = {W.D. Penny and E. Duzel and K.J. Miller and J.G. Ojemann}
}

@ARTICLE{Hulsemann2019,
AUTHOR={Hülsemann, Mareike J.  and Naumann, Ewald  and Rasch, Björn },
TITLE={Quantification of Phase-Amplitude Coupling in Neuronal Oscillations: Comparison of Phase-Locking Value, Mean Vector Length, Modulation Index, and Generalized-Linear-Modeling-Cross-Frequency-Coupling},
JOURNAL={Frontiers in Neuroscience},
VOLUME={13},
YEAR={2019},
DOI={10.3389/fnins.2019.00573},
ISSN={1662-453X}
}

@article{ONSLOW2011,
title = {Quantifying phase–amplitude coupling in neuronal network oscillations},
journal = {Progress in Biophysics and Molecular Biology},
volume = {105},
number = {1},
pages = {49-57},
year = {2011},
note = {BrainModes: The role of neuronal oscillations in health and disease},
issn = {0079-6107},
doi = {10.1016/j.pbiomolbio.2010.09.007},
author = {Angela C.E. Onslow and Rafal Bogacz and Matthew W. Jones}
}

@article{KRAMER2013,
title = {Assessment of cross-frequency coupling with confidence using generalized linear models},
author = {M.A. Kramer and U.T. Eden},
journal = {Journal of Neuroscience Methods},
volume = {220},
number = {1},
pages = {64-74},
year = {2013},
issn = {0165-0270},
doi = {10.1016/j.jneumeth.2013.08.006},
}

@article{Caiola2019,
doi = {10.1371/journal.pone.0219264},
author = {Caiola, Michael AND Devergnas, Annaelle AND Holmes, Mark H. AND Wichmann, Thomas},
journal = {PLOS ONE},
publisher = {Public Library of Science},
title = {Empirical analysis of phase-amplitude coupling approaches},
year = {2019},
month = {07},
volume = {14},
pages = {1-18},
number = {7}
}

@article{Miskovic2019,
author = {Miskovic, Vladimir and MacDonald, Kevin J. and Rhodes, L. Jack and Cote, Kimberly A.},
title = {Changes in EEG multiscale entropy and power-law frequency scaling during the human sleep cycle},
journal = {Human Brain Mapping},
volume = {40},
number = {2},
pages = {538-551},
doi = {10.1002/hbm.24393},
year = {2019}
}

@article{BiyuHE2014,
title = {Scale-free brain activity: past, present, and future},
journal = {Trends in Cognitive Sciences},
volume = {18},
number = {9},
pages = {480-487},
year = {2014},
issn = {1364-6613},
doi = {10.1016/j.tics.2014.04.003},
author = {Biyu J. He}
}

@article{Pettersen2014,
doi = {10.1371/journal.pcbi.1003928},
author = {Pettersen, Klas H. AND Lindén, Henrik AND Tetzlaff, Tom AND Einevoll, Gaute T.},
journal = {PLOS Computational Biology},
publisher = {Public Library of Science},
title = {Power Laws from Linear Neuronal Cable Theory: Power Spectral Densities of the Soma Potential, Soma Membrane Current and Single-Neuron Contribution to the EEG},
year = {2014},
month = {11},
volume = {10},
pages = {1-26},
number = {11}
}

@Article{Ao2025,
author={Ao, Yujia
and Klar, Philipp
and Catal, Yasir
and Wang, Yifeng
and Northoff, Georg},
title={Infra-slow scale-free dynamics modulate the connection of neural and behavioral variability during attention},
journal={Communications Biology},
year={2025},
month={07},
day={16},
volume={8},
number={1},
pages={1057},
issn={2399-3642},
doi={10.1038/s42003-025-08448-3}
}

@article{Chiras2002,
author = {Chiras , N.  and Evans , C.  and Rees, D. },
title = "{Global Nonlinear Modeling of Gas Turbine Dynamics Using NARMAX Structures }",
journal = {Journal of Engineering for Gas Turbines and Power},
volume = {124},
number = {4},
pages = {817-826},
year = {2002},
month = {09},
issn = {0742-4795},
doi = {10.1115/1.1470483}
}

@incollection{WANG2024,
title = {Nonlinear autoregressive exogenous method for structural health monitoring using ultrasonic guided waves},
editor = {Fuh-Gwo Yuan},
booktitle = {Structural Health Monitoring/ Management (SHM) in Aerospace Structures},
publisher = {Woodhead Publishing},
pages = {427-452},
year = {2024},
series = {Woodhead Publishing Series in Composites Science and Engineering},
isbn = {978-0-443-15476-8},
doi = {10.1016/B978-0-443-15476-8.00004-6},
author = {Kangwei Wang and Jie Zhang and Anthony J. Croxford and Yong Yang}
}

@article{ZAINOL2022,
title = {Estimating the incidence of spontaneous breathing effort of mechanically ventilated patients using a non-linear auto regressive (NARX) model},
journal = {Computer Methods and Programs in Biomedicine},
volume = {220},
pages = {106835},
year = {2022},
issn = {0169-2607},
doi = {10.1016/j.cmpb.2022.106835},
author = {Nurhidayah Mohd Zainol and Nor Salwa Damanhuri and Nor Azlan Othman and Yeong Shiong Chiew and Mohd Basri Mat Nor and Zuraida Muhammad and J. Geoffrey Chase}
}

@article{RITZBERGER2017,
title = {Nonlinear data-driven identification of polymer electrolyte membrane fuel cells for diagnostic purposes: A Volterra series approach},
journal = {Journal of Power Sources},
volume = {361},
pages = {144-152},
year = {2017},
issn = {0378-7753},
doi = {10.1016/j.jpowsour.2017.06.068},
author = {D. Ritzberger and S. Jakubek}
}

@Article{Gao2023,
AUTHOR = {Gao, Yi and Yu, Changshuai and Zhu, Yun-Peng and Luo, Zhong},
TITLE = {A {NARX} Model-Based Condition Monitoring Method for Rotor Systems},
JOURNAL = {Sensors},
VOLUME = {23},
YEAR = {2023},
NUMBER = {15},
ARTICLE-NUMBER = {6878},
PubMedID = {37571661},
ISSN = {1424-8220},
DOI = {10.3390/s23156878}
}

@article{HE2016,
title = {Nonlinear interactions in the thalamocortical loop in essential tremor: A model-based frequency domain analysis},
journal = {Neuroscience},
volume = {324},
pages = {377-389},
year = {2016},
issn = {0306-4522},
doi = {10.1016/j.neuroscience.2016.03.028},
author = {F. He and P.G. Sarrigiannis and S.A. Billings and H. Wei and J. Rowe and C. Romanowski and N. Hoggard and M. Hadjivassilliou and D.G. Rao and R. Grünewald and A. Khan and J. Yianni}
}

@article{HE2021,
title = {Nonlinear System Identification of Neural Systems from Neurophysiological Signals},
journal = {Neuroscience},
volume = {458},
pages = {213-228},
year = {2021},
issn = {0306-4522},
doi = {10.1016/j.neuroscience.2020.12.001},
author = {Fei He and Yuan Yang}
}

@article{LIU2024,
title = {Digital twin-based anomaly detection for real-time tool condition monitoring in machining},
journal = {Journal of Manufacturing Systems},
volume = {75},
pages = {163-173},
year = {2024},
issn = {0278-6125},
doi = {10.1016/j.jmsy.2024.06.004},
author = {Zepeng Liu and Zi-Qiang Lang and Yufei Gui and Yun-Peng Zhu and Hatim Laalej}
}

@article{Chen1989a,
author = {S. Chen and S. A. Billings},
title = {Representations of non-linear systems: the NARMAX model},
journal = {International Journal of Control},
volume = {49},
number = {3},
pages = {1013--1032},
year = {1989},
publisher = {Taylor \& Francis},
doi = {10.1080/00207178908559683}
}

@article{CHEN1989b,
author = {S. Chen and S. A. Billings and W. Luo},
title = {Orthogonal least squares methods and their application to non-linear system identification},
journal = {International Journal of Control},
volume = {50},
number = {5},
pages = {1873--1896},
year = {1989},
publisher = {Taylor \& Francis},
doi = {10.1080/00207178908953472}
}

@article{mendes1998a,
author = {Mendes, Eduardo M. A. M. and Billings, S. A.},
title = {On Overparametrization of Nonlinear Discrete Systems},
journal = {International Journal of Bifurcation and Chaos},
volume = {08},
number = {03},
pages = {535-556},
year = {1998},
doi = {10.1142/S0218127498000346}
}

@article{guo2015a,
author = {Yuzhu Guo, L.Z. Guo, S.A. Billings and Hua-Liang Wei},
title = {An iterative orthogonal forward regression algorithm},
journal = {International Journal of Systems Science},
volume = {46},
number = {5},
pages = {776-789},
year = {2015},
publisher = {Taylor & Francis},
doi = {10.1080/00207721.2014.981237}
}

@article{WANG1996,
title = {Use of PRESS residuals in dynamic system identification},
journal = {Automatica},
volume = {32},
number = {5},
pages = {781-784},
year = {1996},
issn = {0005-1098},
doi = {10.1016/0005-1098(96)00003-9},
author = {Liuping Wang and William R. Cluett}
}

@article{hong2003,
title={Automatic nonlinear predictive model-construction algorithm using forward regression and the PRESS statistic},
author={Hong, X and Sharkey, PM and Warwick, K},
journal={IEE Proceedings-Control Theory and Applications},
volume={150},
number={3},
pages={245--254},
year={2003},
publisher={IET},
doi = {10.1049/ip-cta:20030311}
}

@article{Lang1997,
author = {Zi-Qiang Lang and S. A. Billings},
title = {Output frequencies of nonlinear systems},
journal = {International Journal of Control},
volume = {67},
number = {5},
pages = {713-730},
year = {1997},
publisher = {Taylor & Francis},
doi = {10.1080/002071797223965}
}

@article{Lang1996,
author = {Zi-Qiang Lang and S. A. Billings},
title = {Output frequency characteristics of nonlinear systems},
journal = {International Journal of Control},
volume = {64},
number = {6},
pages = {1049-1067},
year = {1996},
publisher = {Taylor & Francis},
doi = {10.1080/00207179608921674}
}

@article{Lang2005,
author = {Z. Q. Lang and S. A. Billings},
title = {Energy transfer properties of non-linear systems in the frequency domain},
journal = {International Journal of Control},
volume = {78},
number = {5},
pages = {345-362},
year = {2005},
publisher = {Taylor & Francis},
doi = {10.1080/00207170500095759}
}

@book{billings2013a,
citation-number = {24},
author = {Billings, S. A.},
title = {Nonlinear System Identification: NARMAX Methods In The Time, Frequency, And Spatio-Temporal Domains},
volume = {13},
publisher = {John Wiley \& Sons, Ltd},
year = {2013},
issue = {4},
language = {en},
address = {Chichester, UK},
doi = {10.1002/9781118535561}
}

@article{LJUNG2010,
title = {Perspectives on system identification},
journal = {Annual Reviews in Control},
volume = {34},
number = {1},
pages = {1-12},
year = {2010},
doi = {10.1016/j.arcontrol.2009.12.001},
author = {Lennart Ljung}
}

@article{KUGIUMTZIS1996,
title = {State space reconstruction parameters in the analysis of chaotic time series — the role of the time window length},
journal = {Physica D: Nonlinear Phenomena},
volume = {95},
number = {1},
pages = {13-28},
year = {1996},
issn = {0167-2789},
doi = {10.1016/0167-2789(96)00054-1},
author = {D Kugiumtzis}
}

@article{Tan2023,
author = {Tan, Eugene and Algar, Shannon and Corrêa, Débora and Small, Michael and Stemler, Thomas and Walker, David},
title = {Selecting embedding delays: An overview of embedding techniques and a new method using persistent homology},
journal = {Chaos: An Interdisciplinary Journal of Nonlinear Science},
volume = {33},
number = {3},
pages = {032101},
year = {2023},
month = {03},
issn = {1054-1500},
doi = {10.1063/5.0137223}
}

@article{AGUIRRE1996,
author = {Aguirre, Luis Antonio and Mendes, Eduardo M. A. M.},
title = {GLOBAL NONLINEAR POLYNOMIAL MODELS: STRUCTURE, TERM CLUSTERS AND FIXED POINTS},
journal = {International Journal of Bifurcation and Chaos},
volume = {06},
number = {02},
pages = {279-294},
year = {1996},
doi = {10.1142/S0218127496000059}
}

@article{AGUIRRE1995,
author = {Luis A. Aguirre and S. A. Billings},
title = {Improved structure selection for nonlinear models based on term clustering},
journal = {International Journal of Control},
volume = {62},
number = {3},
pages = {569--587},
year = {1995},
publisher = {Taylor \& Francis},
doi = {10.1080/00207179508921557}
}

@article{Aguirre1997,
author = {Aguirre, Luis A. and Rodrigues, Giovani G. and Mendes, Eduardo M. A. M.},
title = {Nonlinear Identification and Cluster Analysis of Chaotic Attractors from a Real Implementation of Chua's Circuit},
journal = {International Journal of Bifurcation and Chaos},
volume = {07},
number = {06},
pages = {1411-1423},
year = {1997},
doi = {10.1142/S0218127497001138}
}

@article{SPINELLI2005,
title = {ON THE ROLE OF PRE-FILTERING IN NONLINEAR SYSTEM IDENTIFICATION},
journal = {IFAC Proceedings Volumes},
volume = {38},
number = {1},
pages = {791-796},
year = {2005},
note = {16th IFAC World Congress},
issn = {1474-6670},
doi = {10.3182/20050703-6-CZ-1902.00133},
author = {W. Spinelli and L. Piroddi and M. Lovera}
}

@article{ANDERSON2007,
title = {Modelling and identification of non-linear deterministic systems in the delta-domain},
journal = {Automatica},
volume = {43},
number = {11},
pages = {1859-1868},
year = {2007},
issn = {0005-1098},
doi = {10.1016/j.automatica.2007.03.020},
author = {S.R. Anderson and V. Kadirkamanathan}
}
\appendix
\section{A brief overview of system identification for nonlinear systems} \label{appndx:Brief_sysID_FRA}
%
\subsection{Nonlinear system identification} \label{sec:NARX}
System identification is a machine learning technique for deriving time-series models that describe the dynamic behaviour of linear and nonlinear systems using experimental input–output data. The resulting dynamic models are typically linear or nonlinear autoregressive with exogenous inputs (\ac{ARX} or \ac{NARX}), meaning that the present output is expressed as a function of past input and output values. System identification serves two main objectives:
\vspace{1ex}
\begin{itemize}
  \item Accurately mapping the input(s) to the output(s) of the system, enabling reliable prediction for new, unseen data.
  \vspace{0.5ex}
  \item Faithfully capturing the underlying system dynamics within the model structure.
\end{itemize}
To meet these objectives, system identification seeks a functional relationship mapping past inputs (input-lagged terms),
\begin{multline}\label{eq:Ut_sysid}
    U = \Big\{ u_1(t-1)\ ,\ \cdots,\ u_1(t-n_{b_1}),\\
    u_2(t-1)\ ,\ \cdots,\ u_2(t-n_{b_2}) \Big\}, \ \
\end{multline}
and past outputs (output-lagged terms), 
\begin{equation}\label{eq:Yt_sysid}
    Y = \Big\{ y(t-1)\ ,\ y(t-2)\ ,\ \cdots,\ y(t-n_a) \Big\}, 
\end{equation}
to the present output in time $y(t)$. Equations \eqref{eq:Ut_sysid} and \eqref{eq:Yt_sysid} are an example of lagged input-output data of a two-input single-output system, where $u_1(t)$ and $u_2(t)$ are two different inputs. $n_a$ is the maximum number of past output time instances (output delays) considered. $n_{b_1}$ and $n_{b_2}$ are the maximum number of past input time instances (input delays) considered for $u_1(t)$ and $u_2(t)$, respectively. These values are related to the Lyapunov exponents of the actual system that is being modelled \cite{mendes1998a}. This functional mapping is described by the equation:
\begin{equation}\label{eq:sys_id_func}
y(t) = f^{N_p}\bigl( Y, U \bigr) + \xi(t),
\end{equation}
where $f^{N_p}( \ )$ is a polynomial function with a maximum polynomial degree $N_p \in \mathbb{Z}^{+}$. $Y$ and $U$ refer to the input-output lagged data, while $\xi(t)$ represents the error between the predicted output $f\bigl( Y, U \bigr)$ and the actual output $y(t)$ at time instance $t$. $\xi(t)$ will contain noise and unmodeled dynamics. In the case of polynomial (N)ARX models, equation \eqref{eq:sys_id_func} can be expressed as
\begin{equation}\label{eq:sys_id_func_summation}
y(t) = \sum_{m=1}^{\mathcal{M}} \theta_{m} \times \phi_{m}(t) + \xi(t),
\end{equation}
where $m = 1, \cdots, \mathcal{M}$, $\mathcal{M}$ being the total number of variables or model terms. $\theta_{m}$ are the model parameters or coefficients and $\phi_{m}(t)$ are the corresponding model terms or variables. $\phi_{m}(t)$ are $n$\textsuperscript{th}-order monomials of the polynomial \ac{NARX} model $f^{N_p}( \ )$, where $n = 1, \cdots, N_p$ is the degree of the monomial. $\phi_{m}(t)$ is composed of past output and input time instances from $Y$ and $U$ respectively. An example of a polynomial \ac{NARX} model can be
\begin{multline}\label{eq:narx_exmpl}
y(t) = \theta_{1}y(t-1) + \theta_{2}u_1(t-2) + \\
\theta_{3}y(t-2)^{2}u_2(t-1)^{3} + \xi(t).
\end{multline}
In the above example, $\phi_{1}(t)=y(t-1)$ and $\phi_{2}(t)=u_1(t-2)$ have a degree of 1 and are linear terms (degree or order of monomial is 1) of the model. While $\phi_{3}(t) = y(t-2)^{2}u_2(t-1)^{3}$ is a nonlinear term with a degree of $5$ (degree or order of monomial is 5). The \ac{NARX} model shown in equation \eqref{eq:narx_exmpl} has a polynomial degree $N_p=5$ (highest degree of any monomial). Given that the total number of time samples available is $L$, where $t = 1, \cdots, L$. Equation \eqref{eq:sys_id_func_summation} can be represented in matrix form as
\begin{equation}\label{eq:sys_id_func_mat}
\mathbf{Y} = \mathbf{\Phi} \mathbf{\Theta} + \mathbf{\Xi},
\end{equation}
where $\mathbf{Y} = \left[ y(1), \cdots, y(L) \right]^T$ is the vector containing the output samples $y(t)$. $\mathbf{\Phi} = \left[ \bar{\phi}_{1}, \cdots, \bar{\phi}_{\mathcal{M}} \right]$, where $\bar{\phi}_{m} = \left[ \phi_{m}(1), \cdots, \phi_{m}(L) \right]^T$ is the vector containing all time samples of the model variable $\phi_{m}(t)$. $\mathbf{\Theta} = \left[ \theta_{1}, \cdots, \theta_{\mathcal{M}}  \right]^T$  is the parameter vector and $\mathbf{\Xi} = \left[ \xi(1), \cdots, \xi(L) \right]$ is the vector containing all the error terms $\xi(t)$. 

The main concern in learning a \ac{NARX} model is to identify the polynomial structure of the model, i.e. from a set of candidate model terms (monomials), $D$, which terms should be included in the model. For example, a possible set of candidate terms could be
\begin{flalign}\label{eq:exmpl_D}
    \begin{aligned}
        D = \ 
        \Big\{ 
              &y(t-1), y(t-2), u_1(t-1), u_1(t-2), \\
              &y(t-1)u_1(t-2), y(t-2)u_2(t-1)^{3}, \\
              &y(t-2)^{2}u_1(t-1), y(t-2)^{2}u_2(t-1)^{3}
        \Big\} ,
    \end{aligned}
\end{flalign}
from which a \ac{NARX} model structure such as in equation \eqref{eq:narx_exmpl} could be identified. Once the model structure is identified, the parameters of the model need to be estimated. Given the input-output data of a dynamic system that is to be modelled, the model structure has to be determined so that a parsimonious model is obtained. This is important, especially in the nonlinear case, so that the model identified, because of the unnecessary model terms, does not contain any dynamics that should not be there \cite{mendes1998a}. In addition, a model should generalise well to data (validation) that is not used during the learning or training process (model identification process). This is known as obtaining a bias-variance compromise and is achieved through an appropriate cross-validation strategy \cite{Little2017,Stone1974}. For this purpose, we use the \ac{iFRO} algorithm \cite{guo2015a} for model term selection with the \ac{PRESS} \cite{WANG1996, hong2003} as the term selection criterion. 

As mentioned in Section \ref{sec:canon_apprx_sysid}, in this study, input-only second-order \ac{NARX} models are used for the canonical approximation of \ac{PAC}. This means the maximum polynomial order considered is $N_p=2$ and only past input-lagged terms (equation \eqref{eq:Ut_sysid}) are used to identify the \ac{NARX} model.
\subsection{System identification algorithm} \label{appndx:SysID_alg}
The \ac{iFRO} algorithm \cite{guo2015a} is capable of evaluating the contribution of each candidate model term to the output, independently of the influence exerted by other terms. This property enables efficient term selection within a forward-selection framework, in which model terms are incorporated sequentially according to a specified selection criterion, thereby facilitating the construction of a globally optimal parsimonious model \cite{guo2015a}. When combined with the \ac{PRESS} statistic \cite{WANG1996, hong2003}, \ac{iFRO} preferentially selects model terms that minimise the predicted leave-one-out cross-validation error in a forward-selection manner, thus maximising generalisation performance without the need for a dedicated validation dataset \cite{WANG1996, hong2003}. In addition, the final identified model must be stable in simulation to ensure reliable dynamical reproduction. 

\textit{NonSysID} \cite{Gunawardena2024,nonsysid} is an open-source software framework for system identification employing the \ac{NARX} model structure, integrating \ac{iFRO} with \ac{PRESS}-based term selection. Moreover, \textit{NonSysID} incorporates the procedures outlined in \cite{Wei2004}, which expedite the identification process by initially estimating an \ac{ARX} model and subsequently using its terms to construct candidate nonlinear terms for \ac{NARX} model identification. This strategy is theoretically appropriate and accelerates the canonical approximation process (Section \ref{sec:canon_apprx_sysid}), as elaborated below.

As detailed in Section \ref{sec:Dynm_sys_PAC} and further examined in Section \ref{sec:PAC_models}, the slow and fast oscillations, $x(t)$ and $h(t)$ respectively, interact nonlinearly to generate the amplitude modulation $y(t)$ (see equations \eqref{eq:simple_PAC} and \eqref{eq:nonsine_PAC}). In accordance with \cite{Wei2004}, an initial \ac{ARX} model can be identified to capture the independent dynamics of $x(t)$ and $h(t)$. The identified terms are then multiplicatively combined to form the 2\textsuperscript{nd}-order candidate nonlinear terms, which serve as the basis for identifying a \ac{NARX} model that encapsulates the nonlinear dynamics governing $y(t)$ (Fig. \ref{fig:mthd_exmpl}). Owing to these methodological features, the \textit{NonSysID} package \cite{Gunawardena2024,nonsysid} is particularly well-suited for the canonical approximation of \ac{PAC} via system identification. For the initial \ac{ARX} identification, term selection was stopped using a \ac{PRESS}-based threshold of $10^{-3}$, while the subsequent \ac{NARX} model size was determined automatically from the minimum \ac{PRESS} criterion.

For the input-only model structures used in this study, identification is performed using \textit{NonSysID-i}, a dedicated variant of \textit{NonSysID} that retains the model-term selection framework described above. Its candidate regressors are constructed exclusively from lagged inputs and their nonlinear combinations, with no lagged output terms included.
%
\section{Taylor approximations of complex \ac{PAC}-generating mechanisms}\label{appndx:taylor_approx}
This section shows how complex \ac{PAC}-generating mechanisms, such as non-sinusoidal amplitude modulation (equation \eqref{eq:nonsine_PAC}) and the \ac{NMM} (equation \eqref{eq:NMM_E}), implicitly contain the basic \ac{PAC} model (equation \eqref{eq:simple_PAC}). This is demonstrated by applying Taylor approximations to the governing equations of these complex mechanisms.
\subsection{Taylor approximation of equation \eqref{eq:nonsine_PAC}}\label{appndx:taylor_approx_nonsine}
Using a Taylor expansion of the shifted and scaled sigmoid function about $x(t)=0$, equation \eqref{eq:nonsine_PAC} can be written as
\begin{equation}
     y(t)\approx \Big( 1- q_0 - q_1x(t) - q_2x^2(t) - \cdots \Big) h(t)
\end{equation}
where
\begin{flalign}
    \begin{aligned}
        &q_0 = \frac{1}{1+e^{\alpha c}} \\
        &q_1 = \alpha \frac{e^{\alpha c}}{(1+e^{\alpha c})^2} \\
        &q_2 = \alpha^2 \frac{e^{\alpha c}(e^{\alpha c}-1)}{2(1+e^{\alpha c})^3}
    \end{aligned}
\end{flalign}
Therefore, considering a first-order Taylor series approximation, equation \eqref{eq:nonsine_PAC} can be reduced to a form similar to equation \eqref{eq:simple_PAC}, as shown below
\begin{equation}
     y(t)\approx \Big( (1- q_0) - q_1x(t) \Big) h(t)
\end{equation}
Therefore, substituting this into equation \eqref{eq:PAC_math}, the first-order Taylor approximation of \ac{PAC} with non-sinusoidal amplitude modulation can be expressed as 
\begin{equation}\label{eq:taylor_approx_nonsine_PAC}
    \overline{z}(t) = x(t) + \Big( (1- q_0) - q_1x(t) \Big) h(t),
\end{equation}
which has a similar form to the most basic \ac{PAC} model shown in equations \eqref{eq:PAC_math} and \eqref{eq:simple_PAC}.

\subsection{Taylor approximation of equation \eqref{eq:NMM_E}}\label{appndx:taylor_approx_NMM}
A Taylor expansion of the shifted and scaled sigmoid function, $f(z)$, in equation \eqref{eq:NMM_sig}, about $z=0$, is given as
\begin{equation}\label{eq:taylor_approx_fz}
f(z)\approx c_0 + c_1 z + c_2 z^2 + c_3 z^3 + \cdots
\end{equation}
where
\begin{flalign}
    \begin{aligned}
        &c_0 = \frac{1}{1+e^{b}} \\
        &c_1 = b \ c_0(1-c_0) \\
        &c_2 = \frac{b^2}{2} \ c_0(1-c_0)(1-2c_0) \\
        &c_3 = \frac{b^3}{6} \ c_0(1-c_0)(1-6c_0+6(c_0)^2).
    \end{aligned}
\end{flalign}
Using a second-order Taylor approximation, equation \eqref{eq:NMM_E} can be written as
\begin{equation}\label{eq:NMM_E_taylor}
    \tau_E \dot E(t)\approx -E(t) + c_0 + c_1 z_E(t) +  c_2 {z_E}^2(t),
\end{equation}
where $z_E(t) = x_E(t) + w_{EE}E(t) - w_{IE}I(t)$. Denoting $h(t) = w_{EE}E(t)-w_{IE}I(t)$, equation \eqref{eq:NMM_E_taylor} can be expressed as
\begin{equation}\label{eq:NMM_E_taylor_2}
    \begin{aligned}
        \tau_E \dot{E}(t) \approx & -E(t) + c_0 + c_1x_E(t) + \Big( c_1 + 2 c_2 x_E(t) \Big)h(t)\\
        &+ c_2\Big({x_E}^2(t) + h^2(t) \Big)
    \end{aligned}
\end{equation}
The corresponding terms from equation \eqref{eq:NMM_E_taylor_2}, 
\begin{equation}\label{eq:taylor_approx_NMM_PAC}
    \overline{z}(t) = c_1x_E(t) + \Big( c_1 + 2 c_2 x_E(t) \Big)h(t)
\end{equation}
have a similar form to the most basic \ac{PAC} model shown in equations \eqref{eq:PAC_math} and \eqref{eq:simple_PAC}. 

The \ac{NMM} defined by equations \eqref{eq:NMM_E}–\eqref{eq:NMM_sig} (as illustrated in Fig. \ref{fig:pac_decom}) can be decomposed (using the Taylor approximation in \eqref{eq:NMM_E_taylor_2}) to separate the slow oscillation dynamics, $x(t)$, from the amplitude-modulated fast oscillation dynamics, $y(t)$, such that
\begin{equation}
    x(t) \approx c_1 x_E + c_2 {x_E}^2(t) + \cdots 
\end{equation}
and
\begin{equation}
    y(t) \approx c_1 h(t) + c_2 h^2(t) + 2 c_2 x_E(t) h(t) + \cdots.
\end{equation}
The above decomposition becomes more accurate when higher-order terms from the Taylor expansion of $f(z)$ are included.
%
\section{Supplementary figures} \label{appndx:extra_plts}
This appendix includes all the supplementary figures used in this paper.
\clearpage

\begin{figure*}
  \centering
  \includegraphics[width=\textwidth]{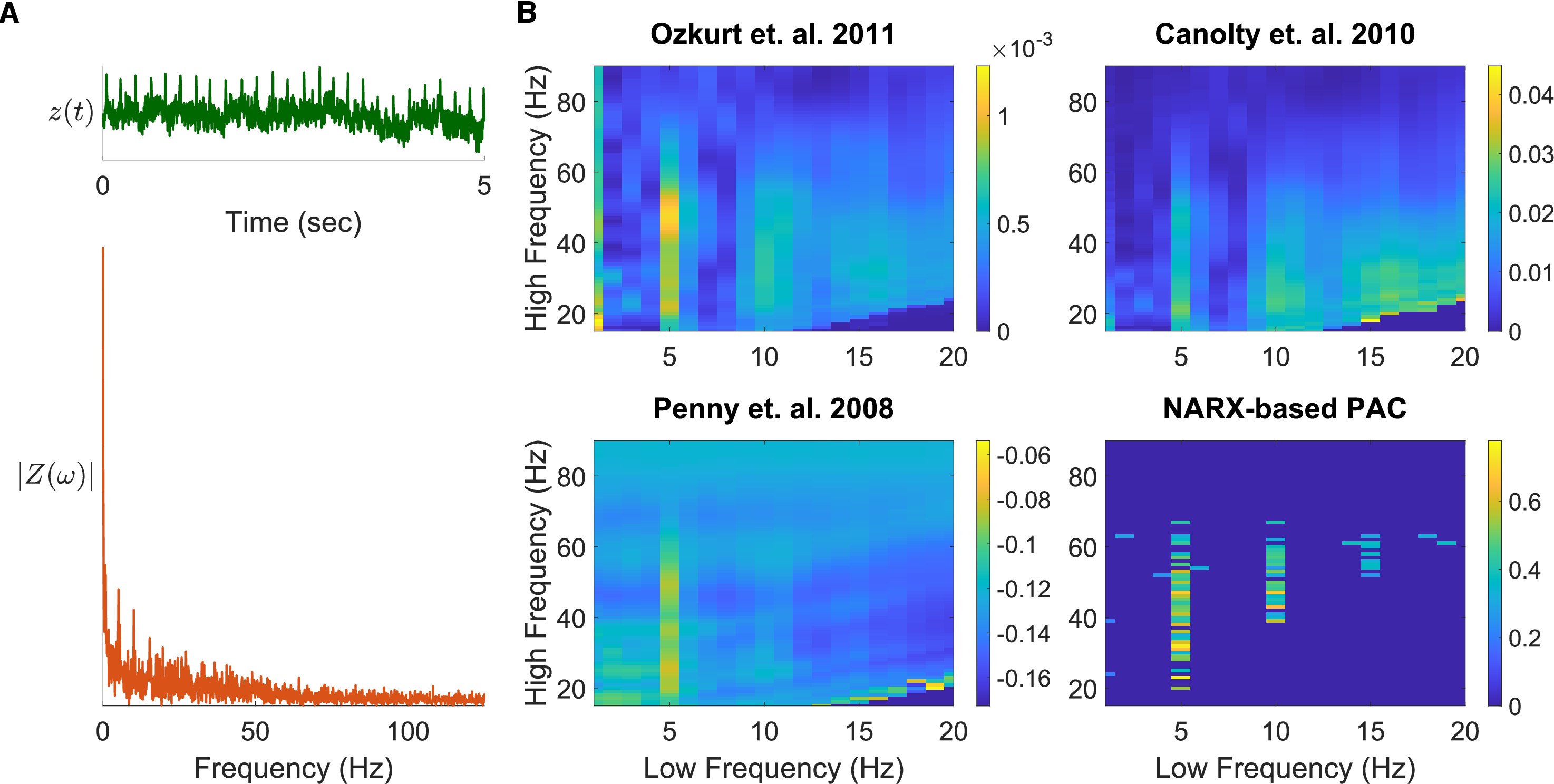} 
  \caption{\textbf{Comparison using a simulated periodic spike-train signal that produces spurious \ac{PAC}}. Periodic spike trains are phase-locked to lower frequencies. Using the procedures described in \cite{Gerber2016}, a 20-second Gaussian-shaped periodic spike-train signal $z(t)$, phase-locked to 5 Hz, is used here to compare how the proposed method represents spike-related spurious \ac{PAC} relative to conventional methods. \textbf{A} shows a snippet of the signal $z(t)$ and its magnitude spectrum $\left| Z(\omega) \right|$. Comparisons between the proposed \ac{NARX}-\ac{PAC} method and other methods are shown in \textbf{B} using the comodulogram evaluated from each method. As seen from the comodulograms, periodic spikes produce a characteristic pattern \cite{Gerber2016}, and this pattern is more clearly resolved in the proposed \ac{NARX}-based \ac{PAC} method. This clearer representation can support diagnostic identification and future rejection procedures for spike-related spurious \ac{PAC}, although a formal spike-train rejection rule is not introduced here. Note that the comodulogram from the method by Penny \textit{et al.} is shown on a natural logarithmic scale.}
  \label{fig:spike_train_cmd}
\end{figure*}
\begin{figure*}
  \centering
  \includegraphics[width=\textwidth]{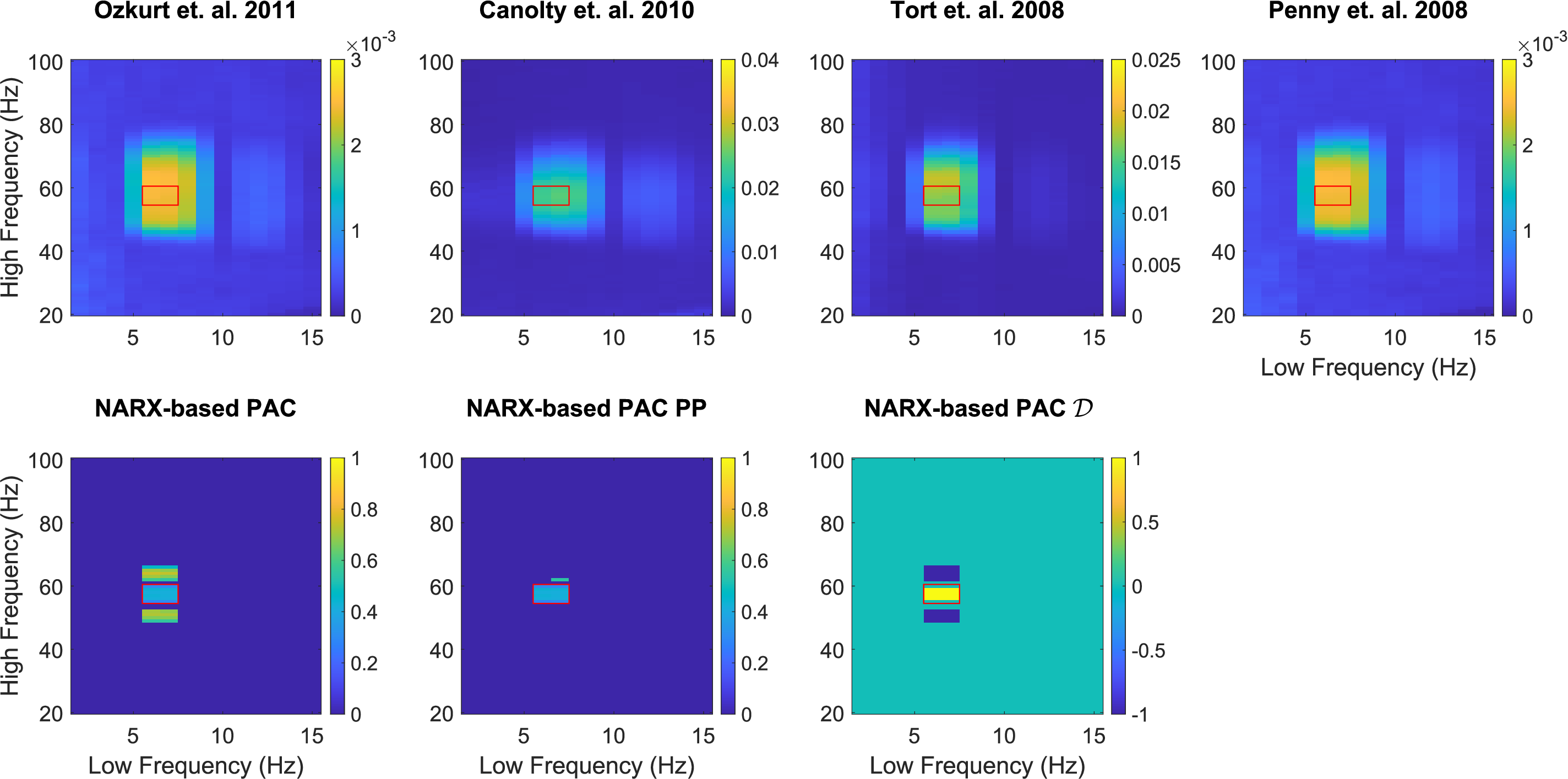} 
  \caption{\textbf{Median of 100 samples: Non-stationary slow oscillation (6–7 Hz) coupled to a non-stationary fast oscillation (55–60 Hz) with a \ac{SNR} of 3 ($9.54\,\mathrm{dB}$).}
  \ac{SNR} is expressed as a ratio of standard deviations.
  This figure compares the median comodulograms obtained from 100 \ac{PAC} signal realisations with identical coupling dynamics. Each \ac{PAC} signal was constructed using a different pink-noise sample (Fig. \ref{fig:pink_pac}). The red box marks the region of the comodulograms where \ac{PAC} is present. Comparisons with alternative methods demonstrate the repeatability and specificity of the proposed \ac{NARX}-\ac{PAC} method. The last row shows the median comodulograms from the proposed method (left to right): the raw output, the post-processed (PP) result with intermodulation-related spurious \ac{PAC} suppressed, and the corresponding discriminator map, $\mathcal{D}$ (equation \eqref{eq:commod_diff}).}
  \label{fig:rng_PinkNoise_cmd_stat}
\end{figure*}
\begin{figure*}
  \centering
  \includegraphics[width=\textwidth]{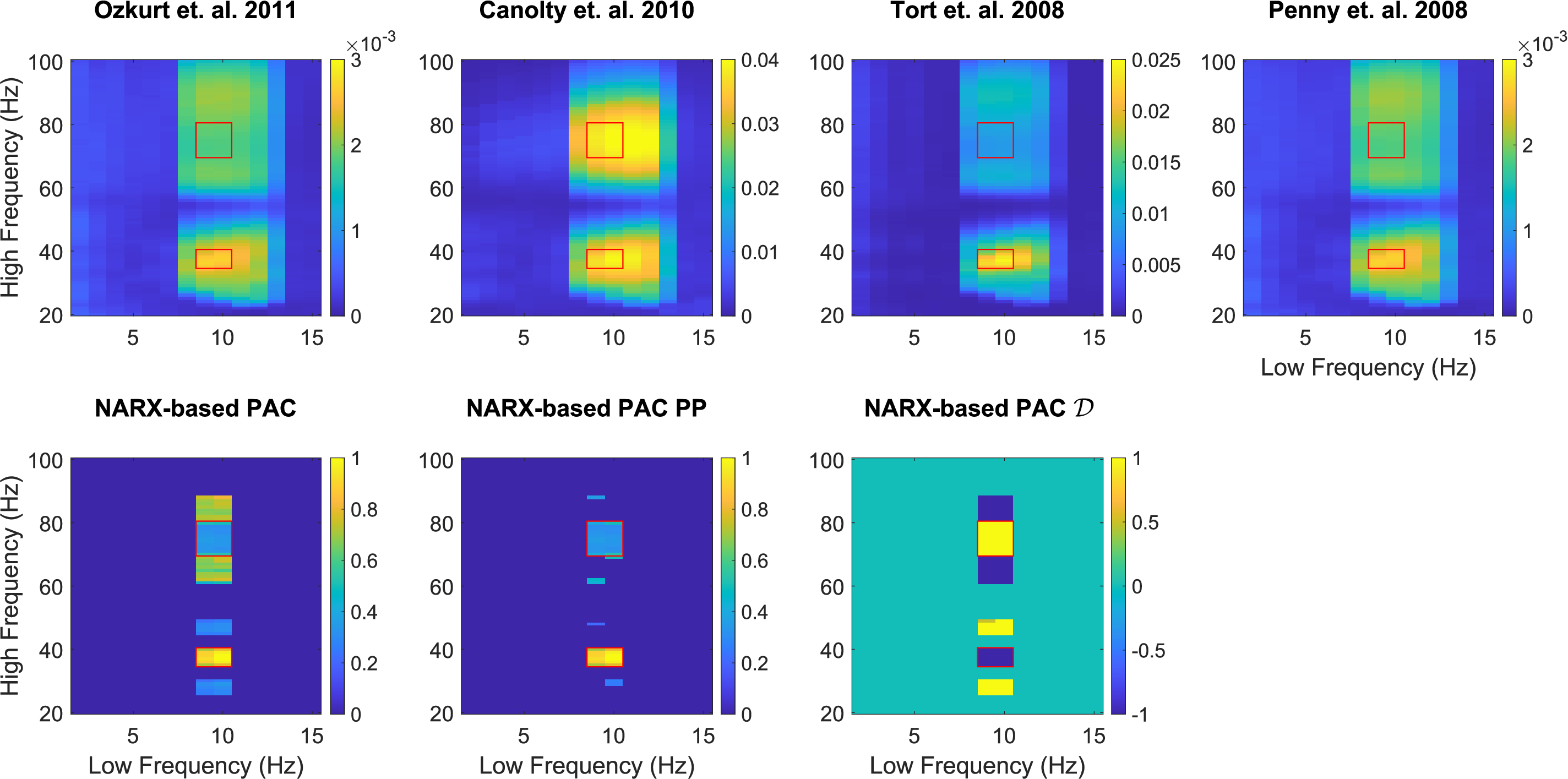} 
  \caption{\textbf{Median of 100 samples: Non-stationary slow oscillation (9–10 Hz) coupled to two non-stationary fast oscillations (35–40 Hz and 70–80 Hz) with a \ac{SNR} of 3 ($9.54\,\mathrm{dB}$).}
  \ac{SNR} is expressed as a ratio of standard deviations.
  This figure compares the median comodulograms obtained from 100 \ac{PAC} signal realisations with identical coupling dynamics. Each \ac{PAC} signal was constructed using a different pink-noise sample (Fig. \ref{fig:pink_pac}). The red boxes mark the regions of the comodulograms where \ac{PAC} is present. Comparisons with alternative methods demonstrate the repeatability and specificity of the proposed \ac{NARX}-\ac{PAC} method. The last row shows the median comodulograms from the proposed method (left to right): the raw output, the post-processed (PP) result with intermodulation-related spurious \ac{PAC} suppressed, and the corresponding discriminator map, $\mathcal{D}$ (equation \eqref{eq:commod_diff}).}
  \label{fig:rng_PinkNoise_multpl_cmd_stat}
\end{figure*}
\begin{figure*}
  \centering
  \includegraphics[width=\textwidth]{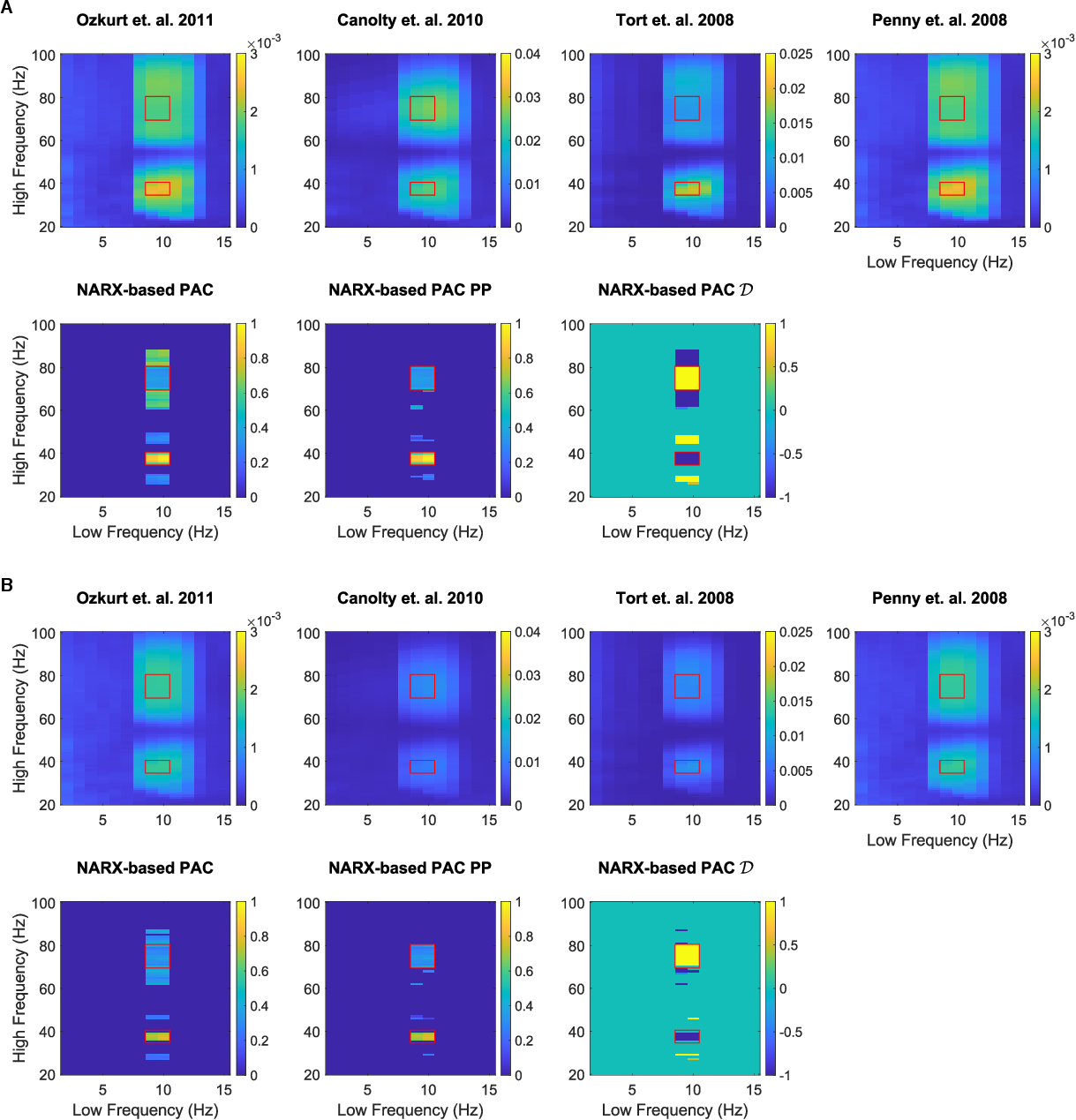} 
  \caption{\textbf{Median of 100 samples: Non-stationary slow oscillation (9--10 Hz) coupled to two non-stationary fast oscillations (35--40 Hz and 70--80 Hz), robustness to higher noise, \ac{SNR} 2 ($6.02\,\mathrm{dB}$) and \ac{SNR} 1 ($0\,\mathrm{dB}$).} \ac{SNR} is expressed as a ratio of standard deviations. Panels \textbf{A} and \textbf{B} are similar to Figure \ref{fig:rng_PinkNoise_multpl_cmd_stat}, but with higher levels of pink noise, showing median results for \ac{SNR} 2 and \ac{SNR} 1, respectively. Red boxes mark regions where \ac{PAC} is present. Compared with conventional methods, the proposed \ac{NARX}-\ac{PAC} method remains robust at \ac{SNR} 2 in terms of modulation-index consistency and comodulogram sharpness and accuracy, and remains fairly robust even at \ac{SNR} 1. In both panels, the last row shows the median comodulograms from the proposed method (left to right): the raw output, the post-processed (PP) result with intermodulation-related spurious \ac{PAC} suppressed, and the corresponding discriminator map, $\mathcal{D}$ (equation \eqref{eq:commod_diff}).}
  \label{fig:rng_PinkNoise_multpl_noise_cmd_stat}
\end{figure*}
\begin{figure*}
  \centering
  \includegraphics[width=\textwidth]{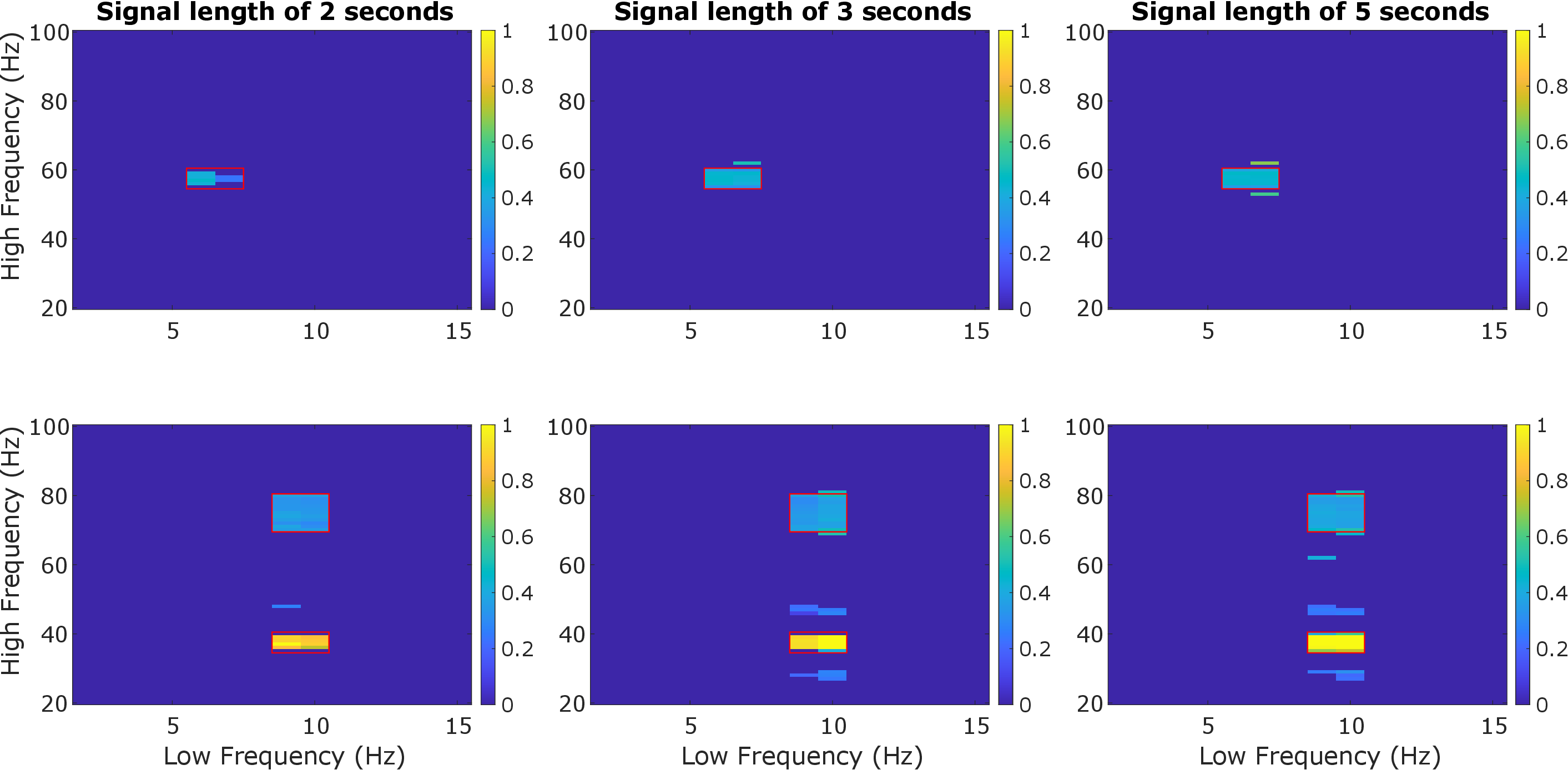} 
  \caption{\textbf{Performance on short signal lengths.}
  \ac{SNR} is expressed as a ratio of standard deviations.
  The top row corresponds to a non-stationary slow oscillation (6--7 Hz) coupled to a non-stationary fast oscillation (55--60 Hz). The bottom row corresponds to a non-stationary slow oscillation (9--10 Hz) coupled to two non-stationary fast oscillations (35--40 Hz and 70--80 Hz). This figure illustrates the performance of the proposed \ac{NARX}-\ac{PAC} method for short signal lengths: 2, 3, and 5 seconds. Results are shown for the same two synthetically generated \ac{PAC} signals presented in Section \ref{sec:Results} (Figs. \ref{fig:rng_PinkNoise_cmd} and \ref{fig:rng_PinkNoise_multpl_cmd}, and supplementary Figs. \ref{fig:rng_PinkNoise_cmd_stat} and \ref{fig:rng_PinkNoise_multpl_cmd_stat}). As in the previous settings, pink noise was added to the generated \ac{PAC} signals to obtain an \ac{SNR} of 3 ($9.54\,\mathrm{dB}$). The panels show median comodulograms computed from 100 realisations of each \ac{PAC} signal with identical coupling dynamics, with each realisation generated using a different pink-noise sample. The comodulograms from \ac{NARX}-\ac{PAC} are the post-processed result with intermodulation-related spurious \ac{PAC} suppressed (Section \ref{sec:avoid_iSC})}
  \label{fig:rng_PinkNoise_multpl_cmd__TmWndw_stat}
\end{figure*}
%
\end{document}